\documentclass{article}
\usepackage{amssymb}

\def\dlbrack{\mathopen{[\mkern-8mu[ \ }}
\def\drbrack{\mathclose{\ ]\mkern-8mu] \,}}

\def\lbracksq{\mathopen{[\mkern-10mu\{ \ }}
\def\rsqbrack{\mathclose{\ \}\mkern-10mu] \,}}

\newcommand{\evaluation}[3]{\dlbrack #1 \drbrack^{\cal #2}_{#3}}
\newcommand{\eval}[3]{\dlbrack #1 \drbrack^{#2}_{#3}}

\newcommand{\modelsem}[1]{\lbracksq #1 \rsqbrack_{LM}}
\newcommand{\semantics}[1]{\lbracksq #1 \rsqbrack_{T}}
\newcommand{\context}[1]{\,\dlbrack #1 \drbrack}

\newcommand{\modelsetsem}[1]{\lbracksq #1 \rsqbrack_{M}}
\newcommand{\cmodelsetsem}[1]{\lbracksq #1 \rsqbrack_{CM}}
\newcommand{\dcmodelsetsem}[1]{\lbracksq #1 \rsqbrack_{D}}
\newcommand{\tidsemantics}[1]{\lbracksq #1 \rsqbrack_{T\sqcup I}}

\newcommand{\ideal}[1]{\langle #1 \rangle}

\newcommand{\function}[3]{#1(\overline{#2})}
\newcommand{\cons}[1]{DS^{#1}_{\Sigma}}
\newcommand{\funs}[1]{FS^{#1}_{\Sigma}}

\newcommand{\alg}[1]{{\cal #1}}
\newcommand{\prg}[1]{{\cal #1}}

\newcommand{\es}{\rightarrow}
\newcommand{\si}{\Leftarrow}

\newtheorem{theorem}{Theorem}[section]
\newtheorem{definition}[theorem]{Definition}
\newtheorem{proposition}[theorem]{Proposition}

\newtheorem{lemma}[theorem]{Lemma}
\newtheorem{corollary}[theorem]{Corollary}

\newtheorem{example}[theorem]{Example}
\newenvironment{proof}{\noindent {\bf Proof.}\\}{\hfill $\square$}

\title
     { Composing Programs in a Rewriting Logic for Declarative Programming
     }

\author
      { J.M. MOLINA-BRAVO
        and
        E. PIMENTEL\\
        Dpto. Lenguajes y Ciencias de la Computaci\'{o}n. \\
        University of M\'{a}laga. Campus de Teatinos. 29071 M\'{a}laga. Spain.\\
        \{jmmb, ernesto\}@lcc.uma.es }

\date{}

\begin{document}

%\title{Modularity in a Rewriting Logic for Declarative
%Programming\thanks{ The authors have been partially supported by
%the Spanish CICYT (project TIC 95-0445-C03-03 ``TREND'').}}
%
%\author{Juan Miguel  Molina-Bravo \and Ernesto Pimentel}
%
%\affiliation{Dpto.\/ Lenguajes y Ciencias de la Computaci\'{o}n.
%    University of M\'{a}laga. Campus de Teatinos. 29071 M\'{a}laga. SPAIN.
%{\bf E-mail}: \{jmmb, ernesto\}{\tt @}{lcc.uma.es}\\ \hbox{$\ $} }

\maketitle
%\bibliographystyle{tlp}
% ====================================================================

\begin{abstract}
Constructor-Based Conditional Rewriting Logic is a general
framework for integrating first-order functional and logic
programming which gives an algebraic semantics for
non-deterministic functional-logic programs.
 In the context of this formalism, we introduce a simple notion of program module
as an open program which can be extended together with several mechanisms to combine
them. These mechanisms are based on a reduced set of operations.
% such that none of them can be expressed in terms of the remaining ones.
However, the high expressiveness of
these operations enable us to model typical constructs for program modularization like
hiding, export/import, genericity/instantiation, and inheritance in a simple way.  We
also deal with the semantic aspects of the proposal by introducing an immediate
consequence operator, and studying several alternative semantics for a program module,
based on this operator, in the line of logic programming: the operator itself, its least
fixpoint (the least model of the module), the set of its pre-fixpoints (term models of
the module), and some other variations in order to find a compositional and fully
abstract semantics wrt\/ the set of operations and a natural notion of observability.


\end{abstract}

% ===================================================================

\section{Introduction}

Constructor-Based Conditional Rewriting Logic (CRWL)\footnote{CRWL must not be confused
with the Rewriting Logic proposed in \cite{meseguer} as a unifying logical framework for
concurrency. CRWL is a particular logic for dealing with indeterminism.}, presented in
\cite{mario:jlp}, is a quite general approach to declarative programming that combines
(first-order) functional and logic paradigms by means of the notion of (possibly) {\it
non deterministic lazy function}. The basic idea is that both relations and deterministic
lazy functions are particular cases of non-deterministic lazy functions. This approach
retains the advantages of deterministic functions while adding the possibility of
modeling non-deterministic functions by means of {\it non-confluent constructor-based
term rewriting systems}, where a given term may be rewritten to constructor terms
(possibly with variables) in more than one way.  Here a fundamental notion is that of
{\it joinability}: two terms {\it a},{\it b} are joinable iff they can be rewritten to a
common
---but not necessarily unique--- constructor term.  In
\cite{mario:jlp}, CRWL is introduced with two equivalent proof calculi that govern
deduction in this logic, an algebraic semantics for programs (theories) based on a freely
generated model, and an operational semantics, based on a lazy narrowing calculus for
solving goals, that is sound and complete wrt the algebraic semantics.

Modularity is a central issue in all programming paradigms motivated by the need of
mastering the complexity inherent in large programs. Modularity related with algebraic
specifications (which, to some extent, can be viewed as a sort of first-order functional
programming) has been extensively studied and all specification languages are extended
for dealing with modules. In this field, a typical module consists of a body,
% which contains the constructions of its own resources,
an export interface,
% which contains those resources realized by the module that can be
% used by other modules,
a list of imports
% which corresponds with concrete resources which are to be provided
% by other modules and used in the module body,
and, possibly, a list of formal parameters,
% which also corresponds with resources which are to be provided by other modules, but are not fixed.
and typical operations with modules have to do with setting up hierarchical relationships
between modules as the union of modules (with some constraints) and the application of a
parameterized module to an actual module, and their semantics are given from a
category-theoretic point of view \cite{goguen/burstall92,ehrig,orejas99}. Nevertheless,
there are other studies of modularity \cite{wirsing} with more flexible sets of
operations semantically defined by means of operations on the sets of models, and also
studies where modularity has been tackled with the tools of algebraic specifications as
\cite{bergstra} where an axiomatic specification is given for an algebra of
non-parameterized modules and it is proved that each expression can be reduced to another
one with, at most, an occurrence of the export (hiding) operator, and \cite{duran99}
where a constructive specification is given for an algebra of parameterized modules
(without hiding) in Maude, and each expression is reduced to a flat module.

In the logic programming field, modularity has been the objective of different proposals
---see \cite{bugliesi93} for a survey about the subject--- which basically have followed
two different guidelines. One, focused on programming-in-the-large, extends logic
programming with modular constructs as a meta-linguistic mechanism \cite{brogi:TCS} and
gives semantics to modules with the aid of the immediate consequence operator. And the
other one, focused on programming-in-the-small, enriches the theory of Horn clauses with
new logical connectives for dealing with modules \cite{miller}.
In the first line, there is the work \cite{brogi:tesis} where an algebra of
logic programs is studied. This algebra is based on three basic operations
(union, intersection and encapsulation) defined at the semantic level and then
translate to the syntactic level. It is proved that each program expression is
equivalent to a, possibly infinite, flat program, and also a transformation is
defined for mapping program expressions into finite programs by introducing
system generated predicates and adding a hidden part to each program. Notions
of module hiding some predicates and module importation are built up with the
aid of the basic operations.

On the other hand, in functional-logic programming we do not know any study of modularity
semantically well founded. With this paper we have tried to contribute to filling this
gap at least in the CRWL context. In this context, we deal with data constructors, as in
logic programming, and functions defined by conditional rewrite rules, instead of
predicates defined by Horn clauses, and we have proved (see section~\ref{operator}) that
an operator, similar to the immediate consequence operator of logic programs, can be
defined to each CRWL-program and its least fixpoint coincides with the freely generated
term-model given in \cite{mario:jlp}. All this has motivated our decision of developing a
study of programs structuring and modularity in CRWL, based on a meta-linguistic
mechanism, similar to the one which appears in \cite{brogi:tesis}. However we have
defined an algebra of program modules based on a different set of operations (union,
deletion of a signature of function symbols, closure wrt a signature and renaming)
defined at the syntactical level in such a way that each program expression can be
reduced to a, possibly infinite, flat program. With these operations we can model as well
as notions of module which hides some functions and module importation, module
parameterization, instantiation and inheritance with overriding. Also, we have introduced
a notion of protected signature labeling symbols with module expressions, which allows to
define structured modules and a representation morphism that maps each program expression
into a finite structured module. We use protected signature, not only for hiding
functions as is done in \cite{brogi:tesis} for predicates, but also for hiding data
constructors.

% we have developed an approach to modularity based on a meta-linguistic mechanism to
%compose modules, distinguishing the module definition from the module composition in two
%different levels, following the current trends in component-based software engineering.

An important aspect to be considered when a language is extended for modular programming
is the sound integration of the behavior of the modular operations into the semantics of
the language.
% Hence, we study the algebraic properties of these operations and some
%semantics for modules based on an immediate consequence operator, looking for
%compositionality and full abstraction.
The compositionality of the semantics of a programming language is particularly relevant
when modularity is involved. In fact, one of the most critical aspects in modular systems
is the possibility of making a separate compilation of modules, and this can only be made
in the presence of this property.
%Thus, we look for a {\it reasonable} semantics for our
%approach (i.e., which captures in a natural way the meaning of components constructed
%from other components), but also making it possible to obtain the meaning of a composed
%module only from the meaning of its subcomponents.
On the other hand, full abstraction measures the implementation details of the semantics
of a programming language. A non-fully abstract semantics makes the intended meaning of a
program to include non relevant aspects, which do not depend on the behavior of the
program but on a particular ``implementation''. In some sense, full abstraction can be
seen as the complementary property of compositionality, and the adequacy of a semantics
is established when both full abstraction and compositionality are obtained.
In \cite{brogi:tesis}, the semantics of a program is given by its immediate
consequence operator which captures the information concerning possible
compositions, this semantics is compositional by construction and it is proved
that also is fully abstract wrt a notion of observable behavior given by the
success sets of programs (least fixpoints of their immediate consequence
operators).
In CRWL-programming, the semantics given by the immediate consequence operator is
compositional but not fully abstract when we take the freely generated term-model as
observable behavior. For this reason, we study several alternative semantics to find one
that is compositional and fully abstract.

We are confident that our work could serve as a reference to other studies of modularity
in functional-logic programming, and, although we are focused on the modular aspects of
the semantics, the results obtained in this paper, as well as the study of a wide range
of other issues concerning semantics, makes the current work also relevant from a purely
semantic point of view, in the context of rewriting logic-based programming languages.
The approach to modularity in CRWL-programming, that we present here,
substantially extends a previous one in \cite{molina} with a more elaborate
notion of program module and a new operation (renaming) that makes clear the
difference between importation and instantiation, and a more recent one
\cite{molina2000} with the notions of structured module and module
representation that allows to express closed modules by means of a finite
number of rules and also to deal with local constructor symbols.

%We also compare different semantics, studying their compositionality and full
%abstraction, and finally obtaining one which satisfies both properties.

The paper is organized as follows: In the next section we introduce the basic features of
the CRWL approach to functional-logic programming and its model-theoretic semantics
---for a detailed presentation we refer to \cite{mario:jlp}.
 In Section~\ref{fixpoint} we introduce an
immediate consequence operator ${\cal T_{R}}$, for each CRWL-program ${\cal R}$, and a
fixpoint semantics that matches the free term-model ${\cal M_{R}}$ proposed in
\cite{mario:jlp}. In Section~\ref{modules} we define a notion of (plain) module together
with a reduced set of operations on program modules, and we express some modular
constructions with these operations.  In Section~\ref{compositionalSem} we give the
$\alg{T}$-semantics that characterizes the meaning of a CRWL-program when we consider
composition of programs and prove that this semantics is compositional but not fully
abstract wrt the set of operations, taking ${\cal M_{R}}$ as the observable behavior of a
program ${\cal R}$.  In Section~\ref{fullabstractSem} we introduce a fully abstract
semantics by denoting a program module with the set of all its consistent term-models
(pre-fixpoints of ${\cal T_{R}}$); but this semantics is not compositional for the
deletion of a signature. In Section~\ref{comfullabstractSem}, we obtain a compositional
and fully abstract semantics as an indexed family of sets of consistent term-models for
single function. In Section~\ref{sec.hiddensymbols}, we introduce the notion of
structured module as a finite representation of expressions made up from finite plain
modules that allows the hiding of constructor symbols. Finally we present a discussion
and some conclusions.

% ===================================================================

\section{CRWL for Declarative Programming}
\label{CRWLapproach}

%In this section we briefly introduce the proof-theoretical presentation of CRWL given in
%\cite{mario:jlp}.

% ....................................................................

\subsection{Signatures, terms and formulas}

A {\it signature with constructors} is a pair $\Sigma = (DS_{\Sigma}, FS_{\Sigma})$,
where $DS_{\Sigma}$ and $FS_{\Sigma}$ are countable disjoint sets of strings $h/n$ with
$n \in {\bf N}$. Each $c$ such that $c/n \in DS_{\Sigma}$ is a {\it constructor symbol}
with arity $n$ and each $f$ such that $f/n \in FS_{\Sigma}$ is a {\it (defined) function
symbol} with arity $n$. The set of all constructor symbols and the set of all function
symbols with arity $n$ are denoted by $DS^{n}_{\Sigma}$ and $FS^{n}_{\Sigma}$,
respectively. Given a signature (with constructors) $\Sigma$ and a set ${\cal V}$ of
variable symbols, disjoint from all of the sets $DS_{\Sigma}^{n}$ and $FS_{\Sigma}^{n}$,
we define {\it $\Sigma$-terms} as follows: each symbol in ${\cal V}$ and each symbol in
$DS_{\Sigma}^{0}\cup FS_{\Sigma}^{0}$ is a $\Sigma$-term, and for each $h \in
DS_{\Sigma}^{n} \cup FS_{\Sigma}^{n}$ and $t_{1}, \ldots, t_{n}$ terms, $h(t_{1}, \ldots,
t_{n})$ is a term. ${\bf Term}_{\Sigma}$ is the set of all $\Sigma$-terms and ${\bf
CTerm}_{\Sigma}$ the subset of those $\Sigma$-terms (called {\it constructor terms})
built up only with symbols in $DS_{\Sigma}$ and ${\cal V}$. In order to cope with partial
definition we add a new 0-arity constructor $\bot$ to each signature $\Sigma$ obtaining
an extended signature $\Sigma_{\bot}$ whose terms are called {\it partial
$\Sigma$-terms}. When the signature $\Sigma$ is clear, we will omit explicit mention of
it, and we will write ${\bf Term}$ and ${\bf CTerm}$ (or ${\bf Term}_{\bot}$ and ${\bf
CTerm}_{\bot}$ for $\Sigma_\bot$) respectively.
Following the approach to non-determinism in \cite{hussmann} we only consider {\it
C-substitutions} $\theta \colon {\cal V} \rightarrow {\bf CTerm}$.  These mappings have
natural extensions $\theta \colon {\bf Term} \rightarrow {\bf Term}$, also noted as
$\theta$, defined in the usual way, and the result of applying $\theta$ to the term $t$
is written $t \theta$.
%$$
%\begin{array}{rcll}
%  X \theta                  & =_{def} & \theta(X), &
%              \forall X \in {\cal V}; \\
%  c \theta                  & =_{def} & c,         &
%              \forall c \in \cons{0} \cup \funs{0}; \\
%  \function{h}{t}{n} \theta & =_{def} & h (t_{1}\theta, \ldots, t_{n}\theta),&
%              \forall h \in \cons{n} \cup \funs{n}, n>0.
%\end{array}
%$$
Analogously, we define {\it partial C-substitutions} as mappings $\theta \colon {\cal V}
\rightarrow {\bf CTerm}_{\bot}$.  The set of all C-substitutions (partial
C-substitutions) is written ${\bf CSubst}$ (${\bf CSubst}_{\bot}$).

A {\it signature morphism} $\rho \colon \Sigma \rightarrow \Sigma'$ from a signature
$\Sigma = (DS_{\Sigma},FS_{\Sigma})$ to a signature $\Sigma' =
(DS_{\Sigma'},FS_{\Sigma'})$ consists of two mappings, that we denote with the same
symbol $\rho \colon DS_{\Sigma} \rightarrow DS_{\Sigma'}$ and $\rho \colon FS_{\Sigma}
\rightarrow FS_{\Sigma'}$, that map strings $h/n$ into strings $h'/n$. By abuse of
notation we will denote $h' = \rho(h)$.  This allows us to define a mapping $\rho \colon
{\bf Term}_{{\Sigma}_{\bot}} \rightarrow {\bf Term}_{{\Sigma'}_{\bot}}$ as follows:
\[
\begin{array}{l}
  \rho(h) =_{def}  h,   \mbox{ for }
            h\in {\cal V}\cup \{\bot\}\cup\cons{0} \cup \funs{0}; \\
  \rho(\function{h}{t}{n}) =_{def} \rho(h) (\rho(t_{1}),\ldots,\rho(t_{n})),
                     \mbox{ for } h \in \cons{n} \cup \funs{n}, n>0.
\end{array}
\]
We will consider signature morphisms $\rho \colon \Sigma \rightarrow \Sigma$ such that
$\rho(h/n) = h/n$ for every string $h/n$ in $DS_{\Sigma}$.  Such morphisms will be called
{\it function symbol renamings}.

Given a signature $\Sigma$ and a set ${\cal V}$ of variable symbols, there are two kinds
of atomic {\it CRWL-formulas} for $a, b \in {\bf Term}_{\bot}$, {\it reduction
statements\/} $a \rightarrow b$,  with the intended meaning ``$a$ can be reduced to
$b$,'' and {\it joinability statements\/} $a \bowtie b$, with the intended meaning ``$a$
and $b$ can be reduced to a common value in {\bf CTerm}''. Terms $t \in {\bf CTerm}$ are
intended to represent totally defined values whereas terms $t \in {\bf CTerm}_{\bot}$
represent partially defined values ---to model the behavior of non-strict functions.
Reduction statements $a \rightarrow t$ with $t \in {\bf CTerm}_{\bot}$, called {\it
approximation} statements, have the intended meaning that $t$ {\it approximates} a
possible value of $a$, whereas $a \rightarrow t$ with $t \in {\bf CTerm}$ have the
intended meaning that $t$ {\it represents} a possible value of $a$
---an expression may denote several values capturing the behavior
of non-deterministic functions. Substitutions $\theta \in {\bf CSubst}_{\bot}$ and
signature morphisms $\rho \colon \Sigma \rightarrow \Sigma'$ apply to formulas in the
obvious way.
 %\[
%\begin{array}{ll}
%  (a \es b)\theta =_{def} a\theta \es b\theta,
%          & (a \bowtie b)\theta =_{def} a\theta \bowtie b\theta; \\
%  \rho(a \es b) =_{def} \rho (a) \es \rho(b),
%          & \rho(a \bowtie b) =_{def} \rho(a) \bowtie \rho(b).
%\end{array}
%\]

% ....................................................................

\subsection{Programs and formal derivation}

A {\it CRWL-program} is a CRWL-theory ${\cal R}$ defined as a signature $\Sigma$ together
with a set of {\it conditional rewrite rules} of the general form
 \[
     f(\overline{t}) \es r \si C,
 \]
where $f(\overline{t})$ is the left hand side (lhs), $r$ the right hand side (rhs), $C$
the condition of the rule, $f$ is a function symbol with arity $n \geq 0$, and $C$
consists of finitely many (possibly zero) joinability statements between fully defined
terms (with no occurrence of $\bot$). When $n>0$, $\overline{t}$ is a linear $n$-tuple
(i.e., without repeated variables) of fully defined constructor terms $t_i\in
\textbf{CTerm}$. When $n=0$ rules take the simpler form $f \es r \si C$. Formal
derivation of CRWL-statements from a given program ${\cal R}$ is governed by two
equivalent calculi (see \cite{mario:jlp}). We present here the so-called Goal-Oriented
Proof Calculus (GPC) which focuses on top-down proofs of reduction and joinability
statements: $$
 \footnotesize
 \renewcommand{\arraystretch}{.75}
\begin{tabular}{lcl}
 {\bf (Bo)} \qquad & $e \rightarrow \bot$, & $\mbox{for } e \in {\bf
Term_{\bot}}$;\\
  & & \\
 {\bf (RR)} &$ e \rightarrow e$, & $\hbox{for } e \in {\cal V} \cup DS^{0}$;\\
  & & \\
 {\bf (DS)} & $\displaystyle
    \frac{ e_{1} \rightarrow t_{1} \ldots e_{n} \rightarrow t_{n} }
         {\function{c}{e}{n} \rightarrow \function{c}{t}{n}}$, &
  $\mbox{for } c \in DS^{n} \mbox{ and } e_{i},t_{i} \in {\bf Term}_{\bot}$;\\
  & & \\
 {\bf (OR)}\qquad & $\displaystyle
    \frac{ e_{1} \rightarrow t_{1} \ldots e_{n} \rightarrow t_{n}
                                          \quad C \quad  r \rightarrow t}
         {\function{f}{e}{n} \rightarrow t}$, &
      $\mbox{if } (\function{f}{t}{n}
        \rightarrow r \Leftarrow C) \in [{\cal R}]_{\bot}$
       $\mbox{and } t \not\equiv \bot $;\\
 {\bf (Jo)} & $\displaystyle
     \frac{ a \rightarrow t \quad  b \rightarrow t }{ a \bowtie b }$, &
      $\mbox{if } t \in {\bf CTerm} \hbox{ and } a,b \in {\bf
Term_{\bot}}$;
\end{tabular}
$$
where $[{\cal R}]_{\bot} = \{ (l \rightarrow r \Leftarrow C)\theta \ \vert
\ (l \rightarrow r \Leftarrow C) \in {\cal R}, \ \theta \in {\bf
CSubst_{\bot}}\}$ is the set of possibly partial {\it constructor
instances} of rewrite rules and C-substitutions apply to
 rules in the obvious way.
%$$
%  (l \es r \si C) \theta  =_{def}  l \theta \es r \theta \si C \theta.
%$$
Rule {\bf (Bo)} shows that a CRWL-reduction is related to the idea of approximation, and
rule {\bf (OR)} states that only constructor instances of rewrite rules are allowed in
this calculus reflecting the so-called ``call-time-choice'' \cite{hussmann} for
non-determinism (values of arguments for functions are chosen before the call is made).
%In \cite{mario:jlp} it is also proved that the reduction relation governed by the
%above calculus is a preorder (reflexive and transitive) in ${\bf Term_{\bot}}$ and all
%defined functions are monotonic wrt this relation.
When a reduction or joinability
statement $\varphi$ is derivable from a program ${\cal R}$ we write ${\cal R}
\vdash_{CRWL} \varphi$ and we say that $\varphi$ is {\it provable} in ${\cal R}$.
{\it Goals} for a program ${\cal R}$ are finite conjunctions of atomic formulas, and {\it
solutions} are C-substitutions that make goals derivable. In \cite{mario:jlp} a sound and
complete lazy narrowing calculus for goal-solving can be found.

% --------------------------------------------------------------------

\subsection{CRWL-Algebras and models}
\label{algebras&terms}

We interpret CRWL-programs over algebraic structures consisting of posets with
bottom as carriers (i.e., sets $D$ with a partial order $\sqsubseteq_{D}$ and a
least element $\bot_{D}$), whose elements are thought of as finite
approximations of possibly infinite values in the poset's ideal completion
\cite{moller}, and monotonic mappings from elements to cones (non-empty subsets
of a poset with bottom, downclosed wrt the partial order of the poset) as
function symbol denotations reflecting possible non-determinism. Such a mapping
$f \colon D \rightarrow {\cal C}(E)$ ---where $D$, $E$ are posets with bottom,
and ${\cal C}(E)$ is the set of cones of $E$--- can be extended to a monotonic
mapping $\hat f \colon {\cal C}(D) \rightarrow {\cal C}(E)$, defined by $\hat
f(C) =_{def} \bigcup_{u \in C} f(u)$ and also noted $f$ by abuse of notation.
In particular, deterministic function symbols are represented by mappings $f
\colon D \rightarrow {\cal I}(E)$ computing directed cones or ideals (i.e.,
cones $\alg{C}$ such that for all $x,y\in \alg{C}$ there exists $z\in\alg{C}$
with $x\sqsubseteq z$ and $y\sqsubseteq z$) where ${\cal I}(E)$ is the set of
ideals of $E$. These mappings become continuous mappings between algebraic cpos
after performing the ideal completion (for a comprehensive exposition of these
notions we refer to \cite{abramsky}). These ideas are behind the notion of
CRWL-algebra.

Given a signature $\Sigma$ and a set ${\cal
V}$ of variable symbols, a {\it CRWL-algebra} of signature
$\Sigma$ is an algebraic structure
 $ {\cal A} = (D_{\cal A},\{ c^{\cal A} \}_{c \in DS_{\Sigma}},
 \{ f^{\cal A} \}_{f \in FS_{\Sigma}})
 $
where the carrier $D_{\cal A}$ is a poset with bottom $\bot_{\cal A}$, $f^{\cal A}$ is a
monotonic mapping $D_{\cal A}^{n} \rightarrow {\cal C}(D_{\cal A})$ for each $f \in
FS^{n}_{\Sigma}$ and $c^{\cal A}$ is a monotonic mapping $D_{\cal A}^{n} \rightarrow
{\cal I}(D_{\cal A})$ for each $c \in DS^{n}_{\Sigma}$. Both $f^{\cal A}$ and $c^{\cal
A}$ reduce to cones when $n = 0$. In order to ensure preservation of finite and maximal
elements in the ideal completion, we require for all $u_{1},\ldots,u_{n} \in D_{\cal A}$
that there exists $v \in D_{\cal A}$ such that $c^{\cal A}(u_{1},\ldots,u_{n}) = \langle
v \rangle$, where $\langle v \rangle$ is the ideal generated by $v$ (i.e., the set
$\{d\in D_{\cal A} \mid d\sqsubseteq v\}$), and if all $u_{i}$ are maximal (totally
defined) then $v$ must also be maximal. The class of all CRWL-algebras of signature
$\Sigma$ is denoted by ${\bf Alg}_{\Sigma}$. We are specially interested in {\it
CRWL-term algebras}, which are CRWL-algebras with carrier ${\bf CTerm_{\bot}}$, ordered
by the {\it approximation ordering} ``$\sqsubseteq$,'' defined as the least partial
ordering satisfying the following properties:
 \[
\begin{array}{lcl}
 (a) & \bot \sqsubseteq t, & \forall t  \in {\bf CTerm_{\bot}}; \\
 (b) & c(\overline{s}) \sqsubseteq c(\overline{t}) & \mbox{ if }
       s_{i}\sqsubseteq t_{i},\ i=1,\dots,n,  \mbox{ for } c \in DS^{n}_{\Sigma}, n\geq 0;
\end{array}
 \]
and fixed interpretation for constructor symbols:
 $c^{\cal A} = \langle c \rangle$, for all $c \in DS^{0}_{\Sigma}$, and
 $c^{\cal A}(\overline{t}) = \langle c(\overline{t}) \rangle$, for all
 $c \in DS^{n}_{\Sigma}$ and $n\geq 0$.
 Therefore, two CRWL-term algebras of the same signature $\Sigma$ will only
differ in their interpretations for the function symbols of $\Sigma$. As a consequence of
the above definition, for $s, t \in {\bf CTerm}_{\bot}$, $s \sqsubseteq t$ implies $s =
\bot$ or $s = c(\overline{s})$ and $t = c(\overline{t})$ for some $c \in \cons{n}$ and
$n\geq 0$ with each component $s_{i} \sqsubseteq t_{i}$. Also, for $s,t \in {\bf
CTerm}_{\bot}$,
\begin{equation}\label{reductionOrdering}
 s\sqsubseteq t \ \Leftrightarrow\ \vdash_{\mathit{CRWL}}t\es s.
\end{equation}
It can be proved, by induction, that every  $\theta \in {\bf CSubst}_{\bot}$ is a
monotonic mapping from ${\bf CTerm}_{\bot}$ to ${\bf CTerm}_{\bot}$, that is:
 $ s \sqsubseteq t \ \Rightarrow \ s \theta \sqsubseteq t \theta$, for all
 $s,t \in {\bf CTerm}_{\bot}$.

A {\it valuation} over a structure ${\cal A} \in {\bf Alg}_{\Sigma}$ is any
mapping $\eta \colon {\cal V} \rightarrow D_{\cal A}$. $\eta$ is {\it totally
defined} when $\eta (X)$ is maximal for all $X \in {\cal V}$. ${\bf Val}({\cal
A})$ is the set of all valuations over ${\cal A}$ and  ${\bf DefVal}({\cal A})$
the set of all totally defined valuations. Given a valuation $\eta$ we can
evaluate each partial $\Sigma$-term in ${\cal A}$ as follows: $$
\begin{array}{rcll}
  \evaluation{\bot}{A}{\eta} & =_{def} & \ideal{\bot_{\cal A}}, & \\
  \evaluation{X}{A}{\eta}    & =_{def} & \ideal{\eta (X)}, &
                                 \forall X \in {\cal V}; \\
  \evaluation{c}{A}{\eta}    & =_{def} & c^{\cal A},  &
     \forall c \in \cons{0} \cup \funs{0}; \\
  \evaluation{\function{h}{e}{n}}{A}{\eta}
                             & =_{def} &
         \hat h^{\cal A}(\, \evaluation{e_{1}}{A}{\eta}, \ldots,
                            \evaluation{e_{n}}{A}{\eta}), &
         \forall h \in \cons{n} \cup \funs{n}, n>0.
\end{array}
$$ In this way each partial $\Sigma$-term is evaluated to a cone. For each CRWL-algebra
${\cal A}$, every $\eta \in {\bf Val}({\cal A})$, and $ e \in {\bf Term_{\bot}}$,  the
following properties are proved in \cite{mario:jlp},
\begin{enumerate}
\item
 $ \evaluation{e}{A}{\eta} \in {\cal C}(D_{\cal A})$.
\item
 $ \evaluation{e}{A}{\eta} \in {\cal I}(D_{\cal A})$,
  if  $e$ is only built from deterministic functions (i.e., function symbols
  interpreted by ideal valued functions).
\item
  $\evaluation{e}{A}{\eta} = \ideal{v}$ for some  $v \in D_{\cal A}$,
   if  $e \in {\bf CTerm}_{\bot}$. Moreover, when  $e \in {\bf CTerm}$ and
   $\eta \in {\bf DefVal}({\cal
    A})$, $v$  is maximal.
\item (Substitution Lemma)
 $\evaluation{e \theta}{A}{\eta} =\evaluation{e}{A}{\rho}$, for
 $\theta \in {\bf CSubst}_{\bot}$,
 where $\rho$ is the uniquely determined valuation that satisfies
 $\ideal{\rho(X)} = \evaluation{X \theta}{A}{\eta}$, for all $X \in {\cal V}$.
\end{enumerate}
From these results and taking into account that each substitution
is equivalent to a valuation over any CRWL-term algebra, we have
the following complementary results for term algebras:

\begin{proposition}   % ..............................................
\label{cterm-evaluation}
 For each CRWL-term algebra ${\cal A}$ and every $\eta \in {\bf Val}({\cal A})$ we have:
 \begin{enumerate}
 \item
  $ \evaluation{t}{A}{\eta} = \ideal{t \eta}$ for every $t \in {\bf CTerm_{\bot}}$;
 \item
  $\evaluation{\function{h}{t}{n}}{A}{\eta} = h^{\alg{A}}(\overline{t}\eta)$
  for all $h \in \cons{n}\cup\funs{n}$, $n>0$, and
  $t_{1},\ldots,t_{n} \in {\bf CTerm_{\bot}}$;
 \item
  $\evaluation{e \theta}{A}{\eta} = \evaluation{e}{A}{\theta\eta}$ for all
  $e \in {\bf Term_{\bot}}$ and $\theta \in {\bf CSubst}_{\bot}$, where
  $\theta\eta$ represents the function composition $\eta\circ\theta$.
 \end{enumerate}
\end{proposition}     % ..............................................
\begin{proof}
 (1) is easily proved by induction on the structure of $t$ and (2) follows
 from (1). By the Substitution~Lemma, $\ \evaluation{e \theta}{A}{\eta} =
 \evaluation{e}{A}{\rho}$ for a valuation $\rho$ uniquely determined by the condition
 $\ideal{\rho(X)} = \evaluation{X \theta}{A}{\eta}, \forall X \in {\cal V}$,
 and by (1),
 $\ \evaluation{X \theta}{A}{\eta} = \ideal{X \theta \eta}; $
 then $\rho = \theta \eta$ and we obtain (3).
\end{proof}

Models in CRWL are introduced from the following notion of satisfiability:
\begin{itemize}
\item
  ${\cal A}$ {\it satisfies a reduction statement} $a \rightarrow b$ under a
  valuation $\eta \in {\bf Val}(D_{\cal A})$,
  or ${\cal A} \models_{\eta} (a \rightarrow b)$, iff
  $\evaluation{a}{A}{\eta} \supseteq \evaluation{b}{A}{\eta}$.
\item
  ${\cal A}$ {\it satisfies a joinability statement} $a \bowtie b$ under a
  valuation $\eta \in {\bf Val}(D_{\cal A})$,
  or ${\cal A} \models_{\eta} (a \bowtie b)$, iff
  $\evaluation{a}{A}{\eta} \cap \, \evaluation{b}{A}{\eta}
  \mbox{ contains a maximal element in } D_{\cal A}$.
\item
  ${\cal A}$ {\it satisfies a rule} $l \rightarrow r  \Leftarrow C$,
  or ${\cal A} \models (l \rightarrow r \Leftarrow C)$, iff
 ${\cal A} \models_{\eta} C \mbox{ implies } {\cal A} \models_{\eta}
  (l \rightarrow r), \mbox{ for every valuation }\eta \in {\bf Val}
  (D_{\cal A})$.
\item
  ${\cal A}$ is a {\it model} of a program ${\cal R}$,
  i.e., ${\cal A} \models {\cal R}$, iff
  ${\cal A}$ satisfies all rules in ${\cal R}$.
\end{itemize}
CRWL-provability is sound and complete wrt this model-theoretic semantics when we
consider totally defined valuations only. In \cite{mario:jlp} is proved that for any
program ${\cal R}$ and any approximation or joinability statement $\varphi$,
 \begin{equation}\label{eq.sound&complete}
  {\cal R} \vdash_{CRWL} \varphi \ \Leftrightarrow \ {\cal A}
  \models_{\eta} \varphi,
  \mbox{ for every ${\cal A}$ model of ${\cal R}$ and }
  \eta \in {\bf DefVal}(D_{\cal A}).
 \end{equation}
This result is achieved with the help of a CRWL-term algebra ${\cal M_{R}}$ characterized
by the following interpretation for any defined function symbol $f \in FS^{n}_{\Sigma}$,
$n \geq 0$,
 \[
  f^{\cal M_{R}}(\overline{t}) =_{def}
  \{ r \in {\bf CTerm}_{\bot} \mid {\cal R} \vdash_{CRWL}
                                f(\overline{t}) \rightarrow r \}.
 \]
${\cal M_{R}}$ is such that
 ${\cal R} \vdash_{CRWL} \varphi \ \Leftrightarrow \
  {\cal  M_{R}} \models_{id} \varphi $
for any approximation or joinability statement $\varphi$.
According to this result, ${\cal M_{R}}$ is taken as the {\it canonical model} of the
program ${\cal R}$.  Also in \cite{mario:jlp} it is proved that this model is freely
generated by ${\cal V}$ in the category of all models of ${\cal R}$. This is the {\it
model-theoretical semantics} of the program $\prg{R}$.

% \begin{definition}(Model-theoretic semantics)\\ % .. .. .. .. .. .. .. .
% \label{model:semantics}
% We define the model-theoretic semantics of
% a program $\prg{R}$ as
% $\modelsem{\prg{R}} =_{def} {\cal  M_{R}}$
% \end{definition}   % .. .. .. .. .. .. .. .. .. .. .. .. .. .. .. .. .

% -------------------------------------------------------------------

% \subsection{Models and Renaming}

Given a signature $\Sigma$ and a function symbol renaming $\rho \colon\Sigma
\rightarrow \Sigma$, for each CRWL-term algebra
 $\ \alg{A} = ({\bf CTerm}_{\bot},
             \{ c^{\alg{A}} \}_{c \in DS_{\Sigma}},
             \{ f^{\alg{A}} \}_{f \in FS_{\Sigma}})\
 $
of this signature we can define another CRWL-term algebra
 \[
\alg{A}_{\rho} = ({\bf CTerm}_{\bot},
                  \{ c^{\alg{A}_{\rho}} \}_{c \in DS_{\Sigma}},
                  \{ f^{\alg{A}_{\rho}} \}_{f \in FS_{\Sigma}})
 \]
such that $f^{\alg{A}_{\rho}} = \rho(f)^{\alg{A}}$. The relation
between evaluation and satisfaction in $\alg{A}$ and evaluation and
satisfaction in $\alg{A}_{\rho}$ is stated by the following
proposition.

\begin{proposition} % .. .. .. .. .. .. .. .. .. .. .. .. .. .. .. ..
\label{modelsandrenaming} Given a signature $\Sigma$, for every CRWL-term algebra
$\alg{A}$ of this signature, every function symbol renaming $\rho \colon \Sigma
\rightarrow \Sigma$, and all $\theta \in {\bf CSubst}_{\bot}$,  we have
\begin{enumerate}
\item
 $(\rho(t))\theta = \rho(t\theta)$, for all $t \in {\bf Term}_{\bot}$.
\item
 $\eval{\rho(t)}{\alg{A}}{\theta} = \eval{t}{\alg{A}_{\rho}}{\theta}$,
 for all $t \in {\bf Term}_{\bot}$.
\item
 $
   \alg{A} \models_{\theta} \rho(\varphi) \  \Leftrightarrow \
   \alg{A}_{\rho} \models_{\theta} \varphi
 $, for any reduction or joinability statement $\varphi$.
\end{enumerate}
\end{proposition}   % .. .. .. .. .. .. .. .. .. .. .. .. .. .. .. ..
\begin{proof}
The two first statements can be proved by induction over the structure of $t$,
whereas the third one is directly derived from (2).

\end{proof}

% ====================================================================

\section{Fixpoint Semantics}
\label{fixpoint}

In this section we will prove, for every CRWL-program ${\cal R}$,
that ${\cal M_{R}}$ is the least fixpoint of an operator defined
over CRWL-term algebras.  The approach we use here is similar to
that applied in the field of logic programming \cite{apt}.
However, the notion of interpretation, and the corresponding
mathematical aspects, have to be reformulated in the context of
CRWL-term algebras. This approach has been also used in
\cite{juancarlos:tesis} in the context of a previous formalism to
model functional-logic programming. However, this work does not
deal with some relevant aspects (e.g., non-determinism) of the
CRWL-programming version we are considering here.

% --------------------------------------------------------------------

\subsection{The lattice of all CRWL-term algebras}

Let ${\bf TAlg}_{\Sigma}$ be the set of all CRWL-term algebras of a signature
$\Sigma$ associated to a CRWL-program ${\cal R}$.  We can define the
relationship ${\cal A} \sqsubseteq {\cal B}$ between two algebras ${\cal A},
{\cal B} \in {\bf TAlg}_{\Sigma}$ in the following way:
\[
  {\cal A} \sqsubseteq {\cal B}\, \Leftrightarrow_{def}\,
  \mbox{ for each } f \in FS^{n}_{\Sigma} \mbox{ and } n>0,\,
  f^{\cal A}(\overline{t}) \subseteq f^{\cal B}(\overline{t}),
\]
when $n=0$, $f^{\cal A} \subseteq f^{\cal B}$.  This relationship is obviously
a partial ordering and $({\bf TAlg}_{\Sigma}, \sqsubseteq)$ is a poset. This
poset has a bottom $\bot_{\Sigma}$ and a top $\top_{\Sigma}$ characterized by
the following interpretations, for each $f \in FS^{n}_{\Sigma}$ and $n \geq 0$,
\[
\begin{array}{rcl}
  f^{\bot_{\Sigma}}(\overline{t}) & =_{def} & \ideal{\bot},\\
  f^{\top_{\Sigma}}(\overline{t}) & =_{def} & {\bf CTerm}_{\bot}.
\end{array}
\]
Given a subset ${\bf S} \subseteq {\bf TAlg}_{\Sigma}$,
the following definitions
\[
\begin{array}{rcl}
   f^{\sqcup{\bf S}}(\overline{t}) & =_{def} &
      \bigcup_{{\cal A}\in {\bf S}} f^{\cal A}(\overline{t}), \\
   f^{\sqcap{\bf S}}(\overline{t}) & =_{def} &
      \bigcap_{{\cal A}\in {\bf S}} f^{\cal A}(\overline{t}),
\end{array}
\]
for each $f \in FS^{n}_{\Sigma}$ and $n\geq 0$, characterize two CRWL-term
algebras, $\sqcup{\bf S}$ and $\sqcap{\bf S}$ respectively, because the union
and intersection of any number of cones are cones also, and the resulting
functions in the above definitions are obviously monotonic if $f^{\cal A}$ is
monotonic for all ${\cal A} \in {\bf S}$. Clearly, $\sqcup{\bf S}$ and
$\sqcap{\bf S}$ are the least upper bound and the greatest lower bound of ${\bf
S}$, respectively. So, $({\bf TAlg}_{\Sigma},\sqsubseteq)$ is a complete
lattice.

Valuations (substitutions) of terms in term algebras can be considered continuous
mappings from algebras to cones in the sense given by the following lemma.

\begin{lemma}[Continuity of valuations in ${\bf TAlg}_{\Sigma}$] % ...
\label{continuity}

For each term $e \in {\bf Term}_{\bot}$ and each substitution $\theta
\in {\bf CSubst}_{\bot}$
\begin{enumerate}
\item
 $\alg{A} \sqsubseteq \alg{B} \ \Rightarrow \
   \evaluation{e}{A}{\theta} \subseteq \evaluation{e}{B}{\theta}$,
   for $\alg{A}, \alg{B} \in {\bf TAlg}_{\Sigma}$.
\item
 $\evaluation{e}{\sqcup {\bf D}}{\theta} = \bigcup_{\alg{A} \in {\bf D}} \
  \evaluation{e}{A}{\theta}$,
 for all directed subsets ${\bf D} \subseteq {\bf TAlg}_{\Sigma}$.
\end{enumerate}
\end{lemma}  % .......................................................
\begin{proof} The first statement is proved by induction on the
structure of $e$. If
 $e \in \{ \bot \} \cup \alg{V} \cup \cons{0}$ then
$\ \evaluation{e}{A}{\theta}$ does not depend on the particular term algebra $\alg{A}$
and $\ \evaluation{e}{A}{\theta} = \evaluation{e}{B}{\theta}$. Else, if $e \in \funs{0}$,
$\alg{A} \sqsubseteq \alg{B}$ implies $e^{\alg{A}} \subseteq e^{\alg{B}}$ and then
$\,\evaluation{e}{A}{\theta} \subseteq \evaluation{e}{B}{\theta}$. Finally, if
 $e = h(\overline{e})$ with $h \in \cons{n} \cup
\funs{n}$ and $n>0$, assuming $\,\evaluation{e_{i}}{A}{\theta} \subseteq
\evaluation{e_{i}}{B}{\theta}$, for $i=1,\ldots,n$, as the induction
hypothesis, for every $t \in \evaluation{e}{A}{\theta}$ we have $t \in
h^{\alg{A}}(\overline{t})$ for some $t_{i} \in \evaluation{e_{i}}{A}{\theta}$,
which implies $t \in h^{\alg{B}}(\overline{t})$ with $t_{i} \in
\evaluation{e_{i}}{B}{\theta}$ as a consequence of $\alg{A} \sqsubseteq
\alg{B}$ and the induction hypothesis. Thus, we get $t \in
\evaluation{e}{B}{\theta}$, and consequently $\evaluation{e}{A}{\theta}
\subseteq \evaluation{e}{B}{\theta}$.

To prove the second statement we only need to prove the following inclusion $
\evaluation{e}{\sqcup {\bf D}}{\theta} \subseteq \bigcup_{\alg{A} \in {\bf D}} \
\evaluation{e}{A}{\theta}$ because the inclusion in the other way is trivially derived
from the first statement. We also proceed by induction on  $e$. If $e \in \{ \bot \} \cup
\alg{V} \cup \cons{0}$ then, as $\ \evaluation{e}{A}{\theta}$ does not depend on
 ${\cal A}$, $\evaluation{e}{\sqcup {\bf D}}{\theta} =
\evaluation{e}{A}{\theta}$ for all $\alg{A} \in {\bf D}$. Else, if $e \in \funs{0}$ then
$\evaluation{e}{\sqcup {\bf D}}{\theta} = e^{\sqcup {\bf D}}$ and, by definition,
$e^{\sqcup {\bf D}} = \bigcup_{\alg{A} \in {\bf D}} e^{\alg{A}}$. So, in all these cases,
$\evaluation{e}{\sqcup {\bf D}}{\theta} = \bigcup_{\alg{A} \in {\bf D}} \
\evaluation{e}{A}{\theta}$. Finally, if $e = h(\overline{e})$ with $h \in \cons{n} \cup
\funs{n}$ and $n>0$, assuming $\evaluation{e_{i}}{\sqcup {\bf D}}{\theta} \subseteq
\bigcup_{\alg{A} \in {\bf D}} \ \evaluation{e_{i}}{A}{\theta}$, $i=1,\ldots,n$, as the
induction hypothesis, for every $t \in \evaluation{e}{\sqcup {\bf D}}{\theta}$ we have $t
\in h^{\sqcup {\bf D}}(\overline{t})$ for some $t_{i} \in \evaluation{e_{i}}{\sqcup {\bf
D}}{\theta}$, $i=1,\dots,n$. By definition $h^{\sqcup {\bf D}}(\overline{t}) =
 \bigcup_{\alg{A} \in {\bf D}} \ h^{\alg{A}}(\overline{t})$
 and from this and the induction hypothesis we can deduce
$t \in h^{\alg{A}_{0}}(\overline{t})$ with
 $t_{i} \in \eval{e_{i}}{\alg{A}_{i}}{\theta}$,  for some
  $\alg{A}_{0}, \alg{A}_{1},\ldots,\alg{A}_{n} \in {\bf D}$.
Since ${\bf D}$ is directed, there exists $\alg{A}\in{\bf D}$,
such that $\alg{A}_i\sqsubseteq\alg{A}$, $i=0,1,\dots,n$, and so
$t\in h^{\alg{A}}(\overline{t})$ with $t_{i} \in
\evaluation{e_{i}}{A}{\theta} $, which implies $t \in
\evaluation{e}{A}{\theta}$ and $ \evaluation{e}{\sqcup {\bf
D}}{\theta} \subseteq \bigcup_{\alg{A} \in {\bf D}} \
\evaluation{e}{A}{\theta}$. \end{proof}

Another interesting result relates satisfiability of joinability statements in the least
upper bound of a directed set of term algebras with satisfiability in, at least, one of
the algebras of the set.

\begin{lemma}  % .....................................................

\label{modelexistence}
Let $C$ be a finite set of joinability statements and ${\bf D}$ a directed
subset of ${\bf TAlg}_{\Sigma}$, then
 $\sqcup {\bf D} \models_{\theta} C$ implies that there exists
 $\alg{A} \in {\bf D}$ such that $\alg{A} \models_{\theta} C$.
\end{lemma}    % .....................................................
\begin{proof} It is sufficient to prove that this lemma holds when $C$ reduces to one
joinability statement $r \bowtie s$, because with more statements we shall obtain
algebras $\alg{A}_{1}, \ldots,\alg{A}_{n}$, one for each joinability statement,  and the
upper bound $\sqcup \{ \alg{A}_{1},\ldots,\alg{A}_{n} \}$ will satisfy all joinability
statements in $C$. By definition, $\sqcup {\bf D} \models_{\theta} r \bowtie s$ implies
that there exists a totally defined term $t \in \evaluation{r}{\sqcup {\bf D}}{\theta}
\cap \evaluation{s}{\sqcup {\bf D}}{\theta}$ and by lemma~\ref{continuity}, $t \in
\evaluation{r}{\sqcup {\bf D}}{\theta} \Rightarrow t \in \eval{r}{\alg{A}_{1}}{\theta}$
for some $\alg{A}_{1} \in {\bf D}$ and $t \in \evaluation{s}{\sqcup {\bf D}}{\theta}
\Rightarrow t \in \eval{s}{\alg{A}_{2}}{\theta}$ for some $\alg{A}_{2} \in {\bf D}$. By
the first statement of lemma~\ref{continuity}, considering $\alg{A}\in{\bf D}$ such that
$\alg{A}_i\sqsubseteq\alg{A}$, $i=1,2$, we have a term algebra such that $t
\in
\evaluation{r}{\alg{A}}{\theta} \cap
\evaluation{s}{\alg{A}}{\theta}$ and consequently $\alg{A}
\models_{\theta} r \bowtie s.$
\end{proof}

% ...................................................................

\subsection{The algebra transformer associated with a program}
\label{operator}

Given a CRWL-program $\prg{R}$, with a signature $\Sigma$, we can define an algebra
transformer ${\cal T_{R}} \colon {\bf TAlg}_{\Sigma} \rightarrow {\bf TAlg}_{\Sigma}$,
similar to the immediate consequences operator used in logic programming, by fixing the
interpretation of each function symbol $f \in FS^{n}_{\Sigma}$,  in a transformed algebra
${\cal T_{R}(A)}$, as the result of one step applications of reduction statements
corresponding to instances ---not necessarily ground--- of those rules of ${\cal R}$,
defining $f$, satisfied in $\alg{A}$. We formalize this idea defining, for each $f\in
\funs{n}$, $n\geq 0$,
 $$
  \function{f^{\cal T_{R}(A)}}{t}{n} =_{def}
   \{ t \mid \exists (\function{f}{s}{n} \rightarrow r \Leftarrow C)
    \in [\prg{R}]_{\bot}, s_{i} \sqsubseteq t_{i},
    \alg{A} \models_{id} C, t \in \evaluation{r}{A}{id} \} \cup \{ \bot \},
 $$
that is basically a union of cones $\evaluation{r}{A}{id}$. This definition corresponds
to a monotonic mapping because all rule instances $(\function{f}{s}{n} \rightarrow r
\Leftarrow C) \in [\prg{R}]_{\bot}$, applicable to arguments $\overline{t'}$ are also
applicable to arguments $\overline{t}$ such that $t'_{i} \sqsubseteq t_{i}$, for
$i=1,\ldots,n$, and so the corresponding interpretation characterizes a CRWL-term
algebra. From this definition of ${\cal T_{R}}$ we can derive the continuity of the
operator in ${\bf TAlg}_{\Sigma}$.
\begin{proposition}  % ...............................................
For each program $\prg{R}$ its associated operator $\prg{T_{R}}$
is continuous.
\end{proposition}    % ...............................................
\begin{proof} $\prg{T_{R}}$ is monotonic. Given $\alg{A}, \alg{B} \in {\bf TAlg}_{\Sigma}$ such
that $\alg{A} \sqsubseteq \alg{B}$, $\alg{A} \models_{id} C \Rightarrow \alg{B}
\models_{id} C$ for every set $C$ of joinability statements, and by
Lemma~\ref{continuity}, $\,\evaluation{e}{A}{id} \subseteq \evaluation{e}{B}{id}$ for
every term $e$; hence, every rule instance $(\function{f}{s}{n} \rightarrow r \Leftarrow
C) \in [\prg{R}]_{\bot}$ applicable to obtain $\function{f^{\cal T_{R}(A)}}{t}{n}$ also
will be applicable to obtain $\function{f^{\cal T_{R}(B)}}{t}{n}$, and therefore
$\prg{T_{R}}(\alg{A}) \sqsubseteq \prg{T_{R}}(\alg{B})$.
$\prg{T_{R}}$ is continuous. For every directed set ${\bf D} \subseteq {\bf
TAlg}_{\Sigma}$, $\prg{T_{R}}(\sqcup {\bf D}) \sqsubseteq \sqcup \{
\prg{T_{R}}(\alg{A}) \vert \alg{A} \in {\bf D} \}$ because each rule instance
$(\function{f}{s}{n} \rightarrow r \Leftarrow C) \in [\prg{R}]_{\bot}$ that is
applicable to obtain $\function{f^{{\cal T_{R}}(\sqcup {\bf D})}}{t}{n}$, by
Lemmas~\ref{continuity} and \ref{modelexistence}, is also applicable to obtain
$\bigcup_{\alg{A} \in {\bf D}} \function{f^{{\cal T_{R}}(\alg{A})}}{t}{n}$, and
this expression is  $\function{f^{\sqcup \{ \prg{T_{R}}(\alg{A}) \vert \alg{A}
\in {\bf D} \}}}{t}{n}$. The inclusion in the other way is trivial.
\end{proof} % ...........................................................

Thus, $\prg{T_{R}}$ has a least fixpoint $\alg{F_{R}}$ given by
$\sqcup{\bf A}_{\prg{R}}$ (that is also the least pre-fixpoint),
where ${\bf A}_{\prg{R}}$ is the chain of CRWL-term algebras
$\alg{A}_{i}, i
\in {\bf N}$, such that
 \[
  \alg{A}_{0} = \bot_{\Sigma} \sqsubseteq  \ldots \sqsubseteq
     \alg{A}_{i+1} = \prg{T_{R}}(\alg{A}_{i}) \sqsubseteq \ldots
 \]
$\alg{F_{R}}$ is also denoted as $\prg{T_{R}}^{\omega}(\bot_{\Sigma})$ (see
\cite{abramsky}). In order to prove that $\alg{F_{R}}$ coincides with $\alg{M_{R}}$ we
need two lemmata, one characterizing the set of term models and other relating
CRWL-provability with ${\bf A}_{\prg{R}}$ satisfiability.
\begin{lemma}[Model characterization] % ..............................
\label{model-charact} Given a program $\prg{R}$,
    $\alg{M}$ is a term model for $\alg{R}$ iff
                   $ \prg{T_{R}}(\alg{M}) \sqsubseteq \alg{M}$
\end{lemma}  % .......................................................
\begin{proof}
 First, we will prove that $\prg{T_{R}}(\alg{M}) \sqsubseteq \alg{M}$ for each term
model $\alg{M}$. Let us consider $\function{f^{{\cal T_R}(\alg{M})}}{t}{n}$ for $f \in
\funs{n}, \ n>0$, with all $t_{i} \in {\bf CTerm_{\bot}}$. If there exists a rule
instance $(\function{f}{s}{n} \rightarrow r \Leftarrow C) \in [\prg{R}]_{\bot}$ with
$r\not= \bot$, $s_{i} \sqsubseteq t_{i}$, and $\alg{M} \models_{id} C$ then as $\alg{M}$
is a model of $\prg{R}$, $\,\evaluation{r}{M}{id} \subseteq
\evaluation{\function{f}{s}{n}}{M}{id}$. By Proposition~\ref{cterm-evaluation}~(2),
$\evaluation{\function{f}{s}{n}}{M}{id} = \function{f^{\alg{M}}}{s}{n}$, and by
 $f^{\alg{M}}$ monotonic, $\function{f^{\alg{M}}}{s}{n} \subseteq
\function{f^{\alg{M}}}{t}{n}$, and $\evaluation{r}{M}{id} \subseteq
\function{f^{\alg{M}}}{t}{n}$. Thus, $\function{f^{{\cal T_R}(\alg{M})}}{t}{n} \subseteq
\function{f^{\alg{M}}}{t}{n}$, and consequently ${\cal T_R}(\alg{M})\sqsubseteq \alg{M}$.
For $f \in \funs{0}$ the proof is similar but somewhat simpler.

Now we will prove that every term algebra $\alg{M}$ such that
${\cal T_R}(\alg{M}) \sqsubseteq \alg{M}$ is a model for
$\prg{R}$. Given a rule $(\function{f}{t}{n} \es r \si C) \in
\prg{R}$, for $\theta \in {\bf CSubst}_{\bot}$ such that $\alg{M}
\models_{id} C \theta$, or equivalently $\alg{M} \models_{\theta}
C$ (by Proposition~\ref{cterm-evaluation}~(3)), we can consider
$f^{\prg{T_{R}}(\alg{M})}(\overline{t}\theta)$, and because of the
instance $(\function{f}{t}{n} \es r \si C)\theta \in
[\prg{R}]_{\bot}$ we have $\evaluation{r \theta}{M}{id} \subseteq
f^{\prg{T_{R}}(\alg{M})}(\overline{t}\theta)$. By hypothesis,
$f^{\prg{T_{R}}(\alg{M})}(\overline{t}\theta) \subseteq
f^{\alg{M}}(\overline{t}\theta)$; by
Proposition~\ref{cterm-evaluation}~(3), $\evaluation{r
\theta}{M}{id} = \evaluation{r}{M}{\theta}$; and by
Proposition~\ref{cterm-evaluation}~(2),
$f^{\alg{M}}(\overline{t}\theta) =
\evaluation{\function{f}{t}{n}}{M}{\theta}$; thus,
$\evaluation{r}{M}{\theta} \subseteq
\evaluation{\function{f}{t}{n}}{M}{\theta}$ which is $\alg{M}
\models_{\theta} \function{f}{t}{n} \es r$, and $\alg{M}$
satisfies the rule $\function{f}{t}{n} \es r \si C$. \end{proof}
% ...........................................................
\begin{lemma} % ......................................................
\label{chainSatisfaction}
Given $e \in {\bf Term}_{\bot}$ and $t \in {\bf CTerm}_{\bot}$, we
have
 \[
 \prg{R} \vdash_{CRWL} e \es t \ \Rightarrow \
   \alg{A}_{i} \models_{id} e \es t, \mbox{ for some }
   \alg{A}_{i} \in {\bf A}_{\prg{R}}.
 \]
\end{lemma}   % ......................................................
\begin{proof} As $\prg{T_{R}}(\sqcup{\bf A}_{\prg{R}})=\sqcup{\bf A}_{\prg{R}}$, by the model
characterization lemma, $\sqcup{\bf A}_{\prg{R}}$ will be a model of $\prg{R}$. Thus, by
equivalence (\ref{eq.sound&complete}), $\prg{R} \vdash_{CRWL} e \es t$ implies
$\sqcup{\bf A}_{\prg{R}} \models_{id} e \es t$ or $\ideal{t} \subseteq
\eval{e}{\sqcup{\bf A}_{\prg{R}}}{id}$ that is equivalent to $t \in \eval{e}{\sqcup{\bf
A}_{\prg{R}}}{id}$. By lemma~\ref{continuity}, $\eval{e}{\sqcup{\bf A}_{\prg{R}}}{id} =
\bigcup_{\alg{A}_{i} \in {\bf A}_{\prg{R}}} \eval{e}{\alg{A}_{i}}{id}$, so there will be
an $\alg{A}_{i}$ such that $t \in \eval{e}{\alg{A}_{i}}{id}$ that means $\alg{A}_{i}
\models_{id} e \es t$.
\end{proof} % ...........................................................

From the above results we obtain the following proposition.
\begin{proposition} % ................................................
For every program $\prg{R}$, $\alg{M_{R}}$ is the least fixpoint (and the least
pre-fixpoint) of $\prg{T_{R}}$.
\end{proposition}   % ................................................
\begin{proof} First we can prove $\sqcup{\bf A}_{\prg{R}}\sqsubseteq\alg{M_{R}}$, from
$\alg{A}_{0}\sqsubseteq\alg{M_{R}}$, $\alg{T_{R}(M_{R})}\sqsubseteq\alg{M_{R}}$ (because
$\alg{M_{R}}$ is a model of $\prg{R}$) and the continuity of $\alg{T_{R}}$ that assures
$\alg{A}_{i} \sqsubseteq \alg{M_{R}}$ for all $i$.
Now we can prove that $\alg{M_{R}}\sqsubseteq\sqcup{\bf
A}_{\prg{R}}$ by proving, for each $f \in \funs{n}$, that
$\function{f^{\alg{M_{R}}}}{t}{n} \subseteq
 \function{f^{\sqcup{\bf A}_{\prg{R}}}}{t}{n}$,
for $t_{1}, \dots, t_{n} \in {\bf CTerm_{\bot}}$, and this
inclusion is proved by reasoning with elements. By definition, $t
\in \function{f^{\alg{M_{R}}}}{t}{n}$ is equivalent to $\prg{R}
\vdash_{CRWL} \function{f}{t}{n} \es t$ and, by
Lemma~\ref{chainSatisfaction}, this implies $\alg{A}_{i}
\models_{id} \function{f}{t}{n} \es t$, for some $\alg{A}_{i} \in
{\bf A}_{\prg{R}}$. Taking into account that, by
Proposition~\ref{cterm-evaluation}~(2)
$\,\eval{\function{f}{t}{n}}{\alg{A}_{i}}{id} =
\function{f^{\alg{A}_{i}}}{t}{n}$ we obtain $t \in
\function{f^{\alg{A}_{i}}}{t}{n}$ and $t\in
\function{f^{\sqcup{\bf A}_{\prg{R}}}}{t}{n}$. \end{proof}

Thus, if we consider the meaning of a program $\prg{R}$ as the least fixpoint of its
associated transformer $\alg{T_{R}}$, then this fixpoint semantics coincides with the
model-theoretic semantics as it happens in logic programming. In fact, this semantics
would correspond to the C-semantics in \cite{falaschi}.
% We will call this semantics the {\it least model semantics.}
\begin{definition}[Least model semantics] % .. .. .. .. .. .. .. .
\label{model:semantics} For each program $\prg{R}$ we define its least model semantics
as: $\,\modelsem{\prg{R}} =_{def} {\cal  M_{R}}$.
\end{definition}   % .. .. .. .. .. .. .. .. .. .. .. .. .. .. .. .. .

% ====================================================================

\section{An Algebra of CRWL-Program Modules}
\label{modules}

For designing large programs it is convenient to separate the whole
task into subtasks of manageable size and construct programs in a
structured fashion by combining and modifying smaller programs.  This
idea has been extended to many programming languages giving rise to
different notions of program module, each one being attached to a programming
paradigm.
In CRWL-programming we are going to follow an approach close to that developed in
\cite{brogi:tesis} for logic programming, where modules are open programs in the sense
that function definitions in a module can be completed with definitions for the same
functions in other modules. We will consider a global signature with bottom
$\Sigma_{\bot} = (DS_{\Sigma_{\bot}}, FS_{\Sigma_{\bot}})$ and a countable set ${\cal V}$
of variable symbols and will construct modules and module expressions with symbols of
these sets. $\Sigma_{\bot}$ and ${\cal V}$ will characterize the environment where
modules are written. Also we will consider all constructor symbols in
$DS_{\Sigma_{\bot}}$ common to all program modules as it is usual in other proposals of
modularity for declarative programming, like \cite{brogi:toplas,orejas}, where
compositionality and full abstraction are dealt with. With this decision we give up any
possibility of data abstraction and the only contribution of a program module to the
environment will be a set of (definition) rules for a subsignature of function symbols.
We will take this subsignature to denote the exportable resources of the module, and the
set of rules as its body.  In a program module, function symbols may appear ---in the rhs
of a rule--- with no definition rule in this module.  Although it may be assumed that all
function symbols are defined in each program module by assuming an implicit rule
$f(\overline{t}) \es \bot$ for each function symbol $f$ with no definition rule, these
symbols will be assumed to be provided by other modules and they will be taken to denote
the resources that have to be imported. They will be the parameters of the module.
From these considerations we propose the following definition for
the notion of module in CRWL-programming

\begin{definition}[Module]  % .. .. .. .. .. .. .. .. ..
\label{module} A \textit{module} in CRWL-programming is a tuple
$<\sigma_{p},\sigma_{e},\prg{R}>$ where
\begin{itemize}
 \item
 $\prg{R}$ is a set of program rules $f(\overline{t}) \es r \si C$ ($r\not= \bot$),
 \item
 $\sigma_{e}$ is the (exported) signature of function symbols with a
 definition rule in $\prg{R}$,
 \item
 $\sigma_{p}$ is the (parameter) signature of those function symbols with
 no definition rule in $\prg{R}$ that appear in any rule
 (i.e., they are invoked but not defined).
 \end{itemize}
\end{definition}
$\prg{R}$ is the \textit{body} of the module and $(\sigma_{p},\sigma_{e})$ its
\textit{interface}. The interface of a module could be inferred from its body if one
knows which are the constructor symbols. However, as we consider all constructor symbols
common to all program modules, we do not include an explicit declaration of these symbols
in any module and have to make explicit parameter signatures in order to distinguish
between function and constructor symbols. In this way, every symbol not occurring in
$\sigma_{e}$ nor $\sigma_{p}$ will be a constructor symbol. Next, we have an example of a
module definition.

\begin{example}
\label{OrdNatList} This example shows a module for constructing ordered lists of natural
numbers with functions for inserting elements, checking the type of an element, and
compare natural numbers.

\begin{footnotesize}
\begin{verbatim}
  OrdNatList =
    < {},                               % Parameter signature
      {isnat/1, leq/2, insert/2},       % Exported signature
      { isnat(zero)    -> true.
        isnat(succ(X)) -> isnat(X).
        leq(zero,zero)       -> true.
        leq(zero,succ(X))    -> isnat(X).
        leq(succ(X),zero)    -> false <= isnat(X) >< true.
        leq(succ(X),succ(Y)) -> leq(X,Y).
        insert(X,[])     -> [X]              <= isnat(X) >< true.
        insert(X,[Y|Ys]) -> [X|[Y|Ys]]       <= leq(X,Y) >< true.
        insert(X,[Y|Ys]) -> [Y|insert(X,Ys)] <= leq(X,Y) >< false.}>
\end{verbatim}
\end{footnotesize}
\end{example}
In this module the parameter signature is empty, and symbols like \verb#zero/0#,
\verb#succ/1#, \verb#[]/0#, \verb#[_|_]/2# with no definition rule are considered
constructor symbols, because they are not included in the parameter signature (and
obviously because they occur in arguments of left hand sides).

We write ${\bf PMod}(\Sigma_{\bot})$ for the class of all program modules which can be
defined with a signature $\Sigma_{\bot}$, ${\bf SubSig} (\Sigma_{\bot})$ for the set of
all subsignatures of a signature $\Sigma_{\bot}$, and ${\bf Prg}(\Sigma_{\bot})$ for the
class of all sets of rules (programs) which can be defined with $\Sigma_{\bot}$. On ${\bf
PMod}(\Sigma_{\bot})$ we define three projections:
\begin{itemize}
 \item
 $\mathit{par} \colon {\bf PMod}(\Sigma_{\bot}) \es  {\bf SubSig}(\Sigma_{\bot})$ such that
 $\mathit{par}(<\sigma_{p},\sigma_{e},\prg{R}>) = \sigma_{p}$,
 \item
 $\mathit{exp} \colon {\bf PMod}(\Sigma_{\bot}) \es  {\bf SubSig}(\Sigma_{\bot})$ such that
 $\mathit{exp}(<\sigma_{p},\sigma_{e},\prg{R}>) = \sigma_{e}$, and
 \item
 $\mathit{rl} \colon {\bf PMod}(\Sigma_{\bot}) \es {\bf Prg}(\Sigma_{\bot})$ such that
 $\mathit{rl}(<\sigma_{p},\sigma_{e},\prg{R}>) = \prg{R}$,
 \end{itemize}
which give respectively the parameter signature, the exported signature, and the body of
a module.

% --------------------------------------------------------------------

\subsection{Basic Operations on Modules}
\label{subsec.basicoperations}

In this section we present a set of basic operations with modules that allows us to
express typical features of modularization techniques such as information
hiding/abstraction, import/export relationships and inheritance related to function
symbols as is done in \cite{brogi:tesis}, but our set of operations is different
%(and more reduced)
and we give syntactic definitions for it. We use three operations: union of
programs, closure wrt\/ a signature and deletion of a signature, that are sufficient to
express the most extended ways of composing modules and their relationships, and we do
not need the intersection of programs, used in \cite{brogi:tesis} to model hiding,
because we directly deal with signatures in the closure. In order to give more
flexibility in expressing importation and instantiation, we also include a renaming
operation. We define our operations in such a way that all module expressions can be
reduced to a flat module $<\sigma_{p},\sigma_{e},\prg{R}>$ ---where $\prg{R}$ could be an
infinite set of rules.  This is something like a \textit{ presentation semantics}
\cite{wirsing}.

First we define the {\it union} of two modules as the module obtained
as the simple union of signatures and rules.

\begin{definition}[Union]  % .. .. .. .. .. .. .. .. .. .. .. .. .. ..
\label{union}
Given two modules
$\alg{P}_{1} = <\sigma^{1}_{p},\sigma^{1}_{e},\alg{R}_{1}>$ and
$\alg{P}_{2} = <\sigma^{2}_{p},\sigma^{2}_{e},\alg{R}_{2}>$,
their union is defined as the module:
$$
  \alg{P}_{1} \cup \alg{P}_{2} =_{def}
  <(\sigma^{1}_{p} \cup \sigma^{2}_{p}) \setminus
                     (\sigma^{1}_{e} \cup \sigma^{2}_{e}),
    \sigma^{1}_{e} \cup \sigma^{2}_{e},
    \alg{R}_{1} \cup \alg{R}_{2}>.
$$
\end{definition}   % .. .. .. .. .. .. .. .. .. .. .. .. .. .. .. . ..
Each argument in this operation is considered an open program that can
be extended or completed with the other argument possibly with
additional rules for its exported function symbols.

\begin{example}
\label{MoneyChange}
Let us consider the following module with a function to give change
for an amount of money. Values for coins are provided by the
non-deterministic function \verb#coin/0#, whereas \verb#getcoin/1#
gives different possibilities to get a coin for a fixed amount.
Finally, the function \verb#change/1# returns a list with the coins
corresponding to the change. In this example, we are assuming a
predefined arithmetic with the usual notation for natural numbers.
This was not the case in Example~\ref{OrdNatList}.

\begin{footnotesize}
\begin{verbatim}
  MoneyChange =
    < {_=<_/2, _-_/2},
      {coin/0,getcoin/1,change/1},
      { coin -> 1. coin -> 5. coin -> 10.
        getcoin(N) -> C <= coin >< C, C =< N >< true.
        change(0) -> [].
        change(N) -> [C|change(N-C)] <= getcoin(N) >< C. } >
\end{verbatim}
\end{footnotesize}

\noindent We can extend this module with another module for
providing new coins:
\begin{footnotesize}
\begin{verbatim}
  NewCoins = <{},{coin/0},{coin -> 15. coin -> 20.}>
\end{verbatim}
\end{footnotesize}
\noindent
 simply by joining them to obtain
 \vspace{3 mm}\\
\begin{footnotesize}
\verb#  MoneyChange# $\cup$ \verb#NewCoins =# \vspace{-2 mm}
\begin{verbatim}
    < {_=<_/2, _-_/2},
      {coin/0,getcoin/1,change/1},
      { coin -> 1. coin -> 5. coin -> 10. coin -> 15. coin -> 20.
        getcoin(N) -> C <= coin >< C, C =< N >< true.
        change(0) -> [].
        change(N) -> [C|change(N-C)] <= getcoin(N) >< C. } >
\end{verbatim}
\end{footnotesize}
\end{example}
Union of modules is idempotent, associative, commutative, and there exists a
null element: the module $\prg{O}=<\sigma_{o},\sigma_{o},\emptyset>$, where
$\sigma_{o}$ is the empty signature of function symbols, representing the
module with no rule.
\begin{proposition}  % .. .. .. .. .. .. .. .. .. .. .. .. .. .. .. ..
\label{unionprop} The union of modules has the following
properties:
\begin{enumerate}
 \item
  $\prg{P} \cup \prg{O}= \prg{P}$, for every module $\prg{P}$.
 \item
  $\prg{P} \cup \prg{P} = \prg{P}$, for every module $\prg{P}$.
 \item
  $(\prg{P} \cup \prg{P}_{1}) \cup \prg{P}_{2} =
   \prg{P} \cup (\prg{P}_{1} \cup \prg{P}_{2})$, for all modules
   $\prg{P}$, $\prg{P}_{1}$ and $\prg{P}_{2}$.
 \item
  $\prg{P}_{1} \cup \prg{P}_{2} = \prg{P}_{2} \cup \prg{P}_{1}$,
  for all modules  $\prg{P}_{1}$ and $\prg{P}_{2}$.
\end{enumerate}
\end{proposition}    % .. .. .. .. .. .. .. .. .. .. .. .. .. .. .. ..
\begin{proof} Obvious from the definition of the union of modules.
\end{proof}

The second operation is the {\it closure} of a module {\it wrt\/ a given signature}
$\sigma$. This operation makes accessible the signature $\sigma$ in an extensional way
(i.\/e. only provable approximations can be used) and hides the rest. To define this
operation, we need to introduce the notion of canonical rewrite rule.
\begin{definition}[Canonical rewrite rule]
\label{canonicalrule}
 Given a term $f(\overline{t})$, with $f\in\funs{n}$ and each $t_i\in {\bf CTerm}_\bot$,
 and $r\in{\bf CTerm}_\bot$, we define the canonical rewrite rule $crr(f(\overline{t}),r)$
 which reduces $f(\overline{t})$ to $r$, as the rule
       $ f(\overline{t}') \es r \si C$,
 constructed by substituting in $\overline{t}$ each occurrence of a repeated variable $X$
 or $\bot$ with a fresh variable $Y$ and adding in $C$ a joinability statement
 $X \bowtie Y$ for each occurrence of a repeated variable $X$, and a statement $X \bowtie X$
 for each variable $X$ in $r$ and each variable with only one occurrence in $\overline{t}$.
\end{definition}

In this way we obtain a program rule (with $\overline{t}'$ linear and each $t'_i \in {\bf
CTerm}$) from which $f(\overline{t}) \es r$ can be proved, because for
$\theta_{\overline{t}}\in {\bf CSubst}_\bot$ such that $\theta_{\overline{t}}(Y) = X$ for
each fresh variable $Y$ that substitutes an occurrence of $X$ in $\overline{t}$,
$\theta_{\overline{t}}(Y) = \bot$ for each fresh variable $Y$ that substitutes an
occurrence of $\bot$, and $\theta_{\overline{t}}(X) = X$ for  all other variables,
$C\theta_{\overline{t}}$ always can be proved and $(f(\overline{t}') \es
r)\theta_{\overline{t}}$ is $f(\overline{t}) \es r$.

\begin{example}
\label{canonicalruleex} The canonical rewrite rule which reduces $f(\bot,b(X,Y),X)$ to
$a(X,Z)$ is:
\begin{eqnarray*}
   f(V,b(X,Y),X1) \es a(X,Z) & \si & \{X1 \bowtie X, Y \bowtie Y, Z \bowtie Z\},
\end{eqnarray*}
and the associated substitution $\theta_{\overline{t}}$ is such that
$\theta_{\overline{t}}(X1)=X$, $\theta_{\overline{t}}(V)=\bot$, and
$\theta_{\overline{t}}(W)=W$ for all other variables $W$. In this case
$C\theta_{\overline{t}} = \{X \bowtie X, Y \bowtie Y, Z \bowtie Z\}$ and all these
joinability statements can be trivially derived from ({\bf RR}) and ({\bf Jo}), and
therefore $f(\bot,b(X,Y),X) \es  a(X,Z)$ by the ({\bf OR}) rule.
\end{example}
 Now, we can define the closure of a module as follows.
\begin{definition}[Closure wrt\/ a signature]  % .. .. .. .. .. .. ..
\label{closure} Given a module $\prg{P} = <\sigma_{p},\sigma_{e},\alg{R}>$, its
closure $\overline{\alg{P}}^{\sigma}$ wrt\/ a signature of function symbols
$\sigma$ is defined as the module:
 $$
  <\sigma_{o}, \sigma'_{e},
   \{crr(f(\overline{t}),r) \mid
        f/n \in \sigma,\ r \in {\bf CTerm}_\bot,\ r\not=\bot,\
         \alg{R} \vdash_{CRWL} f(\overline{t}) \rightarrow r \} >,
 $$
\noindent where $\sigma_{o}$ denotes the empty signature of function symbols, $t_{i} \in
{\bf CTerm}_{\bot}$ for each component of the tuple $\overline{t}$, and $\sigma'_{e}$ is
the corresponding exported signature.
\end{definition}  % .. .. .. .. .. .. .. .. .. .. .. .. .. .. .. .. ..

The closure of a module is a module with a possibly infinite set
of rules (although the exported signature is always finite)
equivalent to the union of the graphs in $\alg{M}_\prg{P}$ of all
functions defined in $\alg{P}$ and contained in $\sigma$. Note
that $\sigma'_{e} \subseteq \sigma_{e} \cap \sigma$ because a
function in $\sigma_{e} \cap \sigma$ that depends on functions in
the parameter signature could remain with no definition rule
---or with the only rule $f(\overline{t}) \es \bot$--- after
closing the module. As a syntactic
simplification we will write $\overline{\alg{P}}$ instead of
$\overline{\alg{P}}^{\sigma_{e}}$ for each module $\alg{P} =
<\sigma_{p},\sigma_{e},\alg{R}>$.

\begin{example}
\label{WeekDays}
Let us consider the following module about week days, where two
functions are defined to get the \verb#next# day and the day
\verb#before# of a given day.

{\footnotesize
\begin{verbatim}
  WeekDays = < {},
               {next/1,before/1},
               { next(mo) -> tu.   next(tu) -> we.   next(we) -> th.
                 next(th) -> fr.   next(fr) -> sa.   next(sa) -> su.
                 next(su) -> mo.
                 before(X) -> Y <= next(Y) >< X. }                     >
\end{verbatim}
}

\noindent The closure of this module wrt its whole exported signature is the
module \vspace{3 mm}\\ {\footnotesize \verb#  # $\overline{\verb#WeekDays#}$
\verb#=# \vspace{-5.32 mm}
\begin{verbatim}
             < {},
               {next/1,before/1},
               { next(mo) -> tu.   next(tu) -> we.   next(we) -> th.
                 next(th) -> fr.   next(fr) -> sa.   next(sa) -> su.
                 next(su) -> mo.
                 before(tu) -> mo.  before(we) -> tu.  before(th) -> we.
                 before(fr) -> th.  before(sa) -> fr.  before(su) -> sa.
                 before(mo) -> su. }            >
\end{verbatim}
 }
\end{example}

Closure wrt  a signature is in some way the counterpart of the encapsulation
operation `$*$' in \cite{brogi:tesis}, but it is more general because it has a
twofold effect: hiding all rules in the module and restricting the visible
signature, so we need no intersection of modules ---as is needed in
\cite{brogi:tesis}--- to restrict visibility in a closed module.  Variables and
bottom can appear in the rules of a closed module, but no functions in the
parameter signature.
\begin{proposition}  % .. .. .. .. .. .. .. .. .. .. .. .. .. .. .. ..
\label{closureprop} Closure of modules has the following properties, where
$\sigma$, $\sigma_1$ and $\sigma_2$ are signatures of function symbols,
\begin{enumerate}
 \item
   $\overline{\prg{P}}^{\sigma} = \prg{O}$, for every module
   $\prg{P}$ and every signature $\sigma$ such that
   $\sigma \cap \mathit{ exp}(\prg{P}) = \sigma_{o}$.
 \item
   $\overline{\prg{O}}^{\sigma} = \prg{O}$, for every signature
    $\sigma$ and the null module $\prg{O}$.
 \item
   $\overline{\prg{P}}^{\sigma_{1} \cup \sigma_{2}} =
    \overline{\prg{P}}^{\sigma_{1}} \cup
    \overline{\prg{P}}^{\sigma_{2}}$, for
   every module  $\prg{P}$ and signatures
   $\sigma_{1}$, $\sigma_{2}$.
 \item
   $\overline{\overline{\prg{P}}^{\sigma_{1}}}^{\sigma_{2}} =
    \overline{\prg{P}}^{\sigma_{1} \cap \sigma_{2}} =
    \overline{\overline{\prg{P}}^{\sigma_{2}}}^{\sigma_{1}}$, for
   every module  $\prg{P}$ and signatures
   $\sigma_{1}$, $\sigma_{2}$.
 \item
   $\overline{\prg{P}_{1} \cup \prg{P}_{2}}^{\sigma} =
    \overline{\prg{P}_{1}}^{\sigma} \cup \overline{\prg{P}_{2}}^{\sigma}$,
    for modules $\prg{P}_{1}$ and $\prg{P}_{2}$ defining disjoint signatures
    and such that neither $\prg{P}_{1}$ nor $\prg{P}_{2}$ use the
    signature defined in the other module.
\end{enumerate}
\end{proposition}    % .. .. .. .. .. .. .. .. .. .. .. .. .. .. .. ..
\begin{proof} Obvious from the definitions of the closure and the
union of modules.
\end{proof}

Our third operation is the {\it deletion of a signature}  in a
module.

\begin{definition}[Deletion of a signature]  % .. .. .. .. .. .. .. ..
\label{deletion}
Given a module $\alg{P} = <\sigma_{p},\sigma_{e},\prg{R}>$,
the deletion in $\alg{P}$ of a signature of function symbols $\sigma$
produces the module:
 \[
  \prg{P} \setminus \sigma =_{def}
  <\sigma'_{p} , \sigma_{e} \setminus \sigma,\prg{R} \setminus \sigma>,
 \]
where $\prg{R} \setminus \sigma$  denotes the set of those rules in $\prg{R}$
defining function symbols not appearing in $\sigma$, and $\sigma'_{p}$ denotes
the corresponding parameter signature.
\end{definition}  % .. .. .. .. .. .. .. .. .. .. .. .. .. .. .. .. ..

We do not give an explicit expression for $\mathit{ par}(\prg{P} \setminus
\sigma)$ in terms of $\mathit{ par}(\prg{P})$ because new parameters can appear
and old ones can disappear with the deletion of rules in $\mathit{
rl}(\prg{P})$. However,  $\mathit{ par}(\prg{P} \setminus \sigma) \subseteq
\sigma_{p} \cup (\sigma_{e} \cap \sigma)$ is satisfied.

\begin{example}
\label{DeletionOrdNatList}
In the module \verb#OrdNatList# of Example~\ref{OrdNatList} we can
delete or abstract the signature \verb#{isnat/1,leq/2}# to obtain
the following parameterized module

{\footnotesize
\begin{verbatim}
  OrdNatList\{isnat/1,leq/2} =
    < {isnat/1,leq/2},
      {insert/2},
      { insert(X,[])     -> [X]              <= isnat(X) >< true.
        insert(X,[Y|Ys]) -> [X|[Y|Ys]]       <= leq(X,Y) >< true.
        insert(X,[Y|Ys]) -> [Y|insert(X,Ys)] <= leq(X,Y) >< false. } >
\end{verbatim}
 }

\noindent The resulting module is now parameterized by the two symbol functions
\verb#isnat/1# and \verb#leq/2#, whereas only the function \verb#insert/2# is
exported.
\end{example}

This operation recalls the {\it undefine} clause in the object-oriented
language Eiffel, and we will use it (combined with the union) to perform
inheritance with overriding. Note the differences between the deletion of a
signature and the closure wrt a signature. The former operation removes rules
defining function symbols in the signature ---but not those rules containing
invocations in their rhs or condition--- whereas the latter only hides the
definitions of the functions in the signature, but maintains their consequences
---hiding all other functions.

\begin{proposition}  % .. .. .. .. .. .. .. .. .. .. .. .. .. .. .. ..
\label{deletionprop} The deletion of a signature (of function symbols) in a
module has the following properties, where $\sigma$, $\sigma_1$ and $\sigma_2$
are signatures of function symbols,
\begin{enumerate}
 \item
   $\prg{P} \setminus \sigma = \prg{O}$, for every module $\prg{P}$ and
   every $\sigma$ such that $\mathit{ exp}(\prg{P}) \subseteq \sigma$.
 \item
   $\prg{P} \setminus \sigma = \prg{P}$, for every module $\prg{P}$ and
   every  $\sigma$ such that $\mathit{ exp}(\prg{P}) \cap \sigma = \sigma_{o}$.
 \item
   $(\prg{P} \setminus \sigma_{1})\setminus \sigma_{2} =
     \prg{P} \setminus (\sigma_{1} \cup \sigma_{2}) =
    (\prg{P} \setminus \sigma_{2})\setminus \sigma_{1}$, for all  modules
   $\prg{P}$ and  $\sigma_{1}$, $\sigma_{2}$.
 \item
   $(\prg{P}_{1} \cup \prg{P}_{2}) \setminus \sigma =
    (\prg{P}_{1} \setminus \sigma) \cup (\prg{P}_{2} \setminus \sigma)$,
   for all modules $\prg{P}_{1}$, $\prg{P}_{2}$ and signatures
   $\sigma$.
 \item
   $(\overline{\prg{P}}^{\sigma_{1}}) \setminus \sigma_{2} =
   \overline{\prg{P}}^{(\sigma_{1} \setminus \sigma_{2})}$,
   for all modules $\prg{P}$ and signatures $\sigma_{1}$, $\sigma_{2}$.
 \item
   $\overline{\prg{P}}^{\sigma} = \overline{\prg{P}} \setminus
   (\sigma_{e} \setminus \sigma)$, for a module $\prg{P}$, with
   exported signature $\sigma_{e}$, and all  $\sigma$.
\end{enumerate}
\end{proposition}    % .. .. .. .. .. .. .. .. .. .. .. .. .. .. .. ..
\begin{proof} Obvious from the definitions of the deletion, union
and closure.
\end{proof}

Finally, we introduce a renaming operation that allows us to change function symbols with
other function symbols of the same arity, in the global signature $\Sigma_{\bot}$.
Therefore, given a module $\prg{P}$ and a function symbols  renaming $\rho$, we define
the renaming of $\prg{P}$ by $\rho$ as a new module $\rho(\prg{P})$ where rules are
conveniently renamed. The following definition formalizes this idea.

\begin{definition}[Renaming]  % .. .. .. .. .. .. .. .. .. .. .. .. ..
\label{renaming}
 Given a module $\alg{P} = <\sigma_{p},\sigma_{e},\prg{R}>$ and
a function symbol renaming $\rho$, $\alg{P}$ renamed by $\rho$ is the module $$
  \rho(\prg{P}) =_{def}
  <\rho^{\ast}(\sigma_{p}) \setminus \rho^{\ast}(\sigma_{e}),
     \rho^{\ast}(\sigma_{e}),\rho^{\ast}(\prg{R})>,
$$ where $\rho^{\ast}(\sigma)$ is the signature resulting from applying $\rho$
to all symbols in $\sigma$, and $\rho^{\ast}(\prg{R})$ is the set of rules
resulting from applying $\rho$ to all rules in $\prg{R}$.
\end{definition}  % .. .. .. .. .. .. .. .. .. .. .. .. .. .. .. .. ..

The following example illustrates the usefulness of this operation
to adequate parameter names of a module.

\begin{example}
\label{RenamingOrdNatList}
In the module \verb#OrdNatList\{isnat/1,leq/2}# of
Example~\ref{DeletionOrdNatList} we can rename the function symbol
\verb#isnat/1# with the new name
\verb#isbasetype/1# to obtain a more appropriate parameterized module

{\footnotesize
\begin{verbatim}
  OrdList = {isnat/1 -> isbasetype/1}(OrdNatList\{isnat/1,leq/2}),
\end{verbatim}
 }
\noindent where we have denoted the corresponding renaming function $\rho$ as
the set of pairs $f/n \es \rho(f/n)$ such that $f/n \not= \rho(f/n)$. This
module has the following appearance

{\footnotesize
\begin{verbatim}
  OrdList =
    <{isbasetype/1,leq/2},
     {insert/2},
     {insert(X,[])     -> [X]              <= isbasetype(X) >< true.
      insert(X,[Y|Ys]) -> [X|[Y|Ys]]       <= leq(X,Y) >< true.
      insert(X,[Y|Ys]) -> [Y|insert(X,Ys)] <= leq(X,Y) >< false.} >
\end{verbatim}
 }
 \noindent
Now, the parameters become \verb#isbasetype/2# and \verb#leq/2#.
\end{example}

We will use this operation to change function names in exportation,
importation and, specially, in instantiation for matching function
names in the parameter signature of a module with function names in
the exported signature of another module. See
Section~\ref{ModularConstructs} for some illustrative examples.

\begin{proposition}  % .. .. .. .. .. .. .. .. .. .. .. .. .. .. .. ..
\label{renamingprop} Renaming of modules has the following properties, where
$\rho$, $\rho_1$ and $\rho_2$ are function symbol renamings,
\begin{enumerate}
 \item
   $\iota(\prg{P}) = \prg{P}$, for every module $\prg{P}$, where $\iota$
   is the identity renaming.
 \item
   $\rho(\prg{O}) = \prg{O}$, for every  $\rho$.
 \item
   $\rho_{2}(\rho_{1}(\prg{P})) = (\rho_{2} \circ \rho_{1})(\prg{P})$,
   for all modules $\prg{P}$ and all $\rho_{1}$, $\rho_{2}$.
 \item
   $\rho(\prg{P}_{1} \cup \prg{P}_{2}) = \rho(\prg{P}_{1})
            \cup \rho(\prg{P}_{2})$, for all modules
   $\prg{P}_{1}$, $\prg{P}_{2}$ and all $\rho$.
 \item
   $\rho(\overline{\prg{P}}^{\sigma}) =
           \overline{\rho(\prg{P})}^{\rho^{\ast}(\sigma)}$,
   for all modules $\prg{P}$, signatures $\sigma$ and injective
   $\rho$.
 \item
   $\rho(\prg{P} \setminus \sigma) = \rho(\prg{P}) \setminus
   \rho^{\ast}(\sigma)$, for all modules $\prg{P}$, signatures $\sigma$ and
   injective $\rho$.
\end{enumerate}
\end{proposition}    % .. .. .. .. .. .. .. .. .. .. .. .. .. .. .. ..
\begin{proof} Obvious from the definitions of deletion, union,
closure and renaming.
\end{proof}

% ====================================================================

\subsection{Other Modular Constructions in CRWL-programming}
\label{ModularConstructs} Our notion of module is basically that of a program
inside a context made up of other programs providing explicit rules for
function symbols and implicit declarations of constructor symbols, all together
defining a global signature $\Sigma_{\bot}$.  In this section, we will show how
the operations that we have defined above can be used to model typical module
interconnections used in conventional modular programming languages. We will
introduce new operations with modules for these relationships, but all these
will be defined as derived expressions from the basic set.  These expressions
will reflect the relationship between the module denoted by the expression and
its component modules, and the resulting modules will be interpreted as flat
modules in all cases.

The closure of a module $\alg{M}$ wrt a signature $\sigma$ gives a form of
encapsulation, hiding those function symbols in $\alg{M}$ that are not in
$\sigma$, and making the function symbols in $\alg{M}$ and $\sigma$ visible but
only in an extensional way, i.e., by the results ---as partial constructor
$\Sigma$-terms--- of the function applications to constructor $\Sigma$-terms
(including variables). Thus, we can provide an {\it export with encapsulation}
operation `$\square$' over modules, in this simple way
\[
 \sigma \square \alg{M} =_{def} \overline{\alg{M}}^{\sigma}.
\]

The union of modules reflects the behavior of some logic programming systems
that allow adding new programs ---saved in separate files--- to the main
database. With this operation, but modifying one of its arguments, we can
express different forms of importation and instantiation. We can define an {\it
import} operation $\ll$ between modules as the union of a module $\alg{M}$
---representing the body of the importing module--- with the
closure of the imported module $\alg{N}$ as follows
 \[
   \alg{M} \ll \alg{N} =_{def} \alg{M} \cup \overline{\alg{N}}.
 \]
Module $\alg{M} \ll \alg{N}$ imports $\alg{N}$,
%or $\alg{N}$ is imported by $\alg{M} \ll \alg{N}$,
which means that only the consequences of the functions defined in $\alg{N}$ are
imported, and not their rules. When $\mathit{ exp}(\alg{M}) \cap \mathit{ exp}(\alg{N}) =
\sigma_{o}$ we have a typical importation because functions defined in $\alg{N}$ are only
reduced in $\alg{N}$.
 We can also express {\it selective importation} of a signature
$\sigma$ from $\alg{N}$ by combining importation with exportation,
in order to restrict the visible signature of the imported module:
 \[
   \alg{M} \ll (\sigma \square \alg{N}), \mbox{ with } \sigma \subseteq \mathit{ exp}(\alg{N})
 \]
This expression is equivalent to $\alg{M} \cup \overline{\alg{N}}^{\sigma}$ by
Proposition~\ref{closureprop}({\it 4}). {\it Multiple importation} or (selective)
importation from several modules can be written as
 \[
  ( \dots (\alg{M} \ll (\sigma_{1} \square \alg{N}_{1})) \ldots )
    \ll (\sigma_{k} \square \alg{N}_{k}),
 \]
where the importation order is not relevant by  Propositions~\ref{unionprop}({\it 3,4})
and~\ref{closureprop}({\it 4,5}). It can be easily proved that this expression is
equivalent to the single importation
 \[
  \alg{M} \ll ((\sigma_{1} \square \alg{N}_{1}) \cup \ldots \cup
  (\sigma_{k} \square \alg{N}_{k})).
 \]
{\it Importation with renaming} can be expressed by an expression of
the form
 \[
    \alg{M} \ll \rho(\sigma \square \alg{N})
 \]
with $\sigma \subseteq \mathit{ exp}(\alg{N})$,  and an injective function
symbol renaming $\rho$ (see Proposition~\ref{renamingprop}({\it 5})). By the
properties of renaming this expression is equivalent to
 \[
    \alg{M} \ll (\rho^{\ast}(\sigma) \square \rho(\alg{N}))
 \]
and can be reduced to $\alg{M} \cup\rho(\overline{\alg{N}}^{\sigma})$.

\begin{example}
\label{Importation}
Let us consider the module \verb#OrdList# in Example~\ref{RenamingOrdNatList}
and the new module

{\footnotesize
\begin{verbatim}
  OrdNat =
    < {},
      {isnat/1, leq/2, geq/2},
      { isnat(zero)    -> true.
        isnat(succ(X)) -> isnat(X).
        leq(zero,zero)       -> true.
        leq(zero,succ(X))    -> isnat(X).
        leq(succ(X),zero)    -> false <= isnat(X) >< true.
        leq(succ(X),succ(Y)) -> leq(X,Y).
        geq(X,Y)             -> leq(Y,X). } >
\end{verbatim}
}

 \noindent
where we define the predicate \verb#isnat/1# and the two order relationships
\verb#leq/2# (less than or equal to) and \verb#geq/2# (greater than or equal
to). The importation
 {\footnotesize
  \[
     \verb#OrdList# \ll \verb#{isnat/1 -> isbasetype/1}(OrdNat)#
  \]
 }
 \noindent
is a module with an infinite number of rules for \verb#isbasetype/1#, \verb#leq/2# and
\verb#geq/2# (all possible reductions to true or false), that behaves as calls to
\verb#isbasetype/1# and \verb#leq/2# are reduced in \verb#OrdNat# as calls to
\verb#isnat/1# and \verb#leq/2# itself respectively.
\end{example}

Thus a typical program $\alg{M}$ with a hierarchical structure in the sense of standard
modular programming, i.e., importing from several modules $\alg{N}_{1},\ldots,
\alg{N}_{k}$, possibly with renaming, can be built up from a plain program $\alg{P}$
---its body--- and the imported modules as
\[
  \alg{M} = \alg{P} \ll (\rho_{1}(\sigma_{1} \square \alg{N}_{1})
  \cup \ldots \cup \rho_{k}(\sigma_{k} \square \alg{N}_{k})),
\]
with $\sigma_{1} \subseteq \mathit{ exp}(\alg{N}_{1})$, \ldots , $\sigma_{k} \subseteq
\mathit{ exp}(\alg{N}_{k})$ and $\mathit{ par}(\alg{P}) \subseteq
(\rho^{\ast}_{1}(\sigma_{1}) \cup \ldots \cup \rho^{\ast}_{k}(\sigma_{k}))$. This
expression can be reduced to $\alg{P} \cup \rho_{1}(\overline{\alg{N}_{1}}^{\sigma_{1}})
\cup \ldots \cup \rho_{k}(\overline{\alg{N}_{k}}^{\sigma_{k}})$.

Because our basic modules can be parameterized, we can {\it instantiate} function symbols
of the parameterized signature of a module $\alg{M}$ with function symbols, of the same
arity but different name, exported by other module $\alg{N}$, simply by renaming suitably
the parameters of $\alg{M}$ to fit (a part of) the exported signature of $\alg{N}$. Thus
we obtain an {\it instantiation} operation that we denote $\alg{M}[\alg{N},{\rho}]$ and
define as
 \[
  \alg{M}[\alg{N},{\rho}] =_{def} \rho(\alg{M}) \ll \alg{N},
 \]
where $\rho$ is the function symbol renaming that characterizes the
instantiation. This operation makes sense when $\rho^{\ast}(\mathit{
par}(\alg{M}))\cap \mathit{ exp}(\alg{N})\not= \sigma_{o}$.
 When $\mathit{ par}(\rho(\alg{M})) \subseteq \mathit{
exp}(\alg{N})$ the instantiation is total and is partial in another case. Note that {\it
instantiation} can be seen as a special form of importation. The difference between a
(renamed) importation $\alg{M} \ll \rho(\alg{N})$ and an instantiation $\rho(\alg{M}) \ll
\alg{N}$ is that in the former, symbols in the parameter signature of $\alg{M}$ refer to
actual names in the exported signature of the imported module $\alg{N}$ (renamed by
$\rho$), whereas in the latter, symbols in the parameter signature of $\alg{M}$ behave as
true parameters being replaced (by $\rho$) with actual values of the exported signature
of $\alg{N}$.

\begin{example}
\label{Parameterization} Let us consider again the module \verb#OrdList# in
Example~\ref{RenamingOrdNatList} and the module \verb#OrdNat# defined in
Example~\ref{Importation}. The instantiation

 {\footnotesize
  \[
   \verb#OrdList[OrdNat,{isbasetype/1 -> isnat/1, leq/2 ->geq/2}]#
  \]
 }

 \noindent
is equivalent to a module, also with an infinite number of rules, but
defining the predicates  \verb#isnat/1# and \verb#geq/2# instead of
\verb#isbasetype/1# and \verb#leq/2# respectively.
\end{example}

Deletion of a signature $\sigma$ in a module removes  all rules defining function symbols
in that signature but maintains the occurrences of these symbols in the rhs of the other
rules. This operation can be used to \textit{ abstract a signature} $\sigma$ from a
module $\alg{M}$ in the following way
 \[
 \alg{M} [\sigma] =_{def} \alg{M} \setminus \sigma.
 \]
This abstraction operation makes sense when $\sigma \subseteq \mathit{
exp}(\alg{M})$ and each function symbol in $\sigma$ appears in some rule of
$\mathit{ rl}(\alg{M} \setminus \sigma)$. This operation is very useful for
making generic modules from concrete ones but unfortunately it is not
implemented in conventional modular programming systems. As an example of the
use of this operation we refer to Example~\ref{DeletionOrdNatList}.
 Also, with the deletion operation, we can model a
sort of inheritance relationship between modules. {\it Inheritance
with overriding} may be captured by means of union and deletion of
a signature in the following way
 \[
   \alg{M} \, \mathit{isa} \, \alg{N} =_{def} \alg{M} \cup (\alg{N}
                    \setminus \mathit{exp}(\alg{M})).
 \]
Module $\alg{M} \, \mathit{isa} \, \alg{N}$ inherits all functions in $\alg{N}$
---with their rules--- not defined in $\alg{M}$ and uses the rules of $\alg{M}$ for all
functions defined in $\alg{M}$, overriding the definition rules in $\alg{N}$, for common
functions. In this case, overriding is carried out by deleting the common signature of
the inherited module before adding it to the derived module.

\begin{example}
\label{Inheritance} Let us consider a module defining some operations on polygonal lines
and parameterized wrt\/ an addition operation \verb#_+_/2#, a predicate \verb#ispoint/1#
to test if something is a point, and operations \verb#distance/2# and
\verb#translatepoint/2# for computing the distance between points and the point resulting
of applying a translation, given by a vector (its second argument), to another point (its
first argument).

{\footnotesize
\begin{verbatim}
  Polygonal =
  <{_+_/2, ispoint/1, distance/2, translatepoint/2 },
   {perimeter/1,translate/2 },
   {perimeter([P1])         -> zero <= ispoint(P1) >< true.
    perimeter([P1|[P2|Ps]]) -> distance(P1,P2)+perimeter([P2|Ps]).
    translate([P1],V)         -> [translatepoint(P1,V)].
    translate([P1|[P2|Ps]],V) -> [translatepoint(P1,V)|translate([P2|Ps],V)].} >
\end{verbatim}
 }

 \noindent
(where we suppose that \verb#distance/2# and \verb#translatepoint/2# check that
their arguments are points). Let us also consider another module defining some
operations on squares and also parameterized wrt\/ a multiplication operation
\verb#_*_/2#, and the above operations \verb#ispoint/1# and \verb#distance/2#.

{\footnotesize
\begin{verbatim}
  Square =
  < {_*_/2, ispoint/1, distance/2},
    {issquare/1, side/1, perimeter/1, surface/1},
    {issquare([P1,P2,P3,P4]) -> true <= distance(P1,P2) >< distance(P2,P3),
                                        distance(P2,P3) >< distance(P3,P4),
                                        distance(P1,P2) >< distance(P3,P4).
      side([P1,P2,P3,P4]) -> distance(P1,P2) <= issquare([P1,P2,P3,P4]) >< true.
      perimeter(C) -> 4*side(C)       <= issquare(C) >< true.
      surface(C)   -> side(C)*side(C) <= issquare(C) >< true.} >.
\end{verbatim}
 }

 \noindent
With these modules we could define a new module
\verb#SquarePolygone# making module \verb#Square# inherit from
\verb#Polygonal#,

{\footnotesize
  \[
            \verb# SquarePolygone = Square isa Polygonal.#
  \]
 }

 \noindent
The resulting module would be

{\footnotesize
\begin{verbatim}
  SquarePolygone =
  < {_+_/2, _*_/2, ispoint/1, distance/2, translatepoint/2},
    {issquare/1, side/1, perimeter/1, surface/1, translate/2},
    {issquare([P1,P2,P3,P4]) -> true <= distance(P1,P2) >< distance(P2,P3),
                                        distance(P2,P3) >< distance(P3,P4),
                                        distance(P1,P2) >< distance(P3,P4).
      side([P1,P2,P3,P4]) -> distance(P1,P2) <= issquare([P1,P2,P3,P4]) >< true.
      perimeter(C) -> 4*side(C)       <= issquare(C) >< true.
      surface(C)   -> side(C)*side(C) <= issquare(C) >< true.
      translate([P1],V)         -> [translatepoint(P1,V)].
      translate([P1|[P2|Ps]],V) -> [translatepoint(P1,V)|translate([P2|Ps],V)].} >.
\end{verbatim}
 }

 \noindent
Note that \verb#perimeter/1#, defined in the module \verb#Polygonal#, has been
redefined with the version of the module \verb#Square#. The function
\verb#translate/2# has been inherited from \verb#Polygonal#.
\end{example}

% --------------------------------------------------------------------

\section{A Compositional Semantics}
\label{compositionalSem}

A module is basically a program because its interface can be
extracted from its set of rules when we know the data constructor
symbols, and operations defined on modules are operations on their
sets of rules, i.e., operations on programs. The difference
between a program and a program module is that a module can be
thought of as a program piece that can be assembled with other
pieces to build larger programs (this is one of the main reasons
of making explicit their interfaces).

With this idea in mind, the model-theoretic semantics defined for
CRWL-programs is not suitable for program modules because it is not
 compositional wrt the operations defined over
modules as we can see in the following example.

\begin{example}
\label{noCompSem} Let $\Sigma$ be a signature $(\{ a/0, b/0, c/0
\},\{ p/1, r/1 \})$, and modules $\alg{P}_{1}$ and $\alg{P}_{2}$
with the following sets of rules:
 \[
 rl(\alg{P}_{1}) = \{ p(a) \rightarrow c \} \ \ \
 rl(\alg{P}_{2}) = \{  p(a) \rightarrow c, \,
                       r(b) \rightarrow c \Leftarrow p(b) \bowtie c\}.
 \]
These modules have the same model-theoretic semantics, ${\cal M}_{\alg{P}_1}={\cal
M}_{\alg{P}_2}$, which is the CRWL-algebra $\alg{A}$ with functions $p^{\alg{A}}$ and
$r^{\alg{A}}$ such that
 $$
  \begin{array}{l}
  p^{\alg{A}}(a) = \{c,\bot\},\  p^{\alg{A}}(b) =  p^{\alg{A}}(c) =
  p^{\alg{A}}(\bot) = \{\bot\},\  p^{\alg{A}}(X) =\{\bot\}, \; \forall X\in\alg{V} \\

  r^{\alg{A}}(a) =\{\bot\},\quad\ r^{\alg{A}}(b) = r^{\alg{A}}(c) =
  r^{\alg{A}}(\bot) =\{\bot\},\ r^{\alg{A}}(X) =\{\bot\}, \; \forall X\in\alg{V}.
  \end{array}
%\begin{array}{ll}
%  p^{\alg{A}}(a) = \{c,\bot\},   &    r^{\alg{A}}(a) =\{\bot\}, \\
%  p^{\alg{A}}(b) = \{\bot\},     &    r^{\alg{A}}(b) =\{\bot\}, \\
%  p^{\alg{A}}(c) = \{\bot\},     &    r^{\alg{A}}(c) =\{\bot\}, \\
%  p^{\alg{A}}(\bot) = \{\bot\},  &    r^{\alg{A}}(\bot) =\{\bot\}, \\
%  p^{\alg{A}}(X) =\{\bot\}, \; \forall X\in\alg{V} \ \ \ \ \ \
%                  & r^{\alg{A}}(X) =\{\bot\}, \; \forall X\in\alg{V}
%\end{array}
 $$
However, their unions with  $\alg{Q}$, such that
 $rl(\alg{Q})=\{ p(b) \rightarrow c \}$, have different model-theoretic
semantics. The intended model of $\alg{P}_{1} \cup \alg{Q}$ has a
function $r^{{\cal M}_{\alg{P}_1 \cup \alg{Q}}}$ such that
$r^{{\cal M}_{\alg{P}_1 \cup \alg{Q}}}(b) = \{\bot\}$,
whereas $r^{{\cal M}_{\alg{P}_2 \cup \alg{Q}}}(b) = \{c,\bot\}$. So,
${\cal M}_{\alg{P}_1 \cup \alg{Q}}\neq{\cal M}_{\alg{P}_2 \cup \alg{Q}}$.
\end{example}

The compositionality of the semantics of a programming language is
particularly relevant when modularity is involved. In fact, one of
the most critical aspects in modular systems is the possibility of
making a separate compilation of modules, and this can only be
made in the presence of some kind of compositionality.

\subsection{Compositionality and Full Abstraction}
\label{composfullabst}
In order to study the compositionality and full abstraction of a
semantics, we have to clearly set out these notions. We will adopt
the approach proposed in \cite{brogi:TCS}, where compositionality and
full abstraction are defined in terms of the equivalence relation
between programs induced by the semantics.

\begin{definition}[Compositional relation]   % .. .. .. .. .. .. .. .. .. .. .. .. .. .. .. ..
Given an equivalence relation $\equiv$ defined between programs, an observation function
{\it Ob} defined for programs, and a set {\it Oper} of operations with programs, we say
that
\begin{enumerate}
\item
  $\equiv$ {\it preserves Ob} iff for all programs $\prg{P}$ and $\prg{Q}$,
  $
        \prg{P} \equiv \prg{Q} \ \Rightarrow \
                Ob(\prg{P}) = Ob(\prg{Q});
  $
\item
  $\equiv$ is a {\it congruence} wrt {\it Oper} iff for all programs $\prg{P}_i$ and $\prg{Q}_i$
  and all  $O \in \mathit{ Oper}$,
  $\prg{P}_i \equiv \prg{Q}_i$, for $i=1,\dots, n$, implies
       $ O(\prg{P}_1,\dots,\prg{P}_n) \equiv O(\prg{Q}_1,\dots,\prg{Q}_n)$;
\item
  $\equiv$ is {\it compositional} wrt $(\mathit{Ob},\mathit{Oper})$
  iff it is a congruence wrt {\it Oper} and preserves $Ob$.
\end{enumerate}
\end{definition}
% .. .. .. .. .. .. .. .. .. .. .. .. .. .. .. ..

To set the notion of full abstraction for an equivalence relation, we need some way of
distinguishing programs and for that reason we introduce the notion of \textit{context}.
Given a set of operations on programs {\it Oper}, and a metavariable $\alg{X}$, we define
contexts $C\context{\alg{X}}$ inductively as follows: $\alg{X}$ and each program is a
context, also for each operation $O\in\mathit{ Oper}$ with $n$ program arguments and
$C_1,\ldots,C_n$ contexts, $O(C_1,\ldots,C_n)$ is a context.
Two programs ${\cal P}$ and ${\cal Q}$ are {\it distinguishable} under
$(\mathit{Ob},\mathit{Oper})$ if there exists a context $C\context{\alg{X}}$ such that
$C\context{{\cal P}}$ and $C\context{{\cal Q}}$ have different external behavior, i.\/e.
    $ Ob(C\context{{\cal P}}) \neq Ob(C\context{{\cal Q}})$.
When ${\cal P}$ and ${\cal Q}$ are indistinguishable under $(\mathit{Ob},\mathit{Oper})$
we will write ${\cal P} \cong_\mathit{{Ob},\mathit{Oper}} {\cal Q}$, i.e.\/ for all
contexts $C$, $\mathit{Ob}(C\context{{\cal P}}) = \mathit{Ob}(C\context{{\cal Q}})$.

\begin{definition}[Fully abstract relation]   % .. .. .. .. .. .. .. .. .. .. .. .. .. .. .. ..
An equivalence relation $\equiv$ is {\it fully abstract} wrt\/
$(\mathit{Ob},\mathit{Oper})$ iff for all programs ${\cal P}$ and ${\cal Q}$,
$
  {\cal P} \cong_\mathit{{Ob},\mathit{Oper}} {\cal Q} \ \Rightarrow \ {\cal P} \equiv {\cal Q}.
$
\end{definition}
% .. .. .. .. .. .. .. .. .. .. .. .. .. .. .. ..

A semantics ${\cal S}$ for a programming language provides a meaning for programs and
also induces an equivalence relation $\equiv_{\alg{S}}$ between programs: two programs
are equivalent iff they have the same meaning in this semantics. This equivalence
relation is used for defining compositionality and full abstraction for semantics.
\begin{definition}[Compositional and fully abstract semantics]  % .. .. .. .. .. .. .. .. .. .. .. .. .. .. .. ..
\label{compositional} A semantics ${\cal S}$ is compositional or fully abstract wrt
$(\mathit{Ob},\mathit{Oper})$ iff its corresponding relation $\equiv_{\alg{S}}$ is
compositional or fully abstract, respectively, wrt $(\mathit{Ob},\mathit{Oper})$.
\end{definition}
% .. .. .. .. .. .. .. .. .. .. .. .. .. .. .. ..

Obviously, for each pair $(\mathit{Ob},\mathit{Oper})$ there exits a compositional and
fully abstract relation between programs, the relation $$
    \prg{P} \equiv_{(\mathit{Ob},\mathit{Oper})} \prg{Q} \ \Leftrightarrow_{def} \
       \mathit{ Ob}(C\context{\prg{P}})=\mathit{ Ob}(C\context{\prg{Q}}),
       \ \ \mbox{for every context}\ C\context{\cal X}.
$$ For each compositional equivalence relation $\equiv$, it is easy to see that
  $ \prg{P}\equiv \prg{Q} \Rightarrow
   \prg{P} \equiv_{(\mathit{Ob},\mathit{Oper})} \prg{Q}$,
and for each fully abstract equivalence relation $\equiv$,
   $\prg{P} \equiv_{(\mathit{Ob},\mathit{Oper})} \prg{Q} \Rightarrow
   \prg{P}\equiv \prg{Q}$.
Thus, $\equiv_{(\mathit{Ob},\mathit{Oper})}$ will be the only equivalence relation which
is both compositional and fully abstract wrt\/ $(\mathit{Ob},\mathit{Oper})$. And the
more adequate semantics for programs (wrt $(\mathit{Ob},\mathit{Oper})$) will be a
semantics that induces this relation.

% ------------------------------------------------------------------------
\subsection{The $\alg{T}$-Semantics}
\label{tsemantics} To find a compositional semantics we may think about programs as open
in the sense that we can build up programs from other programs adding rules for new
functions and also for already defined functions (of the signature $\Sigma$ we were in)
and imagine them as algebra transformers as is done in \cite{mancarella} and
\cite{brogi:tesis}. The operator $\alg{T_{P}}$ considered as a function ${\bf
TAlg}_{\Sigma} \rightarrow {\bf TAlg}_{\Sigma}$ is a good candidate for the intended
meaning of a program $\alg{P}$.
First, we have to note that the set
$[{\bf TAlg}_{\Sigma} \rightarrow {\bf TAlg}_{\Sigma}]$ of all
continuous functions from ${\bf TAlg}_{\Sigma}$ to ${\bf TAlg}_{\Sigma}$,
ordered by the relation
\[
  \alg{T}_{1} \sqsubseteq  \alg{T}_{2} \ \Leftrightarrow_{def} \
  \forall \alg{A} \in {\bf TAlg}_{\Sigma}  \cdot
    (\alg{T}_{1}(\alg{A}) \sqsubseteq \alg{T}_{2}(\alg{A})),
\]
with the least upper bound and the greatest lower bound of a finite set $\{ \alg{T}_{i}
\}_{i \in I}$ of functions pointwise defined as
\[
  (\sqcup_{i \in I} \alg{T}_{i})(\alg{A}) =
                            \sqcup_{i \in I}(\alg{T}_{i}(\alg{A}))
  \mbox{\ \ \ and\ \ \ }
  (\sqcap_{i \in I} \alg{T}_{i})(\alg{A}) =
                            \sqcap_{i \in I}(\alg{T}_{i}(\alg{A}))
\]
respectively, and with bottom ${\Bbb T}_{\bot}$ and top ${\Bbb T}_{\Sigma}$  such that
\[
 {\Bbb T}_{\bot}(\alg{A}) = \bot_{\Sigma} \mbox{\ \ \ and \ \ \ }
  {\Bbb T}_{\Sigma}(\alg{A}) = \top_{\Sigma}, \ \ \forall
   \alg{A} \in {\bf TAlg}_{\Sigma},
\]
is a complete lattice as a consequence of $({\bf TAlg}_{\Sigma}, \sqsubseteq)$ being a
complete lattice.
Now, we can associate a program with the corresponding
immediate consequence operator, instead of its least fixpoint.

\begin{definition}[${\cal T}$-semantics]  % .. .. .. .. .. .. .. .. ..
\label{Tsemantics} We define the ${\cal T}$-semantics by denoting the meaning of a
program $\alg{P}$ by its algebra transformer $\ \semantics{\alg{P}} =_\mathit{ def}
\alg{T}_\alg{P}$, where $\alg{T_{\alg{P}}}$ is intended as $\alg{T}_{rl(\alg{P})}$.
\end{definition}  % .. .. .. .. .. .. .. .. .. .. .. .. .. .. .. .. ..

This semantics entails the following equivalence relation on programs:
        $\prg{P} \equiv_{T} \prg{Q} \Leftrightarrow_{def}
                \alg{T}_\alg{P} = \alg{T}_\alg{Q}$.
Thus, two programs are $\equiv_{T}$-equivalent if both define the same immediate
consequences operator. In this context, and coinciding with logic programming, a natural
choice of the observable behavior of a program $\prg{R}$ is its model-theoretic
semantics. So we will adopt as observation function
   $Ob(\prg{R}) =_\mathit{ def} \alg{M}_{\prg{R}}$.
Notice that $\alg{M}_{\prg{R}}$ captures the graphs of all functions defined in
$\prg{R}$, whereas functions not included in the program are considered totally undefined
(their images only can be reduced to $\bot$). The semantics $\ \semantics{\cdot}$  is
compositional wrt this observation function and the set of operations $Oper =
\{\cup,\overline{(\cdot)}^{\sigma},(\cdot)\!\setminus\!\sigma,
   \rho(\cdot)\}$.
We can prove this fact by proving that $\ \semantics{\cdot}$ is
homomorphic in the following sense.

\begin{theorem} % ........
\label{homomorphism}
Given a global signature $\Sigma$ and a countable set of variable
symbols $\alg{V}$, for all programs $\alg{P}$, $\alg{P}_{1}$ and
$\alg{P}_{2}$ defined over $\Sigma$, every subsignature of function
symbols $\sigma \subseteq FS_{\Sigma}$, and every function symbol
renaming $\rho$, we have the following results
 \[
\begin{array}{lrcl}
(a) \qquad
     & \semantics{\prg{P}_{1} \cup \prg{P}_{2}} & = &
      \semantics{\prg{P}_{1}} \sqcup \semantics{\prg{P}_{2}};\\
(b)  & \semantics{\overline{\prg{P}}^{\sigma}}  & = &
      \lambda \alg{A}  \cdot  (\ \semantics{\prg{P}}^{\omega} (\bot_\Sigma))|_{\sigma};\\
(c)  & \semantics{\prg{P} \setminus \sigma}     & = &
      \semantics{\prg{P}} \sqcap {\Bbb T}_{exp(\prg{P}) \setminus \sigma};\\
(d)  & \semantics{\rho(\prg{P})} & = &
      \prg{T}_{\rho^{-1}} {\footnotesize \circ} \ \semantics{\prg{P}} \
         {\footnotesize \circ} \prg{T}_{\rho};\\
\end{array}
 \]
where, for every algebra $\alg{A} \in {\bf TAlg}_{\Sigma}$ and
every subsignature $\sigma \subseteq FS_{\Sigma}$,
$\alg{A}|_{\sigma}$ is the term algebra characterized by
 \[
  \begin{array}{rcll}
     \function{f^{\alg{A}|_{\sigma}}}{t}{n} & = & \function{f^{\alg{A}}}{t}{n},
        & \mbox{ for all } \ t_{1},\ldots,t_{n} \in {\bf CTerm}_{\bot},
           \mbox{ if } f/n \in \sigma, \\
     \function{f^{\alg{A}|_{\sigma}}}{t}{n} & = & \{\bot\},
        & \mbox{ for all } \ t_{1},\ldots,t_{n} \in {\bf CTerm}_{\bot},
           \mbox{ otherwise.}
  \end{array}
 \]
For each subsignature $\sigma \subseteq FS_{\Sigma}$, ${\Bbb
T}_{\sigma}$ is the constant algebra transformer that, for all
$\alg{A}\in {\bf TAlg}_{\Sigma}$ produces the same term algebra
$\top_{\sigma}$ characterized by
 \[
  \begin{array}{rcll}
    \function{f^{\top_{\sigma}}}{t}{n} & = & {\bf CTerm}_{\bot},
             & \mbox{ for all } \ t_{1},\ldots,t_{n} \in {\bf CTerm}_{\bot},
                \mbox{ if } f/n\in\sigma, \\
    \function{f^{\top_{\sigma}}}{t}{n} & = & \{\bot\},
             & \mbox{ for all } \ t_{1},\ldots,t_{n} \in {\bf CTerm}_{\bot},
                \mbox{ otherwise.}
  \end{array}
 \]
And, for each rename $\rho$, $\prg{T}_{\rho}$ and $\prg{T}_{\rho^{-1}}$ are the algebra
transformers defined by $\prg{T}_{\rho}(\alg{A}) = \alg{A}_{\rho}$ and
$\prg{T}_{\rho^{-1}}(\alg{A}) = \alg{A}_{\rho^{-1}}$ where $\alg{A}_{\rho}$ and
$\alg{A}_{\rho^{-1}}$ are the term algebras characterized by $$
  f^{\alg{A}_{\rho}} = \rho(f)^{\alg{A}} \ \ \mbox{ and } \ \
  f^{\alg{A}_{\rho^{-1}}} =
     \left\{ \begin{array}{l l}
             \sqcup \{ g^{\alg{A}} \ \vert \ f=\rho(g) \}, &
                     \mbox{when this set is not empty,} \\
             f^{\bot_{\Sigma}} & \mbox{otherwise,}
             \end{array}
     \right.
$$
for every function symbol $f$ in $FS_{\Sigma}$.
\end{theorem} % ......................................................
\begin{proof}
 $(a)$ For the first result we have to prove that
 $\prg{T}_{\prg{P}_{1} \cup\prg{P}_{2}}(\alg{A}) =
  \prg{T}_{\prg{P}_{1}}(\alg{A}) \sqcup \prg{T}_{\prg{P}_{2}}(\alg{A})$,
for all $\alg{A} \in {\bf TAlg}_{\Sigma}$. For each $f \in\funs{n}$, with $n \geq 0$, and
$t_{1},\ldots,t_{n} \in {\bf CTerm}_{\bot}$,
 $
  \function{f^{\alg{T}_{\prg{P}_{1} \cup \prg{P}_{2}}(\alg{A})}}{t}{n} =
  \function{f^{\alg{T}_{\prg{P}_{1}}(\alg{A})}}{t}{n} \cup
  \function{f^{\alg{T}_{\prg{P}_{2}}(\alg{A})}}{t}{n},
 $
because every rule in $\prg{P}_{1} \cup \prg{P}_{2}$ with an instance that can be used in
the construction of $\function{f^{\alg{T}_{\prg{P}_{1} \cup
\prg{P}_{2}}(\alg{A})}}{t}{n}$ is also a rule in $\prg{P}_{1}$ or $\prg{P}_{2}$, and the
same instance can be used to construct
$\function{f^{\alg{T}_{\prg{P}_{1}}(\alg{A})}}{t}{n}$ or
$\function{f^{\alg{T}_{\prg{P}_{2}}(\alg{A})}}{t}{n}$ respectively, because the
applicability of this instance only depends on its arguments and the term algebra
$\alg{A}$. Reciprocally, every rule in $\prg{P}_{1}$ or $\prg{P}_{2}$ with an instance
applicable to construct $\function{f^{\alg{T}_{\prg{P}_{1}}(\alg{A})}}{t}{n}$ or
$\function{f^{\alg{T}_{\prg{P}_{2}}(\alg{A})}}{t}{n}$ is a rule in $\prg{P}_{1} \cup
\prg{P}_{2}$ with the same instance applicable to construct
$\function{f^{\alg{T}_{\prg{P}_{1} \cup \prg{P}_{2}}(\alg{A})}}{t}{n}$ for the same
reason. Finally, by definition of the operation $\sqcup$ between term algebras,
 $
  \function{f^{\alg{T}_{\prg{P}_{1}}(\alg{A})}}{t}{n} \cup
  \function{f^{\alg{T}_{\prg{P}_{2}}(\alg{A})}}{t}{n} =
  \function{f^{\alg{T}_{\prg{P}_{1}}(\alg{A}) \sqcup \alg{T}_{\prg{P}_{2}}(\alg{A})}}{t}{n},
 $
and therefore
 $
    \function{f^{\alg{T}_{\prg{P}_{1} \cup \prg{P}_{2}}(\alg{A})}}{t}{n} =
    \function{f^{\alg{T}_{\prg{P}_{1}}(\alg{A}) \sqcup \alg{T}_{\prg{P}_{2}}(\alg{A})}}{t}{n}
 $.

% ....... ......... ......... ............... .........
$(b)$ In order to prove the second result, as $\prg{T}_{\prg{P}}^{\omega}(\bot_\Sigma) =
\alg{M_{P}}$, we only have to prove $\prg{T}_{\overline{\prg{P}}^{\sigma}}(\alg{A}) =
\alg{M_{P}}|_{\sigma}$ for all $\alg{A} \in {\bf TAlg}_{\Sigma}$. For $f\in\funs{n}$ and
$t_{1},\ldots,t_{n} \in {\bf CTerm}_{\bot}$, if $f/n \not\in \sigma$ then there is no
rule for $f$ in $\overline{\prg{P}}^{\sigma}$ and
$f^{\prg{T}_{\overline{\prg{P}}^{\sigma}}(\alg{A})}(\overline{t}) = \{ \bot \} =
f^{\alg{M_{P}}|_{\sigma}}$, and if $f/n \in \sigma$ then we will prove that
$\function{f^{\prg{T}_{\overline{\prg{P}}^{\sigma}}(\alg{A})}}{t}{n} =
 \function{f^{\alg{M_{P}}}}{t}{n}$. For
$t\in \function{f^{\prg{T}_{\overline{\prg{P}}^{\sigma}}(\alg{A})}}{t}{n}$ there exists
$crr(f(\overline{s}),r) = f(\overline{s}') \to r \si C$ in $\overline{\prg{P}}$ and a
substitution $\theta$ such that $t\in\eval{r\theta}{A}{id} = \ideal{r\theta}$ with
$\overline{s}'\theta \sqsubseteq \overline{t}$ and $\alg{A}\models_{id}C\theta$, what
means that $\theta$ is totally defined in variables of $\overline{s}$ and $r$. By the
special joinability statement of $C$ we can obtain a total substitution $\theta'$ (by
considering only the part of $\theta$ involving the variables of $\overline{s}$ and $r$)
such that $\overline{s}\theta' \sqsubseteq \overline{s}'\theta \sqsubseteq \overline{t}$
and $r\theta' = r\theta$. Since $\prg{P}\vdash_{\mathit{CRWL}} f(\overline{s}) \to r$,
$r\in\function{f^{\alg{M_{P}}}}{s}{n}$, and as $\alg{M_{P}}$ is consistent (see
Section~\ref{CTAlgebras}), $r\theta'\in f^{\alg{M_{P}}}(\overline{s}\theta')$ and by the
monotonicity of $f^{\alg{M_{P}}}$, $r\theta\in f^{\alg{M_{P}}}(\overline{t})$ and so,
$t\in f^{\alg{M_{P}}}(\overline{t})$. Reciprocally, $t\in f^{\alg{M_{P}}}(\overline{t})$
implies $\prg{P}\vdash f(\overline{t})\to t$ or $crr(f(\overline{t}),r)\in
\overline{\prg{P}}$ and, by considering $\theta_{\overline{t}}$, we have
$\eval{t\theta_{\overline{t}}}{A}{id} = \ideal{t} \subseteq
\function{f^{\prg{T}_{\overline{\prg{P}}^{\sigma}}(\alg{A})}}{t}{n}$ for every term
algebra $\alg{A}$ and so, $t\in
\function{f^{\prg{T}_{\overline{\prg{P}}^{\sigma}}(\alg{A})}}{t}{n}$

%$\overline{\prg{P}}^{\sigma}$ with instances applicable to obtain
%$\function{f^{\prg{T}_{\overline{\prg{P}}^{\sigma}}(\alg{A})}}{t}{n}$ will be rules
%$\function{f}{s}{n} \es r$, with $r \in {\bf CTerm}_{\bot}$, derivable in $\prg{P}$, with
%instances $f(\overline{s}\theta) \es r\theta$, such that $r\theta \not= \bot$ and
%$s_{i}\theta \sqsubseteq t_{i}$, for $i=1,\ldots,n$, that also will be derivable in
%$\prg{P}$ by application of the rule ({\bf OR}) of the GPC (taking into account that, by
%Equivalence~(\ref{reductionOrdering}), $s_{i}\theta \sqsubseteq t_{i}$ is equivalent to
%$\vdash_{CRWL} t_{i} \es s_{i}\theta$), and so $r\theta \in
%\function{f^{\alg{M_{P}}}}{t}{n}$. With this reasoning we have proved the inclusion
%$\function{f^{\prg{T}_{\overline{\prg{P}}^{\sigma}}(\alg{A})}}{t}{n} \subseteq
% \function{f^{\alg{M_{P}}}}{t}{n}$.
%Conversely, for all $r \in \function{f^{\alg{M_{P}}}}{t}{n}$, the
%approximation statement $\function{f}{t}{n} \es r$ will be
%derivable in $\prg{P}$; therefore it will be a rule in
%$\overline{\prg{P}}^{\sigma}$ and, by considering the identity
%substitution, it is obtained $r \in
%\function{f^{\prg{T}_{\overline{\prg{P}}^{\sigma}}(\alg{A})}}{t}{n}$
%and we have the other inclusion
%$\function{f^{\prg{T}_{\overline{\prg{P}}^{\sigma}}(\alg{A})}}{t}{n}
%\supseteq
% \function{f^{\alg{M_{P}}}}{t}{n}$.

$(c)$ For this result we have to prove that
$\prg{T}_{\prg{P} \setminus \sigma}(\alg{A}) =
      \prg{T}_{\prg{P}}(\alg{A}) \sqcap
      {\Bbb T}_{exp(\prg{P}) \setminus \sigma}(\alg{A})$,
for all $\alg{A} \in {\bf TAlg}_{\Sigma}$, and this is equivalent to $f^{\prg{T}_{\prg{P}
\setminus \sigma}(\alg{A})} = f^{\prg{T}_{\prg{P}}(\alg{A})}$, for all $f/n \in
exp(\prg{P})\setminus\sigma$, and $\function{f^{\prg{T}_{\prg{P} \setminus
\sigma}(\alg{A})}}{t}{n} = \{ \bot \}$, for all $f/n \in \sigma$. The first equality is
easily proved by taking into account that $\prg{P}$ and $\prg{P}\setminus\sigma$ have the
same rules for each $f/n\in exp(\prg{P})$ and remembering that the applicability of every
instance of these rules to construct
$\function{f^{\prg{T}_{\prg{P}\setminus\sigma}(\alg{A})}}{t}{n}$ and
$\function{f^{\prg{T}_{\prg{P}}(\alg{A})}}{t}{n}$ only depends on its arguments and the
term algebra $\alg{A}$. The second equality is trivial because there is no rule in
$\prg{P} \setminus \sigma$ for $f/n \in \sigma$.

$(d)$ For the last result we have to prove that
$\prg{T}_{\rho(\prg{P})}(\alg{A})=
\prg{T}_{\rho^{-1}}(\prg{T}_{\prg{P}}(\prg{T}_{\rho}(\alg{A})))$
for all $\alg{A} \in {\bf TAlg}_{\Sigma}$. On the one hand
$\function{f^{\prg{T}_{\rho(\prg{P})}(\alg{A})}}{t}{n}$ is
constructed from all rules $\function{g}{s}{n} \es r \si C$ in
$\prg{P}$, with $\rho(g) = f$, such that, for any $\theta \in {\bf
CSubst}_{\bot}$, $(\rho(s_{i}))\theta \sqsubseteq t_{i}$, for $i =
1,\ldots,n$, and $\alg{A} \models_{id} (\rho(C))\theta$, by
considering the union of the corresponding cones
$\evaluation{(\rho(r))\theta}{A}{id}$. On the other hand,
$\function{f^{\prg{T}_{\rho^{-1}}(\prg{T_{P}}(\prg{T}_{\rho}(\alg{A})))}}{t}{n}$
is
$\function{f^{\prg{T}_{\rho^{-1}}(\prg{T_{P}}(\alg{A}_{\rho}))}}{t}{n}$
and by the definition of $\prg{T}_{\rho^{-1}}$ this is equal to
 $
  \function{(\sqcup \{ g^{\prg{T_{P}}(\alg{A}_{\rho})} \
  \vert \ \rho(g) = f \} )}{t}{n}
 $
which is the union of the cones $\function{g^{\prg{T_{P}}(\alg{A}_{\rho})}}{t}{n}$, and
each cone is constructed from all rules $\function{g}{s}{n} \es r \si C$ in $\prg{P}$
such that, for any $\theta \in {\bf CSubst}_{\bot}$, $s_{i}\theta \sqsubseteq t_{i}$, for
$i = 1,\ldots,n$, and $\alg{A}_{\rho} \models_{id} C\theta$, by considering the union of
the corresponding cones $\eval{r\theta}{\alg{A}_{\rho}}{id}$. But, as the function
renaming $\rho$ does not affect constructor terms or variables we have
$(\rho(s_{i}))\theta = s_{i}\theta$; from Proposition~\ref{modelsandrenaming}({\it 1}),
$(\rho(C))\theta = \rho(C\theta)$; and from Proposition~\ref{modelsandrenaming}({\it 3}),
$\alg{A} \models_{id} (\rho(C\theta)) \Leftrightarrow \alg{A}_{\rho} \models_{id}
C\theta$. So,
 the same rules of $\prg{P}$ are used to construct
$\function{f^{\prg{T}_{\rho(\prg{P})}(\alg{A})}}{t}{n}$ and
$\function{f^{\prg{T}_{\rho^{-1}}(\prg{T_{P}}(\prg{T}_{\rho}(\alg{A})))}}{t
}{n}$, and from Proposition~\ref{modelsandrenaming}({\it 1,2}), we
conclude that both cones coincide.
\end{proof}

Thus, the meaning of the union of two programs $(a)$ can be extracted from the meaning of
each one, the meaning of the closure of a program $(b)$ is obtained from the fixpoint of
the program semantics, and deleting a signature from a program $(c)$ is semantically
equivalent to the intersection of the program semantics with an algebra transformer which
depends on the exported signature of the program. Nevertheless, the intersection we are
mentioning here is not an operation over programs (as in \cite{brogi:tesis}) but an
operation on algebra transformers. The meaning of a renamed program $(d)$ can be obtained
as the composition of the meaning of the program with two algebra transformers associated
with the renaming and its reverse.

\begin{corollary}[Compositionality of $ \ \semantics{\cdot}$] % ..................................................
\label{compositionality}
The semantics $\semantics{\cdot}$ is compositional with respect to
$(Ob,\{\cup,\overline{(\cdot)}^{\sigma},(\cdot)\!\setminus\!\sigma,
   \rho(\cdot)\})$.
\end{corollary} % ....................................................
\begin{proof} The notion of observable, which coincides with the least fixpoint of the
semantics, is obviously preserved by this semantics. On the other
hand, the congruence property is directly derived from the
previous theorem. We only need to justify that
$\semantics{\prg{P}} \sqcap {\Bbb T}_{exp(\prg{P}) \setminus
\sigma} = \semantics{\prg{Q}} \sqcap {\Bbb T}_{exp(\prg{Q})
\setminus \sigma}$, for every subsignature $\sigma\subseteq
FS_\Sigma$, when $\semantics{\prg{P}} = \semantics{\prg{Q}}$,
independently on whether $exp(\prg{P})$ is equal to $exp(\prg{Q})$
or not. In fact, if $f/n\in exp(\prg{P})\setminus\sigma$ and
$f/n\not\in exp(\prg{Q})\setminus\sigma$ then $f/n\not\in
exp(\prg{Q})$, which implies
$\function{f^{\prg{T}_{\prg{Q}}(\alg{A})}}{t}{n}=\{\bot\}$, for
all $t_1,\ldots,t_n\in{\bf CTerm}_\bot$ and
$\function{f^{\prg{T}_{\prg{P}}(\alg{A})}}{t}{n}=\{\bot\}$ because
$\alg{T}_\prg{P}=\alg{T}_\prg{Q}$. The same result is obtained if
we suppose $f/n\not\in exp(\prg{P})\setminus\sigma$ and $f/n\in
exp(\prg{Q})\setminus\sigma$. Therefore,
$\function{f^{(\prg{T}_{\prg{P}}\sqcap {\Bbb T}_{exp(\prg{P})
\setminus \sigma})(\alg{A})}}{t}{n}=
\function{f^{(\prg{T}_{\prg{Q}}\sqcap {\Bbb T}_{exp(\prg{Q})
\setminus \sigma})(\alg{A})}}{t}{n}$. \end{proof}

As the above corollary states, $\ \semantics{\cdot}$ is compositional wrt union,
closure, deletion and renaming, when the canonic model of a program is taken as
its observable behavior. However, the following example shows that it is not
fully abstract.

\begin{example}
Let $\Sigma$ be a signature $(\{c/0,d/0\},\{f/0\})$ and let $\prg{P}$ and $\prg{Q}$ be
the modules such that
 $ rl(\prg{P}) = \{ f \es c, \ f \es d \}$ and
  $ rl(\prg{Q}) = \{ f \es c, \ f \es d \si f\bowtie c\}$.
They are indistinguishable under
$\{\cup,\overline{(\cdot)}^{\sigma},(\cdot)\!\setminus\!\sigma,\rho(\cdot)\}$,
but they are not $\equiv_{T}$-equivalent. In fact,
${\cal T}_{\prg{P}}(\bot_{\Sigma}) \not= {\cal T}_{\prg{Q}}(\bot_{\Sigma})$
because $f^{{\cal T}_{\prg{P}}(\bot_{\Sigma})}=\{c,d,\bot\}$
whereas $f^{{\cal T}_{\prg{Q}}(\bot_{\Sigma})}=\{c,\bot\}$.
\end{example}

The ${\cal T}$-semantics distinguishes more than the model-theoretic semantics, since the
immediate consequence operator captures what is happening in each reduction step, but the
non-full abstraction result means that this semantics distinguishes more than necessary.
It is too fine. In the next section we will try a coarser semantics ---also studied in
logic programming \cite{brogi:TCS}--- defined from the sets of pre-fixpoints of ${\cal
T}$.

% ====================================================================

\section{A Fully Abstract Semantics}
\label{fullabstractSem}

In this section, a fully abstract semantics is presented, which is also compositional
except for the deletion operation. For a better motivation, we will not introduce this
semantics directly. Instead, we will define a first approximation, the so-called \textit{
term model semantics} (Definition~\ref{modelsequivalence}), which only is compositional
(wrt the union, closure and renaming operations), and then we will obtain the full
abstraction property by restricting the term models
(Definition~\ref{consistentmodelsequivalence}).

% --------------------------------------------------------------------

\subsection{The Term Model Semantics}
\label{M-semantics}

 Formally, we will introduce the first semantics by directly
 considering the corresponding equivalence relation.

\begin{definition}[Model equivalence] % .. .. .. .. .. .. .. .. .. ..
\label{modelsequivalence}
 Two programs ${\cal P}$ and ${\cal Q}$ are model-equivalent iff their
algebra transformers have the same pre-fixpoints
\[
    {\cal P} \equiv_M {\cal Q} \Leftrightarrow
        \forall \alg{A}\in {\bf TAlg}_{\Sigma} \cdot
                               (\prg{T_P}(\alg{A}) \sqsubseteq \alg{A}
                                \Leftrightarrow
                                \prg{T_Q}(\alg{A}) \sqsubseteq \alg{A}).
\]
By Lemma~\ref{model-charact} this means that two programs are equivalent iff they have
the same term models.
\end{definition}  % .. .. .. .. .. .. .. .. .. .. .. .. .. .. .. .. ..
This equivalence relation corresponds to the following semantics:
\[
     \modelsetsem{{\cal P}} =_\mathit{ def}
        \{\alg{M}\; \vert \; \alg{M} \mbox{ is a term model of } \prg{P}\}
\]
which will be called \textit{ loose model-theoretic semantics}, or simply \textit{term
model semantics}.
In order to derive the corresponding result about compositionality,
we need an auxiliary property about ${\cal T}_\rho$ and ${\cal
T}_{\rho^{-1}}$.

\begin{lemma}
\label{Trho1}
Given two term algebras $\alg{A},\alg{B}\in{\bf TAlg}_{\Sigma}$,
for every function symbol renaming $\rho$,
\[
     \alg{A}_{\rho^{-1}} \sqsubseteq \alg{B} \Leftrightarrow
     \alg{A} \sqsubseteq \alg{B}_\rho
     \mbox{ or, equivalently, }
     \alg{T}_{\rho^{-1}}(\alg{A}) \sqsubseteq \alg{B} \Leftrightarrow
     \alg{A} \sqsubseteq \alg{T}_{\rho}(\alg{B}).
\]

\end{lemma}
\begin{proof} Let $\alg{A}$ and $\alg{B}$ be two term algebras such that $\alg{A}_{\rho^{-1}}
\sqsubseteq \alg{B}$. Then, for all function symbols $f$,
$\function{f^{\alg{A}_{\rho^{-1}}}}{t}{n} \subseteq \function{f^{\alg{B}}}{t}{n}$, for
$t_1,\ldots,t_n\in{\bf CTerm}_\bot$. This is equivalent to
  $ \bigcup \{\function{g^{\alg{A}}}{t}{n}\ |\ f = \rho(g)\} \subseteq
   \function{f^{\alg{B}}}{t}{n}$.
Thus, for all function symbols $g$, by considering their images $\rho(g)=f$, we obtain
   $\function{g^\alg{A}}{t}{n} \subseteq \function{\rho(g)^\alg{B}}{t}{n} =
                                        \function{g^{\alg{B}_\rho}}{t}{n}$
or, equivalently, $\alg{A} \sqsubseteq \alg{B}_\rho$. The
implication in the other way is obtained by reversing this
reasoning.
\end{proof}

This lemma claims that ${\cal T}_{\rho^{-1}}$ is, essentially, the reverse operator
for ${\cal T}_\rho$.

\begin{theorem}[Compositionality of $\ \modelsetsem{\cdot}$] % .........
\label{modelsetcompos}
For all programs $\prg{P}, \prg{Q}, \prg{P}_i, \prg{Q}_i$,
\begin{enumerate}
\item $\prg{P} \equiv_M \prg{Q}$ implies $Ob(\prg{P})=Ob(\prg{Q})$.

\item $\prg{P}_i \equiv_M \prg{Q}_i$ for $i=1,2$, implies
      $\prg{P}_1 \cup \prg{P}_2 \equiv_M \prg{Q}_1 \cup \prg{Q}_2$.
\item $\prg{P} \equiv_M \prg{Q}$ implies
      $\overline{\prg{P}}^\sigma \equiv_M \overline{\prg{Q}}^\sigma$,
      for every signature $\sigma$.
\item $\prg{P} \equiv_M \prg{Q}$ implies
      $\rho(\prg{P}) \equiv_M \rho(\prg{Q})$, for every function symbol
      renaming $\rho$.
\end{enumerate}
Therefore, the semantics $\ \modelsetsem{\cdot}$ is compositional wrt $(\mathit{ Ob},
\{\cup,\overline{(\cdot)}^{\sigma},\rho(\cdot)\})$.
\end{theorem} % ......................................................
\begin{proof}
 {\it 1.} If $\prg{P} \equiv_M \prg{Q}$ then $\prg{P}$ and $\prg{Q}$ have the same
set of term models and, in particular, they have the same least term model. So
$Ob(\prg{P})=Ob(\prg{Q}).$

{\it 2.} Let $\alg{A}$ be a term model of $\prg{P}_1 \cup
\prg{P}_2$, then ${\cal T}_{\prg{P}_1\cup\prg{P}_2}(\alg{A}) \sqsubseteq
\alg{A}$ and, by Theorem~\ref{homomorphism}(a),
 $
  {\cal T}_{\prg{P}_1}(\alg{A}) \sqcup {\cal T}_{\prg{P}_2}(\alg{A}) =
  ({\cal T}_{\prg{P}_1} \sqcup {\cal T}_{\prg{P}_2})(\alg{A})
   = {\cal T}_{\prg{P}_1 \cup \prg{P}_2}(\alg{A}) \sqsubseteq \alg{A}
 $,
therefore ${\cal T}_{\prg{P}_i}(\alg{A}) \sqsubseteq
\alg{A}$, for $i=1,2$. From $\prg{P}_i \equiv_M \prg{Q}_i$, we obtain ${\cal
T}_{\prg{Q}_i}(\alg{A}) \sqsubseteq \alg{A}$, for $i=1,2$, and
again by Theorem~\ref{homomorphism}(a)
 $
  {\cal T}_{\prg{Q}_1 \cup \prg{Q}_2}(\alg{A})
    = ({\cal T}_{\prg{Q}_1} \sqcup {\cal T}_{\prg{Q}_2})(\alg{A}) =
    {\cal T}_{\prg{Q}_1}(\alg{A}) \sqcup {\cal T}_{\prg{Q}_2}(\alg{A})
    \sqsubseteq \alg{A}
 $,
and $\alg{A}$ will be a term model of $\prg{Q}_1 \cup \prg{Q}_2$.
By reasoning in a similar way, it can be obtained that all term
models of $\prg{Q}_1 \cup \prg{Q}_2$ are also term models of
$\prg{P}_1 \cup \prg{P}_2$ and this proves that $\prg{P}_1 \cup
\prg{P}_2 \equiv_M \prg{Q}_1 \cup \prg{Q}_2$.

{\it 3.} To prove the third statement we only need to take into
account that, by the first statement, $\prg{P} \equiv_M \prg{Q}$
implies $\alg{M}_\prg{P} = \alg{M}_\prg{Q}$ and therefore
$\alg{M}_\prg{P} |_{\sigma} = \alg{M}_\prg{Q} |_{\sigma}$, for all
$\sigma \subseteq FS_\Sigma$. And, by
Theorem~\ref{homomorphism}(b), this implies
$\alg{T}_{\overline{\prg{P}}^\sigma} =
\alg{T}_{\overline{\prg{Q}}^\sigma}$. Therefore, they will have the same
pre-fixpoints and consequently $\overline{\prg{P}}^\sigma \equiv_M
\overline{\prg{Q}}^\sigma$.

{\it 4.} Finally, for each term model $\alg{A}$ of
$\rho(\prg{P})$, ${\cal T}_{\rho(\prg{P})}(\alg{A}) \sqsubseteq
\alg{A}$ or $({\cal T}_\prg{P}(\alg{A}_\rho))_{\rho^{-1}}
\sqsubseteq \alg{A}$, by Theorem~\ref{homomorphism}(d). From this,
by Lemma~\ref{Trho1}, we obtain ${\cal T}_\prg{P}(\alg{A}_\rho)
\sqsubseteq \alg{A}_\rho$. Thus, if $\prg{P} \equiv_M \prg{Q}$ we
have ${\cal T}_\prg{Q}(\alg{A}_\rho) \sqsubseteq \alg{A}_\rho$,
and again, by applying Lemma~\ref{Trho1} and
Theorem~\ref{homomorphism}(d), we derive ${\cal
T}_{\rho(Q)}(\alg{A}) \sqsubseteq \alg{A}$. So $\alg{A}$ is a term
model of $\rho(\prg{Q})$. By reasoning in a similar way, it can be
proved that all term models of $\rho(\prg{Q})$ are also term
models of $\rho(\prg{P})$ which proves that $\rho(\prg{P})
\equiv_M \rho(\prg{Q})$. \end{proof}

Unfortunately, this semantics is not compositional wrt\/ deletion.
% as we can see in the following example.
%
\begin{example}
\label{nonCompositional} Let $\Sigma$ be the signature $(\{a/0,b/0\},\{f/0,g/0\})$ and
let $\prg{P}$ and $\prg{Q}$ be two modules with rules
  $rl(\prg{P}) = \{ f \es a, \ g \es b \}$ and
  $rl(\prg{Q}) = \{ f \es a, \ g \es b \si f\bowtie a\}$.
 Both modules have the same term models, those term algebras
$\alg{A}$ with $a\in f^{\alg{A}}$ and $b\in g^{\alg{A}}$. But by deleting $f/0$ in each
module we have $\prg{P} \setminus \{f/0\}$ and $\prg{Q} \setminus \{f/0\}$ with
  $rl(\prg{P}\setminus \{f/0\}) = \{g \es b \}$ and
  $rl(\prg{Q}\setminus \{f/0\}) = \{g \es b \si f\bowtie a\}$,
and now $\bot_{\Sigma}$ is a model of $\prg{Q}\setminus \{f/0\}$
whereas it is not a model of $\prg{P}\setminus \{f/0\}$.
\end{example}
For a different reason, the semantics $\ \modelsetsem{\cdot}$ is not fully abstract.
% as the following example shows.
%
\begin{example}
\label{nonFullyAbstract} Let $\Sigma$ be the signature $(\{a/0\},\{f/0,g/1\})$ and let
$\prg{P}$ and $\prg{Q}$ be two modules with rules
  $rl(\prg{P}) = \{ f \es a \si g(a)\bowtie a\}$ and
  $rl(\prg{Q}) = \{ f \es a \si g(X)\bowtie a\}$,
 where the rule in $\prg{P}$ is an instance of the rule in $\prg{Q}$. Obviously, both
modules are indistinguishable but they do not have the same term models. In fact, if we
consider the algebra $\alg{A}$ such that: $f^\alg{A} =\{\bot\}$, $g^\alg{A}(X) =
\{a,\bot\}$ and $g^\alg{A}(a) = \{\bot\}$, $\alg{A}$ is a model of $\prg{P}$ but it is
not a model of $\prg{Q}$.
\end{example}

% --------------------------------------------------------------------

\subsection{Consistent Term Algebras}
\label{CTAlgebras}

To prove the full abstraction property we need to consider a different equivalence
relation (i.e.  semantics).  If we observe the above counter-example, we can see that,
for the term algebra $\alg{A}$ used to distinguish $\ \modelsetsem{\prg{P}}$ from $\
\modelsetsem{\prg{Q}}$, $g^\alg{A}(X) = \{a,\bot\}$ and $g^\alg{A}(a) = \{\bot\}$; that
is, $\alg{A}$ is such that the instantiation of the variable $X$ derives in a loss of
information for the interpretation of $g$ because $g^\alg{A}(X\theta)$ is smaller than
$(g^\alg{A}(X))\theta$, for $\theta=\{X/a\}$.  In general, the notion of term algebra
(see Section~\ref{algebras&terms}) does not impose any relation between
$g^\alg{A}(\bar{t}\theta)$ and $(g^\alg{A}(\bar{t}))\theta$.  This is not reasonable if
we take into account the role of term algebras when they are used to model programs. On
the contrary, the interpretation of a function symbol (in a term algebra) applied to
arguments with variables must be related to the interpretation of the same function
symbol when these variables are instantiated. With this idea in mind, we introduce the
notion of \textit{ consistency} in a term algebra.

\begin{definition}[Consistency of term algebras]
\label{consistency} A term algebra $\alg{A}\in{\bf TAlg}_\Sigma$ is consistent iff for
every $f\in FS^n_\Sigma$ and $t_i\in{\bf CTerm}_\bot$ $(i=1,\dots,n)$,
 $f^\alg{A}(\overline{t}\theta) \supseteq (f^\alg{A}(\overline{t}))\theta$ for all
 $\theta\in{\bf CSubst}$, where $(f^\alg{A}(\overline{t}))\theta$ stands for the set
 $\{u\theta \ | \ u\in f^\alg{A}(\overline{t}) \}$.
\end{definition}
 We will denote by ${\bf CTAlg}_\Sigma$ the family of all
consistent term algebras. Note that consistency is only required for total substitutions
(i.e. substitutions which do not include partial constructor terms). This is due to the
special treatment of $\bot$, which is considered as lack of information. The notion of
consistency here introduced is close to that of \textit{closure under substitutions}
defined for interpretations in \cite{apt96}, and is also related with the notion of
C-interpretation considered in \cite{falaschi}, but our requirements are weaker than
those ones.
To justify the reasonable nature of consistent term algebras we will prove several
desirable properties. For instance, the immediate consequences operator maps consistent
algebras into consistent algebras, and the canonical model of a program is consistent.

\begin{lemma}
\label{cvaluations}
For every $\alg{A}\in{\bf CTAlg}_\Sigma$, $r\in{\bf Term}_{\bot}$,
and $\theta\in{\bf CSubst}$,
 $\ \evaluation{r}{\alg{A}}{id}\theta \subseteq
                  \evaluation{r\theta}{\alg{A}}{id}$.
\end{lemma}
\begin{proof}
 The proof is by induction on the structure of $r$. There are several base
cases: $r \in \{\bot\} \cup \cons{0}$, $r\in\alg{V}$, or $r\in\funs{0}$. In the first
case, $\evaluation{r}{\alg{A}}{id} = \evaluation{r\theta}{\alg{A}}{id}$ and these cones
have no terms with variables. In the second case, $\evaluation{r}{\alg{A}}{id}\theta =
\{\theta(r), \bot\}$, and this is a subset of $\evaluation{r\theta}{\alg{A}}{id}$. In the
third case,  $\evaluation{r}{\alg{A}}{id}\theta \subseteq
\evaluation{r\theta}{\alg{A}}{id}$ because $r\theta = r$ and $r^\alg{A}\theta \subseteq
r^\alg{A}$ for $\alg{A}$ consistent. In the general case, $r=h(\overline{e})$, with $h\in
\cons{n} \cup \funs{n}$, $e_i\in {\bf Term}_\bot$ $(i=1,\dots,n)$, and $n>0$. Then,
  $
   \evaluation{h(\overline{e})\theta}{\alg{A}}{id} =
   \evaluation{h(\overline{e}\theta)}{\alg{A}}{id} =
   \bigcup_{u_i\in\,\evaluation{e_i\theta}{\alg{A}}{id}} {\function{h^\alg{A}}{u}{n}}
  $.
Assuming $\,\evaluation{e_i}{\alg{A}}{id}\theta \subseteq
      \evaluation{e_i\theta}{\alg{A}}{id}$ $(i=1,\dots,n)$,
as the induction hypothesis, we obtain
 $
   \bigcup_{u_i\in\,\evaluation{e_i\theta}{\alg{A}}{id}} {\function{h^\alg{A}}{u}{n}}
   \supseteq
   \bigcup_{u_i\in\,\evaluation{e_i}{\alg{A}}{id}\theta} {\function{h^\alg{A}}{u}{n}} =
   \bigcup_{v_i\in\,\evaluation{e_i}{\alg{A}}{id}} {h^\alg{A}(\overline{v}\theta)}
 $.
Since $\alg{A}$ is consistent, $h^{\alg{A}}(\overline{v}\theta) \supseteq
\function{h^\alg{A}}{v}{n}\theta$, and therefore
  $
   \bigcup_{v_i\in\,\evaluation{e_i}{\alg{A}}{id}} {h^\alg{A}(\overline{v}\theta)}
    \supseteq
   \bigcup_{v_i\in\,\evaluation{e_i}{\alg{A}}{id}} {\function{h^\alg{A}}{v}{n}\theta}
     =
   (\bigcup_{v_i\in\,\evaluation{e_i}{\alg{A}}{id}} {\function{h^\alg{A}}{v}{n}})\theta
    =
   \evaluation{h(\overline{e})}{\alg{A}}{id} \theta
 $.
So, $\,\evaluation{h(\overline{e})\theta}{\alg{A}}{id} \supseteq
\evaluation{h(\overline{e})}{\alg{A}}{id} \theta$.
 \end{proof}
% -----------------------------
\begin{proposition}
\label{coperator}
Given a program $\prg{P}$, if $\alg{A}\in{\bf CTAlg}_\Sigma$, then
$\alg{T}_\prg{P}(\alg{A})\in {\bf CTAlg}_\Sigma$.
\end{proposition}
% -----------------------------
 \begin{proof}
Let $f\in\funs{n}$ and $t_1,\dots,t_n\in{\bf CTerm_\bot}$. If $u\in
\function{f^{\alg{T}_\prg{P}(\alg{A})}}{t}{n}$ then there exists a rule
$\function{f}{s}{n} \es r \si C$ in $[\prg{P}]_\bot$ such that $s_i \sqsubseteq t_i$
$(i=1,\dots,n)$, $\alg{A}\models_{id} C$, and $u\in\evaluation{r}{\alg{A}}{id}$. For
$\theta\in {\bf CSubst}$,  $u\theta\in\evaluation{r}{\alg{A}}{id}\theta
\subseteq\,\evaluation{r\theta}{\alg{A}}{id}$ by Lemma~\ref{cvaluations}, and
$\alg{A}\models_{id} C\theta$ because, if $a \bowtie b\in C$ then there exists
$t\in\evaluation{a}{\alg{A}}{id} \cap \,\evaluation{b}{\alg{A}}{id}$, with $t\in{\bf
CTerm}$, and $\evaluation{a\theta}{\alg{A}}{id} \supseteq
\evaluation{a}{\alg{A}}{id}\theta$ and $\evaluation{b\theta}{\alg{A}}{id} \supseteq\,
\evaluation{b}{\alg{A}}{id}\theta$ again by Lemma~\ref{cvaluations}, so $t\theta\in
\evaluation{a}{\alg{A}}{id}\theta \cap\, \evaluation{b}{\alg{A}}{id}\theta$, and
$t\theta\in{\bf CTerm}$ because $\theta\in{\bf CSubst}$. Thus, we can consider the rule
$f(\overline{s}\theta) \es r\theta \si C\theta$, which is also in $[\prg{P}]_\bot$, with
$s_i\theta \sqsubseteq t_i\theta$ $(i=1,\dots,n)$, to derive that $u\theta\in
f^{\alg{T}_\prg{P}(\alg{A})}(\overline{t}\theta)$.
 \end{proof}
% ---------------------------------
\begin{proposition}
\label{ctermmodel}
Given a program $\prg{P}$, the canonical term model
$\alg{M}_\prg{P}$ is consistent.
\end{proposition}
% -------------------------------------
\begin{proof}
 Clearly,  $\bot_\Sigma$ is consistent. Thus, by Proposition~\ref{coperator},
$\alg{T}^n_\prg{P}(\bot_\Sigma) \in {\bf CTAlg}_\Sigma$, for all $n\geq 0$, and then
$\alg{T}^\omega_\prg{P}(\bot_\Sigma) \in {\bf CTAlg}_\Sigma$ since for every $f/n\in
FS_\Sigma$,
  $
   (f^{\alg{T}_\prg{P}^\omega(\bot_\Sigma)}(\overline{t}))\theta  =
   (\bigcup_{n\geq 0}{f^{\alg{T}_\prg{P}^{n}(\bot_\Sigma)}(\overline{t})})\theta =
   \bigcup_{n\geq 0}{(f^{\alg{T}_\prg{P}^{n}(\bot_\Sigma)}(\overline{t}))\theta}
    \subseteq
   \bigcup_{n\geq 0}{f^{\alg{T}_\prg{P}^{n}(\bot_\Sigma)}(\overline{t}\theta)} =
   f^{\alg{T}_\prg{P}^{\omega}(\bot_\Sigma)}(\overline{t}\theta)
  $.
\end{proof}

% --------------------------------------------------------------------

\subsection{The Consistent Term Model Semantics}
\label{CM-semantics}

Now, we may define an equivalence relation based only on consistent term models.

\begin{definition}[Consistent model equivalence]
\label{consistentmodelsequivalence} For programs $\prg{P}$ and $\prg{Q}$, we define the
consistent model equivalence as
\[
    {\cal P} \equiv_{CM} {\cal Q} \Leftrightarrow_{def}
        \forall \alg{A}\in {\bf CTAlg}_{\Sigma} \cdot
                               (\prg{T_P}(\alg{A}) \sqsubseteq \alg{A}
                                \Leftrightarrow
                                \prg{T_Q}(\alg{A}) \sqsubseteq \alg{A}).
\]
\end{definition}  % .. .. .. .. .. .. .. .. .. .. .. .. .. .. .. .. ..

This equivalence is clearly weaker than the model equivalence and corresponds to the
following semantics
\[
    \cmodelsetsem{\prg{P}} = \{ \alg{M} \ | \
                    \alg{M} \mbox{ is a consistent term model of }\prg{P} \}
\]
which will be called {\it loose consistent model-theoretic
semantics}, or simply {\it consistent term model semantics}.
Obviously,
 $\ \cmodelsetsem{\prg{P}} = \modelsetsem{\prg{P}} \cap {\bf
 CTAlg}_\Sigma$, and the compositionality property of this
 semantics may be obtained in a similar way as the compositionality
 of the term model semantics.

\begin{theorem}[Compositionality of $\ \cmodelsetsem\cdot$] %... ... ..
\label{cmodelsetcompos}
For all programs $\prg{P}, \prg{Q}, \prg{P}_i, \prg{Q}_i$,
\begin{enumerate}
\item $\prg{P} \equiv_{CM} \prg{Q}$ implies $Ob(\prg{P})=Ob(\prg{Q})$.
\item $\prg{P}_i \equiv_{CM} \prg{Q}_i$ for $i=1,2$, implies
      $\prg{P}_1 \cup \prg{P}_2 \equiv_{CM} \prg{Q}_1 \cup \prg{Q}_2$.
\item $\prg{P} \equiv_{CM} \prg{Q}$ implies
      $\overline{\prg{P}}^\sigma \equiv_{CM} \overline{\prg{Q}}^\sigma$,
      for every signature $\sigma$.
\item $\prg{P} \equiv_{CM} \prg{Q}$ implies
      $\rho(\prg{P}) \equiv_{CM} \rho(\prg{Q})$, for every function symbol
      renaming $\rho$.
\end{enumerate}
Therefore, the semantics $\ \cmodelsetsem{\cdot}$ is compositional wrt $(\mathit{
Ob}, \{\cup,\overline{(\cdot)}^{\sigma},\rho(\cdot)\})$.
\end{theorem} % ......................................................
\begin{proof} We can repeat the proof of Theorem~\ref{modelsetcompos} but considering
pre-fixpoints in ${\bf CTAlg}_{\Sigma}$ and taking into account
that the least model of a program is  consistent. \end{proof}

Example~\ref{nonCompositional} also illustrates the non-compositionality of $\
\cmodelsetsem{\cdot}$ wrt the deletion operation because the programs $\prg{P}$ and $\prg{Q}$
only define functions without arguments. However, this semantics is fully abstract; to
prove this fact, we need an auxiliary result, showing how a (\textit{minimal}) program
$\prg{P}$ can be constructed from a consistent term algebra $\alg{A}$ and an element
$t\in\eval{r}{\alg{A}}{id}$ such that $\alg{A}$ is a model of $\prg{P}$ and
 $t\in\eval{r}{\alg{M}_\prg{P}}{id}$.
Proposition~\ref{minimalprogram} formalizes this idea. In order to simplify the proof of
this result, we will prove some properties about the notion of \textit{canonical rewrite
rule} already introduced in Definition~\ref{canonicalrule}.
\begin{lemma}\label{lemmacanonicalrule}
 For each canonical rewrite rule $crr(e,r)$, $\alg{T}_{\{crr(e,r)\}}$ is constant and
  if $e=f(\overline{t})$ then, for every term algebra $\alg{A}$,
 \[
   h^{\alg{T}_{\{crr(e,r)\}}(\alg{A})}(\overline{s}) = \left\{
        \begin{array}{ll}
           \bigcup_{\eta\in{\bf CSubst}}{\{\ \eval{r\eta}{\alg{A}}{id} \ | \
                                        \overline{t}\eta \sqsubseteq \overline{s} \}
                                        \cup \{\bot\}}
                           & \mbox{if } h=f, \\
           \{\bot\}        & \mbox{otherwise}
        \end{array}
        \right.
 \]
 \end{lemma}
\begin{proof}
 In fact, when we apply $\alg{T}_{\{crr(e,r)\}}$ to
a term algebra $\alg{A}$, only the interpretation of $f$ is affected. It is obvious that
  $f^{\alg{T}_{\{crr(e,r)\}}(\alg{A})}(\overline{s})$  contains
  $\bigcup_{\eta\in{\bf CSubst}}  \{\ \eval{r\eta}{\alg{A}}{id} \ | \
                                    \overline{t}\eta \sqsubseteq \overline{s} \}$.
On the other hand, if $u\in f^{\alg{T}_{\{crr(e,r)\}}(\alg{A})}(\overline{s})$ then there
exists an instantiation $f(\overline{t'}\eta') \es r\eta' \si C\eta'$ of the rule
$crr(e,r)$, with $\eta'\in{\bf CSubst}_\bot$, such that
$\overline{t}'\eta'\sqsubseteq\overline{s}$, $\alg{A}\models_{id}C\eta'$ and
$u\in\eval{r\eta'}{\alg{A}}{id}$. The definition of $C$ forces $\eta'$ to be total for
all variables of $crr(e,r)$ that do not replace occurrences of $\bot$ in $\overline{t}$.
Now, we can define the total substitution $\eta\in{\bf CSubst}$ as $X\eta = X'\eta'$ for
each variable $X$ such that $X = X'\theta_{\overline{t}}$, for any variable $X'$ and
$\theta_{\overline{t}}$ being the substitution considered in the
Definition~\ref{canonicalrule}, and $Y\eta = Y$ for all other variables $Y$. Note that
the definition of $\eta$ is correct because if $X_1\theta_{\overline{t}} =
X_2\theta_{\overline{t}}$ then $X_1\eta' = X_2\eta'$ since $X_1\bowtie X_2\in C$ and
$\alg{A}\models_{id}C\eta'$. Moreover $r\eta=r'\eta'$ and $\overline{t}\eta \sqsubseteq
\overline{t}'\eta'$, and so $\overline{t}\eta \sqsubseteq \overline{s}$ and
$u\in\eval{r\eta}{\alg{A}}{id}$.
As $\eval{r\eta}{\alg{A}}{id}$ has the same value for all algebras
$\alg{A}\in {\bf TAlg}_\bot$, $\alg{T}_{\{crr(e,r)\}}$ will be
constant. \end{proof}
\begin{proposition}
\label{minimalprogram} Let $\alg{A}\in{\bf CTAlg}_\Sigma$ be a consistent term algebra,
and $r \in {\bf Term}_{\bot}$. Then, for every $t\in\evaluation{r}{A}{id}$, a program
$\prg{R}_t$ exists such that
 $
        t\in\eval{r}{\alg{M}_{\prg{R}_{t}}}{id} \ \mbox{ and } \
                        {\alg{T}_{\prg{R}_{t}}}(\alg{A}) \sqsubseteq \alg{A}
 $.
Moreover, $\alg{T}_{\prg{R}_{t}}$ is constant.
\end{proposition}
\begin{proof}
 We will proceed by induction on the structure of $r$. We can distinguish two
base cases: $r\in {\cal V}\cup \cons{0} \cup \{ \bot \}$ and $r\in \funs{0}$.
 In the first case, $\eval{r}{\alg{A}}{id} = \ideal{r} =
\eval{r}{\alg{M}_{\prg{R}_{t}}}{id}$, for every program $\prg{R}_t$, in particular for
$\prg{R}_t = \emptyset$, and $\alg{T}_\emptyset$ is constant with
$\alg{T}_\emptyset(\alg{A}) = \bot_\Sigma \sqsubseteq \alg{A}$.
 In the second case, if $r=f$ and $\bot\neq t\in f^{\alg{A}}$ let be
$\prg{R}_t = \{ crr(f,t) \}$, by Lemma~\ref{lemmacanonicalrule}({\it 2}),
$\alg{T}_{\prg{R}_{t}}$ is constant and
 $
  f^{\alg{T}_{\prg{R}_t}(\alg{A})} =
  \bigcup_{\eta\in{\bf CSubst}}{\eval{t\eta}{\alg{A}}{id}} =
  \bigcup_{\eta\in{\bf CSubst}}{\ideal{t\eta}}
 $
(since $f$ has no arguments,). Obviously $t\in f^{\alg{T}_{\prg{R}_t}(\alg{A})}$ and
 $
  t\in f^{\alg{T}_{\prg{R}_t}(\bot_\Sigma)} \subseteq
  f^{\alg{T}^\omega_{\prg{R}_t}(\bot_\Sigma)} = f^{\alg{M}_{\prg{R}_t}} =
  \eval{r}{\alg{M}_{\prg{R}_t}}{id}
 $.
Since $\alg{A}$ is consistent, $t\eta\in f^\alg{A}\eta \subseteq (f\eta)^\alg{A} =
f^\alg{A}$ for all $\eta\in{\bf CSubst}$ and $f^{\alg{T}_{\prg{R}_t}(\alg{A})}\subseteq
f^\alg{A}$, and as the rest of function symbols are non-defined in
$\alg{T}_{\prg{R}_t}(\alg{A})$, we obtain $\alg{T}_{\prg{R}_t}(\alg{A})\sqsubseteq
\alg{A}$.

In the general case, $r=h(\overline{e})$ with $h\in \funs{n}\cup\cons{n}$ and $e_i\in{\bf
Term}_\bot$ ($i=1,\dots,n$). As $t\in\evaluation{h(\overline{e})}{A}{id}$ implies that
there exist $v_i\in\evaluation{e_i}{A}{id}$ $(i=1,\dots,n)$ such that $t\in
\function{h^\alg{A}}{v}{n}$, by applying the induction hypothesis to each pair $v_i$,
$e_i$, we have programs $\prg{R}_i$ such that
 $v_i \in\eval{e_i}{\alg{M}_{\prg{R}_i}}{id}$ with $\alg{T}_{\prg{R}_i}$ constant and
 $\alg{T}_{\prg{R}_i}(\alg{A}) \sqsubseteq \alg{A}$.
Now.
\begin{enumerate}
 \item
  If $h\in\cons{n}$ let be $\prg{R}_t=\bigcup_{i=1}^{n}{\prg{R}_i}$. As $\prg{R}_i
  \subseteq \prg{R}_t$, we have $\alg{M}_{\prg{R}_i} \sqsubseteq \alg{M}_{\prg{R}_t}$
  and, by Lemma~\ref{continuity},
  $\ \eval{e_i}{\alg{M}_{\prg{R}_i}}{id} \subseteq \eval{e_i}{\alg{M}_{\prg{R}_t}}{id}$
  what implies $v_i\in \eval{e_i}{\alg{M}_{\prg{R}_t}}{id}$ ($i=1,\dots,n$) and
  $
        \function{h^{\alg{M}_{\prg{R}_t}}}{v}{n}
                 \subseteq \bigcup_{u_i\in\ \eval{e_i}{\alg{M}_{\prg{R}_t}}{id}}
                              {\function{h^{\alg{M}_{\prg{R}_t}}}{u}{n}}
                 =  \eval{\function{h}{e}{n}}{\alg{M}_{\prg{R}_t}}{id}
                 =  \eval{r}{\alg{M}_{\prg{R}_t}}{id}
  $;
  but $t\in\function{h^\alg{A}}{v}{n}=\function{h^{\alg{M}_{\prg{R}_t}}}{v}{n}$, since
  $\function{h}{v}{n}\in{\bf CTerm}_\bot$, and so $t\in \eval{r}{\alg{M}_{\prg{R}_t}}{id}$.
  On the other hand, by Theorem~\ref{homomorphism}~({\it a}),
  $
   \alg{T}_{\prg{R}_t}(\alg{A})
                = \bigsqcup_{i=1}^{n}\alg{T}_{\prg{R}_i}(\alg{A})
                \sqsubseteq \alg{A}
  $
  and $\alg{T}_{\prg{R}_t}$ is constant.

 \item
  If $h\in\funs{n}$ let be $\prg{R}_t = (\bigcup_{i=1}^{n}{\prg{R}_i})\cup
  \{crr(h(\overline{v}),t)\}$. If $crr(h(\overline{v}),t) = h(\overline{v}')\es t\si C$;
  then, for the substitution $\theta_{\overline{v}}$ (see Definition~\ref{canonicalrule}),
  $\overline{v}=\overline{v}'\theta_{\overline{v}}$ and $C\theta_{\overline{v}}$ only
  contains joinability statements $X\bowtie X$, which are entailed by every
  term algebra, so
       $ t\in \function{h^{\alg{T}_{\{crr(h(\overline{v}),t)\}}(\bot_\Sigma)}}{v}{n}
        \subseteq \function{h^{\alg{T}_{\prg{R}_t}(\bot_\Sigma)}}{v}{n}
        \subseteq \function{h^{\alg{M}_{\prg{R}_t}}}{v}{n}
       $.
  As $\ \eval{e_i}{\alg{M}_{\prg{R}_i}}{id} \subseteq
  \eval{e_i}{\alg{M}_{\prg{R}_t}}{id}$, $v_i\in \eval{e_i}{\alg{M}_{\prg{R}_t}}{id}$
  ($i=1,\dots,n$) and $\function{h^{\alg{M}_{\prg{R}_t}}}{v}{n} \subseteq
  \eval{\function{h}{e}{n}}{\alg{M}_{\prg{R}_t}}{id} =\eval{r}{\alg{M}_{\prg{R}_t}}{id}$,
  and therefore $t\in\eval{r}{\alg{M}_{\prg{R}_t}}{id}$.
  As, by Theorem~\ref{homomorphism}~({\it a}), $\alg{T}_{\prg{R}_t}
   = (\bigsqcup_{i=1}^{n}\alg{T}_{\prg{R}_i})\sqcup \alg{T}_{crr(h(\overline{v}),t)}$,
   $\alg{T}_{\prg{R}_t}$ is constant.
  By Lemma~\ref{lemmacanonicalrule}({\it 2}),
  \[
   \function{h^{\alg{T}_{\{crr(r,t)\}}(\alg{A})}}{w}{n} =
   \bigcup_{\eta\in{\bf CSubst}}\{\ \eval{t\eta}{\alg{A}}{id}\ |\
                                 \overline{v}\eta\sqsubseteq \overline{w}\ \}
                                 \cup \{\bot\},
  \]
  $t\eta\in h^\alg{A}(\overline{v}\eta)$ for every $\eta\in{\bf CSubst}_\bot$ and
  $\alg{A}\in{\bf CTAlg}_\bot$, and $h^\alg{A}(\overline{v}\eta)\subseteq
  \function{h^\alg{A}}{w}{n}$ when $\overline{v}\eta\sqsubseteq \overline{w}$
  by the monotonicity of $h^\alg{A}$, therefore
  $
  \function{h^{\alg{T}_{\{crr(r,t)\}}(\alg{A})}}{w}{n} \subseteq
  \function{h^\alg{A}}{w}{n}
  $;
  moreover, for every function symbol $g\neq h$,
  $\function{g^{\alg{T}_{\{crr(r,t)\}}(\alg{A})}}{w}{n} = \{\bot\}
  \subseteq
  \function{h^\alg{A}}{w}{n}$.
  Therefore $\alg{T}_{\{crr(r,t)\}}(\alg{A})\sqsubseteq\alg{A}$. As by the induction
  hypothesis $\alg{T}_{\prg{R}_i}(\alg{A})\sqsubseteq\alg{A}$ ($i=1,\dots,n$), it results
  $\alg{T}_\prg{R}(\alg{A})\sqsubseteq\alg{A}$.
\end{enumerate}
\end{proof}

Now, we can obtain the full abstraction property for $\,\cmodelsetsem{\cdot}$.
% ---------------------------
\begin{theorem}[Full abstraction of $\ \cmodelsetsem{\cdot}$]
\label{fullabstraction} The semantics $\ \cmodelsetsem{\cdot}$ is fully abstract wrt
$(\mathit{Ob}$, $\{\cup,\overline{(\cdot)}^{\sigma},(\cdot)\!\setminus\!\sigma,
   \rho(\cdot)\})$
\end{theorem}
% ---------------------------
\begin{proof}
 We will prove that programs $\prg{P}$ and $\prg{Q}$ such that $\prg{P} \not\equiv_{CM}
 \prg{Q}$ are always distinguishable, so non-distinguishability of programs has to imply
 semantics equivalence. If $\prg{P} \not\equiv_{CM} \prg{Q}$ we can assume, without loss
 of generality, that there exists $\alg{A}\in{\bf CTAlg}_\Sigma$ such that $\prg{T_P}(\alg{A})
\sqsubseteq \alg{A}$ and $\prg{T_Q}(\alg{A}) \not\sqsubseteq \alg{A}$, what means that
there exist $f\in \funs{n}$ and $t_i \in {\bf CTerm}_{\bot}\ (i = 1 \dots n)$ such that
$t \in \function{f^{{\cal T_Q}(\alg{A})}}{t}{n}$ and $t \not\in
\function{f^{\alg{A}}}{t}{n}$. By the definition of ${\cal T_Q}(\alg{A})$, we have
$\function{f}{s}{n} \es r \si C \ \in \ [{\cal Q}]_{\bot} $ such that $\overline{s}
\sqsubseteq \overline{t}$, $\alg{A} \models_{id} C$, and $t\in\evaluation{r}{A}{id}$. If
$C = \{a_j\bowtie b_j\}_{j=1}^{m}$ then there exists a maximal
$l_j\in\evaluation{a_j}{A}{id}\cap\evaluation{b_j}{A}{id}$, for $j=1,\dots,m$, and we may
consider the programs $\prg{R}_t$ for $t\in \evaluation{r}{A}{id}$, $\prg{R}_{l_j}$ for
$l_j\in\evaluation{a_j}{A}{id}$ and $\prg{R}'_{l_j}$ for $l_j\in\evaluation{b_j}{A}{id}$
($j=1,\dots,m$), as in the previous proposition, and $\prg{R} = \prg{R}_t \cup
(\bigcup_{j=1}^{m}\prg{R}_{l_j}) \cup (\bigcup_{j=1}^{m}\prg{R'}_{l_j})$. Obviously,
$\alg{T}_\prg{R}(\alg{A}) \sqsubseteq \alg{A}$ and $\alg{M}_{\prg{R}_t},
\alg{M}_{\prg{R}_{l_j}}, \alg{M}_{\prg{R}'_{l_j}} \sqsubseteq \alg{M}_{\prg{R}}$. If we
define the context $C\context{\prg{X}} = \prg{X} \cup \prg{R}$ it can be proved that
$Ob(C\context{\prg{P}}) \neq Ob(C\context{\prg{Q}})$. In fact, as it will be shown, $t\in
\function{f^{\alg{M}_{C\context{\prg{Q}}}}}{t}{n}$, but $t\not\in
\function{f^{\alg{M}_{C\context{\prg{P}}}}}{t}{n}$. Note that $t\in
\function{f^{\alg{T}_\prg{Q}(\alg{M}_{\prg{R}})}}{t}{n}$ because
$t\in\evaluation{r}{M_R}{id}$ and $l_j\in \evaluation{a_j}{M_R}{id} \cap
\evaluation{b_j}{M_R}{id}$, and
 $\function{f^{\alg{T}_\prg{Q}(\alg{M}_{\prg{R}})}}{t}{n} =
 \function{f^{\alg{T}_\prg{Q}(\alg{T}_{\prg{R}}^\omega (\bot_\Sigma))}}{t}{n} \subseteq
 \function{f^{\alg{T}_{\prg{Q}\cup\prg{R}}^\omega (\bot_\Sigma)}}{t}{n}   =
 \function{f^{\alg{M}_{C\context{\prg{Q}}}}}{t}{n}
 $ because $\alg{T}_\prg{Q}, \alg{T}_\prg{R} \sqsubseteq \alg{T}_{\prg{Q}\cup\prg{R}}$, so
 $t\in \function{f^{\alg{M}_{C\context{\prg{Q}}}}}{t}{n}$.
But $\alg{T}_{C\context{\prg{P}}}^{k}(\bot_\Sigma) \sqsubseteq \alg{A}$, for all $k \geq
0$, because it is trivially true for $k=0$ and if we assume
$\alg{T}_{C\context{\prg{P}}}^{k}(\bot_\Sigma) \sqsubseteq \alg{A}$ then, by the
monotonicity of $\alg{T}_{C\context{\prg{P}}}$ and the properties of $\prg{R}$,
%we have
 $
  \alg{T}_{C\context{\prg{P}}}^{k+1}(\bot_\Sigma)
     =
        \alg{T}_{C\context{\prg{P}}}(\alg{T}_{C\context{\prg{P}}}^{k}(\bot_\Sigma))
     \sqsubseteq  \alg{T}_{C\context{\prg{P}}}(\alg{A})
     =  \alg{T}_\prg{P}(\alg{A}) \cup \alg{T}_{\prg{R}}(\alg{A})
     \sqsubseteq  \alg{A}
 $
and thus, $\alg{M}_{C\context{\prg{P}}} \sqsubseteq \alg{A}$. As
$t\notin \function{f^{\alg{A}}}{t}{n}$ also $t \notin
\function{f^{\alg{M}_{C\context{\prg{P}}}}}{t}{n}$. \end{proof}

% ====================================================================

\section{A Compositional and Fully Abstract Semantics}
\label{comfullabstractSem}

The fact that the consistent term model semantics is fully abstract but not
compositional wrt\/ the deletion of a subsignature means that this semantics is
more abstract than necessary. We need a finer semantics but not as fine as the
${\cal T}$-semantics.  One way of obtaining such a semantics is by increasing
the number of pre-fixpoints (related to the ${\cal T}$-operator) to be
considered when we compare two programs, and, in order to obtain
compositionality wrt\/ the deletion operation, we may consider the consistent
term models of all programs obtained by deleting a subsignature. With this idea
we define the following equivalence between programs

\begin{definition}[Deletion equivalence] %........................
For programs $\prg{P}$ and $\prg{Q}$, we define the deletion equivalence as
\[
  \prg{P} \equiv_{D} \prg{Q} \Leftrightarrow_{def}
     \forall \sigma\subseteq FS_{\Sigma} \ \cdot \
        (\prg{P}\setminus\sigma \equiv_{CM} \prg{Q}\setminus\sigma).
\]
\end{definition} %...................................................
This equivalence is finer than the consistent model equivalence and coarser
than the equivalence induced by the ${\cal T}$-semantics. In fact, $\prg{P}
\equiv_{D} \prg{Q}$ implies $\prg{P} \equiv_{CM} \prg{Q}$ because this
relationship coincides with $\prg{P}\setminus\sigma_{0} \equiv_{CM}
\prg{Q}\setminus\sigma_{0}$, where $\sigma_{0}$ is the empty signature. And if
$\prg{P} \equiv_{T} \prg{Q}$, or equivalently ${\cal T}_{\prg{P}}={\cal
T}_{\prg{Q}}$, it can be proved that ${\cal T}_{\prg{P}\setminus\sigma}={\cal
T}_{\prg{Q}\setminus\sigma}$, for all $\sigma\subseteq FS_{\Sigma}$, and then
$\prg{P}\setminus\sigma \equiv_{CM} \prg{Q}\setminus\sigma$, for all
$\sigma\subseteq FS_{\Sigma}$, which is $\prg{P} \equiv_{D} \prg{Q}$. The
deletion equivalence is compositional wrt all operations.

\begin{theorem}[Compositionality of $\equiv_{D}$] % ... ...
\label{dcmodelsetcompos}
For all programs $\prg{P}, \prg{Q}, \prg{P}_i, \prg{Q}_i$,
\begin{enumerate}
\item $\prg{P} \equiv_D \prg{Q}$ implies $Ob(\prg{P})=Ob(\prg{Q})$.
\item $\prg{P}_i \equiv_D \prg{Q}_i$ for $i=1,2$, implies
      $\prg{P}_1 \cup \prg{P}_2 \equiv_D \prg{Q}_1 \cup \prg{Q}_2$.
\item $\prg{P} \equiv_D \prg{Q}$ implies
      $\overline{\prg{P}}^\sigma \equiv_D \overline{\prg{Q}}^\sigma$,
      for every signature $\sigma\subseteq FS_{\Sigma}$.
\item $\prg{P} \equiv_D \prg{Q}$ implies
      $\prg{P}\setminus\sigma \equiv_D \prg{Q}\setminus\sigma$,
      for every signature $\sigma\subseteq FS_{\Sigma}$.
\item $\prg{P} \equiv_D \prg{Q}$ implies
      $\rho(\prg{P}) \equiv_D \rho(\prg{Q})$, for every function symbol
      renaming $\rho$.
\end{enumerate}
Thus, the equivalence $\equiv_D$ is compositional wrt $(\mathit{
Ob},\{\cup,\overline{(\cdot)}^{\sigma}, (\cdot)\setminus\sigma,\rho(\cdot)\})$.
\end{theorem} % ... ... ... ... ... ... ... ... ... ... ... ... ... ..
\begin{proof}

{\it 1.}
Trivial because $\prg{P} \equiv_D \prg{Q}$ implies
$\prg{P}\setminus\sigma_0 \equiv_{CM} \prg{Q}\setminus\sigma_0$,
for the empty signature $\sigma_0$, so $\prg{P}\equiv_{CM}\prg{Q}$
and $\alg{M}_\prg{P} = \alg{M}_\prg{Q}$ or $Ob(\prg{P}) =
Ob(\prg{Q})$.

{\it 2.} $\prg{P}_{i} \equiv_D \prg{Q}_{i}$ implies
$\prg{P}_{i}\setminus\sigma \equiv_{CM}
\prg{Q}_{i}\setminus\sigma$, for all $\sigma\subseteq FS_{\Sigma}$,
and $\prg{P}_{i}\setminus\sigma \equiv_{CM}
\prg{Q}_{i}\setminus\sigma$ ($i=1,2$) implies $(\prg{P}_1\setminus\sigma)
\cup (\prg{P}_2\setminus\sigma)
\equiv_{CM} (\prg{Q}_1\setminus\sigma) \cup (\prg{Q}_2\setminus\sigma)$,
by Theorem~\ref{cmodelsetcompos}({\it 2}). But, by
Proposition~\ref{deletionprop}({\it 4}),
$(\prg{P}_1\cup\prg{P}_2)\setminus\sigma =
(\prg{P}_1\setminus\sigma) \cup (\prg{P}_2\setminus\sigma)$. So
$\prg{P}_{i} \equiv_D \prg{Q}_{i}$, for $i=1,2$, implies
$(\prg{P}_1\cup\prg{P}_2)\setminus\sigma \equiv_{CM}
(\prg{Q}_1\cup\prg{Q}_2)\setminus\sigma$, for all $\sigma\subseteq
FS_{\Sigma}$, which means $\prg{P}_1 \cup \prg{P}_2 \equiv_D
\prg{Q}_1
\cup \prg{Q}_2$.

{\it 3.}
$\prg{P} \equiv_D \prg{Q}$ implies
$\prg{P}\equiv_{CM}\prg{Q}$. By Theorem~\ref{cmodelsetcompos}({\it
3}), this implies $\overline{\prg{P}}^{\sigma\setminus\sigma'}
\equiv_{CM} \overline{\prg{Q}}^{\sigma\setminus\sigma'}$,
for all signatures $\sigma,\sigma'\subseteq FS_{\Sigma}$. And, by
Proposition~\ref{deletionprop}({\it 5}),
$\overline{\prg{P}}^{\sigma\setminus\sigma'} =
\overline{\prg{P}}^\sigma\setminus\sigma'$. Thus,
$\prg{P} \equiv_D \prg{Q}$ implies
$\overline{\prg{P}}^\sigma\setminus\sigma' \equiv_{CM}
\overline{\prg{Q}}^\sigma\setminus\sigma'$, for all signature
$\sigma'\subseteq FS_{\Sigma}$, which is
$\overline{\prg{P}}^\sigma \equiv_D \overline{\prg{Q}}^\sigma$.

{\it 4.}
By definition, $\prg{P} \equiv_D \prg{Q}$ implies
$\prg{P}\setminus(\sigma\cup\sigma') \equiv_{CM}
\prg{Q}\setminus(\sigma\cup\sigma')$, for all signatures
$\sigma,\sigma'\subseteq FS_{\Sigma}$. By
Proposition~\ref{deletionprop}({\it 3}),
$\prg{P}\setminus(\sigma\cup\sigma')
= (\prg{P}\setminus\sigma)\setminus\sigma'$. Thus,
$\prg{P} \equiv_D \prg{Q}$ implies
$(\prg{P}\setminus\sigma)\setminus\sigma' \equiv_{CM}
(\prg{Q}\setminus\sigma)\setminus\sigma'$, for all
$\sigma'\subseteq FS_{\Sigma}$, which is
$\prg{P}\setminus\sigma \equiv_D \prg{Q}\setminus\sigma$.

{\it 5.} Given a function symbol renaming $\rho$ and a signature
$\sigma\subseteq FS_\Sigma$, let $\sigma_\rho$ be the signature
$\{f\in FS_\Sigma\ |\ \rho(f)\in\sigma\}$. By definition, $\prg{P}
\equiv_D \prg{Q}$ implies $\prg{P}\setminus\sigma_\rho \equiv_{CM}
\prg{Q}\setminus\sigma_\rho$. By Theorem~\ref{cmodelsetcompos}({\it
4}), this implies $\rho(\prg{P}\setminus\sigma_\rho) \equiv_{CM}
\rho(\prg{Q}\setminus\sigma_\rho)$. It can also be proved easily that
$\rho(\prg{P}\setminus\sigma_\rho) = \rho(\prg{P})\setminus\sigma$.
So, $\prg{P} \equiv_D \prg{Q}$ implies
$\rho(\prg{P})\setminus\sigma \equiv_{CM}
\rho(\prg{Q})\setminus\sigma$, for all $\sigma\subseteq FS_{\Sigma}$,
which means that $\rho(\prg{P}) \equiv_D \rho(\prg{Q})$.
\end{proof}

\begin{theorem}[Full abstraction of $\equiv_{D}$]
\label{dfullabstraction} The equivalence $\equiv_D$ is fully abstract wrt
$(\mathit{Ob}$, $\{\cup,\overline{(\cdot)}^{\sigma},(\cdot)\!\setminus\!\sigma,
   \rho(\cdot)\})$
\end{theorem}
\begin{proof} We only need prove that $\prg{P} \not\equiv_D \prg{Q}$ implies that there exists a
context where we can discriminate the observable behavior of both programs. But $\prg{P}
\not\equiv_D \prg{Q}$ implies that there exists a signature $\sigma\subseteq FS_\Sigma$
such that $\prg{P}\setminus\sigma \not\equiv_{CM} \prg{Q}\setminus\sigma$, and this
implies that there exists a context $C'\context{\prg{X}}$ such that
$Ob(C'\context{\prg{P}\setminus\sigma}) \not= Ob(C'\context{\prg{Q}\setminus\sigma})$
because the equivalence $\equiv_{CM}$ is fully abstract. Thus by considering the context
$C\context{\prg{X}} = C'\context{\prg{X}\setminus\sigma}$ we have that
$Ob(C\context{\prg{P}}) \not= Ob(C\context{\prg{Q}})$.
\end{proof}

\begin{definition}[Deletion semantics] %.......................
\label{deletionsemantics}
We define the deletion semantics of a program $\prg{P}$ as
\[
  \dcmodelsetsem{\prg{P}} = \{{\bf M}_{f/n}(\prg{P}) \mid f/n\in FS_\Sigma \},
\]
where ${\bf M}_{f/n}(\prg{P})$ is the set of all consistent term
models of the rules of $\prg{P}$ that define $f/n$.
\end{definition} % ...............................................

The deletion semantics induces the deletion equivalence.
\begin{proposition} %.............................................
\[
   \prg{P} \equiv_{D} \prg{Q} \Leftrightarrow_{def}
                  \dcmodelsetsem{\prg{P}} = \dcmodelsetsem{\prg{Q}}
\]
\end{proposition} %...............................................
\begin{proof} If $\prg{P} \equiv_{D} \prg{Q}$ then, for each $f/n\in FS_\Sigma$, we have
$\prg{P}\setminus\sigma_{f/n} \equiv_{CM} \prg{Q}\setminus\sigma_{f/n}$ for $\sigma_{f/n}
= (exp(\prg{P}) \cup exp(\prg{Q}))\setminus\{f/n\}$, which means ${\bf M}_{f/n}(\prg{P})
= {\bf M}_{f/n}(\prg{Q})$, for each $f/n\in FS_\Sigma$; thus $\ \dcmodelsetsem{\prg{P}} =
\dcmodelsetsem{\prg{Q}}$.
Reciprocally, if $\ \dcmodelsetsem{\prg{P}} =
\dcmodelsetsem{\prg{Q}}$ then, for each $\sigma\subseteq FS_\Sigma$,
$
\,\cmodelsetsem{\prg{P}\setminus\sigma}$ = $\bigcap \{{\bf M}_{f/n}(\prg{P}) \ | \ f/n\in
(exp(\prg{P}) \cup exp(\prg{Q}))\setminus\sigma\}
$
(where the intersection reduces to ${\bf CTAlg}_\Sigma$ when the
signature $(exp(\prg{P}) \cup exp(\prg{Q}))\setminus\sigma$ is
empty), and because ${\bf M}_{f/n}(\prg{P}) = {\bf
M}_{f/n}(\prg{Q})$, for all $f/n\in FS_\Sigma$, we have
$\prg{P}\setminus\sigma \equiv_{CM}
\prg{Q}\setminus\sigma$ for each $\sigma\subseteq FS_\Sigma$, and
consequently, $\prg{P} \equiv_{D} \prg{Q}$.
\end{proof}

 Thus, the deletion semantics is compositional and fully abstract wrt
 ({\it Ob},$\{\cup$, $\overline{(\cdot)}^{\sigma}$,
$(\cdot)\setminus\sigma$, $\rho(\cdot)\}$).

% ====================================================================

% ===================================================================================
\section{Introducing Hidden Symbols}\label{sec.hiddensymbols}

In this section we explore an alternative to modules with an
infinite number of rules, generated by the closure operation, that
also supports local constructor symbols. For this aim we will
consider a global or \textit{visible} signature $\Sigma$ and a set
${\cal V}$ of variable symbols together with a new set $\Omega$ of
labels that we identify with the set of module names and module
expressions. With this set we obtain a \textit{labeled signature}
$\Omega\times\Sigma = (\Omega\times DS_\Sigma, \Omega\times
FS_\Sigma)$ which we will consider as protected or non accessible
for users and writers of modules, that is, \textit{hidden}. This
signature will be only managed by the module system for internal
representation of module expressions. Pairs $(M,f)$ of
$\Omega\times\Sigma$, called \textit{labeled} symbols, will be
denoted by $M\mbox{.}f$.

As we have seen in Section~\ref{subsec.basicoperations} the
purpose of the closure of a module is to hide the definitions of
function symbols, making only their results visible. To this aim,
the rules of a module are replaced with all (possibly infinite)
approximations that can be derived from them. But we can obtain an
\textit{internal representation} of the closure operation, with a
finite number of rules, with the aid of labeled symbols, following
an idea that appears in \cite{brogi:tesis} applied to the hiding
of predicate definitions in logic programs. We go further into
this idea applying it to deal with local constructor symbols.

% -----------------------------------------------------------------------------------
\subsection{A Finite Representation of Closure}

Let ${\cal P}=<\sigma_p,\sigma_e,{\cal R}>$ a module of
\textbf{PMod}($\Sigma_\bot$) with a finite set of rules. We can
protect its rules translating them to a protected signature by
labeling all function symbols with the module's name and
introducing a bridge rule $f(\overline{X}) \es {\cal
P}\mbox{.}f(\overline{X})$ for each function symbol
$f/n\in\sigma_e$. In this way we obtain a module ${\cal P}^\ast$
in the signature
 $
 \overline{\Sigma}_\bot = (DS_{\Sigma_\bot},FS_{\Sigma_\bot} \cup
                            (\Omega\times FS_{\Sigma_\bot}))
 $
with an isolated (hidden) part ${\cal R}_H$, made up of all
translated rules, and a bridge part ${\cal R}_B$ for accessing the
isolated part, made up of all bridge rules. Obviously with this
module we can derive the same approximations, for visible function
symbols, as with $\overline{\cal P}$ in every context. We will
call these modules \textit{structured modules} to distinguish them
from \textit{plain modules} used up to now. In general, a
structured module will be a module
\[
    <\sigma_p,\sigma_e,\prg{R}_V\cup\prg{R}_B\cup\prg{R}_H>
\]
with a visible parameter signature $\sigma_p$, a visible exported
signature $\sigma_e$, and a set of rules with three ---possibly
empty--- parts, a visible part $\prg{R}_V$ made up of rules only
with function symbols in $FS_\Sigma$, a hidden part $\prg{R}_H$
made up of rules only with function symbols in $\Omega\times
FS_\Sigma$, and a bridge part $\prg{R}_B$ made up of bridge rules
$f(\overline{X}) \es {\cal P}\mbox{.}g(\overline{X})$, for any
label $\prg{P}\in\Omega$, such that each symbol ${\cal
P}\mbox{.}g$ has a definition rule in $\prg{R}_H$. Also,
$\sigma_e$ is made up of all function symbols with a definition
rule in $\prg{R}_V$ or $\prg{R}_B$, and $\sigma_p$ is made up of
all parameter function symbols which appear in $\prg{R}_V$. We
define union, deletion of functional signature and renaming in the
same way as we did in Section~\ref{subsec.basicoperations}, but we
will use deletion and renaming involving only visible signature,
and, instead of closure, we define a \textit{structured closure}
for a structured module $\prg{P} =
<\sigma_p,\sigma_e,\prg{R}_V\cup\prg{R}_B\cup\prg{R}_H>$ as the
module $\prg{P}^\ast =
<\emptyset,\sigma_e,\prg{R}^\ast_B\cup\prg{R}^\ast_H>$ obtained by
applying the renaming $\tau(\prg{P})$, that transforms each
visible function symbol $f$ of $\prg{R}_V$ and $\prg{R}_B$ into
${\cal P}\mbox{.}f$ and maintains all labeled symbols, and adding
new bridge rules corresponding to the function symbols of
$\sigma_e$. Now, we can define a representation morphism from
modular expressions made up from finite plain modules to
structured modules in the following way:
\begin{itemize}
 \item
  $\iota(\prg{P}) = \prg{P}$, for each finite plain module $\prg{P}$;
 \item
  $\iota(\prg{P}\cup\prg{Q}) = \iota(\prg{P}) \cup \iota(\prg{Q})$, for module
  expressions $\prg{P}$ and $\prg{Q}$;
 \item
   $\iota(\prg{P}\setminus\sigma) = \iota(\prg{P})\setminus\sigma$, for each module
   expression $\prg{P}$ and visible signature $\sigma$;
 \item
  $\iota(\rho(\prg{P})) = \rho(\iota(\prg{P}))$, for each module expression $\prg{P}$ and
  visible signature renaming $\rho$;
 \item
  $\iota(\overline{\prg{P}}) = (\iota(\prg{P}))^\ast$, for each module expression
  $\prg{P}$.
\end{itemize}

\begin{example}
Let \verb#OrdList# and \verb#OrdNat# be the modules defined in the
example~\ref{RenamingOrdNatList} and \ref{Importation},
respectively, and let \verb#P# be the name of the module
$\iota(\verb#OrdNat#)$. The representation of \verb#OrdList#
$\cup$
 \verb#{isnat/1->isbasetype/1}#($\overline{\verb#OrdNat#}$) will be the structured module
$\iota(\verb#OrdList#) \cup
\verb#{isnat/1->isbasetype/1}#(\verb#P#^\ast),$ with the following
aspect

 {\footnotesize
\begin{verbatim}
    <{},
     {isbasetype/1,leq/2,geq/2,insert/2},
     {   % visible rules
      insert(X,[])     -> [X]              <= isbasetype(X) >< true.
      insert(X,[Y|Ys]) -> [X|[Y|Ys]]       <= leq(X,Y) >< true.
      insert(X,[Y|Ys]) -> [Y|insert(X,Ys)] <= leq(X,Y) >< false.
         % bridge rules
      isbasetype(X) -> P.isnat(X).
      leq(X,Y) -> P.leq(X,Y).
      geq(X,Y) -> P.geq(X,Y).
         % hidden rules
      P.isnat(zero)    -> true.
      P.isnat(succ(X)) -> P.isnat(X).
      P.leq(zero,zero)       -> true.
      P.leq(zero,succ(X))    -> P.isnat(X).
      P.leq(succ(X),zero)    -> false <= P.isnat(X) >< true.
      P.leq(succ(X),succ(Y)) -> P.leq(X,Y).
      P.geq(X,Y)             -> P.leq(Y,X). } >
\end{verbatim}
 }
\end{example}

\noindent The behaviour of a structured module $\prg{P} =
<\sigma_p,\sigma_e,\prg{R}_V\cup\prg{R}_B\cup\prg{R}_H>$ wrt the
visible signature can be expressed with the aid of the algebra
transformer
\[
   \prg{U_P}\colon \mathbf{CTAlg}_{\Sigma} \to \mathbf{CTAlg}_{\Sigma}
\]
defined, for each $\alg{A}$, as
 $
  \prg{U_P}(\alg{A}) =
  \prg{T}_{\prg{R}_V\cup\prg{R}_B}(\prg{T}^\omega_{\prg{R}_H}(\bot_{\overline{\Sigma}})
    \sqcup \overline{\alg{A}}) |_\Sigma
 $,
where $\overline{\alg{A}}$ is the extension of $\alg{A}$ to an
algebra of $\mathbf{CTAlg}_{\overline{\Sigma}}$ obtained by adding
functions $\prg{P}\mbox{.}f^{\overline{\alg{A}}}$ defined as
$\prg{P}\mbox{.}f^{\overline{\alg{A}}}(\overline{t}) =
\ideal{\bot}$, for each $f/n \in FS_\Sigma$ and $\prg{P}\in
\Omega$, and $\alg{B}|_\Sigma$ means the reduct of the algebra
$\alg{B}\in \mathbf{CTAlg}_{\overline{\Sigma}}$ obtained by
forgetting all functions denoting labeled function symbols. In
this expression, $\prg{T}^\omega_{\prg{R}_H}(
\bot_{\overline{\Sigma}})$ represents all the information which
can be obtained from the hidden rules; this information is added
to the extended algebra because this information has to be
disponible for the immediate consequences operator corresponding
to the visible and bridge rules in order to obtain the
approximations for the visible functions. The relationship, at the
semantical level, between program modules and structured modules
is given in the following theorem.
\begin{theorem}
 For each modular expression $\prg{E}$, made up from finite plain programs, and its
 implementation $\iota(\prg{E})$, we have $\prg{T_E} =\prg{U}_{\iota(\prg{E})}$.
\end{theorem}
\begin{proof}
 This theorem can be proved by induction on the structure of $\prg{E}$.

 (1) If $\prg{E}$ is a simple expression (a module name) then $\iota(\prg{E})_B = \emptyset
 = \iota(\prg{E})_H$ and
  $ \alg{U}_{\iota(\prg{E})} = \alg{T}_{\iota(\prg{E})_V}(\overline{\alg{A}})|_\Sigma =
   \alg{T_E}(\alg{A})$
 because $\alg{T}_{\iota(\prg{E})_V}$ neither uses nor produces any information about
 labeled signature.

 (2) If $\prg{E} = \prg{P}\cup\prg{Q}$ and $\prg{T_P} = \prg{U_{\iota(P)}}$ and
 $\prg{T_Q} = \prg{U_{\iota(Q)}}$, by {\it Theorem~\ref{homomorphism}}~({\it a}) we have
 $\prg{T_{P\cup Q}}(\alg{A}) = \prg{T_P}(\alg{A}) \sqcup \prg{T_Q}(\alg{A}) =
    \prg{U_{\iota(P)}}(\alg{A}) \sqcup  \prg{U_{\iota(Q)}}(\alg{A})$.
 Also, by definition of $\iota$, $\prg{U_{\iota(P\cup Q)}} =
 \prg{U_{\iota(P)\cup\iota(Q)}}$ and
 \[
    \prg{U_{\iota(P)\cup\iota(Q)}}(\alg{A}) =
    \prg{T}_{(\iota(\prg{P})_V \cup \iota(\prg{P})_B) \cup (\iota(\prg{Q})_V \cup \iota(\prg{Q})_B)}
    (\prg{T}^\omega_{\iota(\prg{P})_H \cup \iota(\prg{Q})_H}(\bot_{\overline{\Sigma}}) \cup
    \overline{\alg{A}})|_\Sigma\mbox{.}
 \]
 By {\it Theorem~\ref{homomorphism}}~({\it a}) and taken into account that $\iota(\prg{P})_H$
 contains all possible rules about its (labeled) function symbols and
 $\prg{T}^\omega_{\iota(\prg{Q})_H}(\bot_{\overline{\Sigma}})$ only contains relevant
 information about function
 symbols in $\iota(\prg{Q})_H$, the expession above is equal to
 \[
    \prg{T}_{(\iota(\prg{P})_V \cup \iota(\prg{P})_B)}
    (\prg{T}^\omega_{\iota(\prg{P})_H}(\bot_{\overline{\Sigma}}) \sqcup
    \overline{\alg{A}})|_\Sigma  \,\sqcup\,
    \prg{T}_{(\iota(\prg{Q})_V \cup \iota(\prg{Q})_B)}
    (\prg{T}^\omega_{\iota(\prg{Q})_H}(\bot_{\overline{\Sigma}}) \sqcup
    \overline{\alg{A}})|_\Sigma
 \]
 that is, $\prg{U_{\iota(P)}}(\alg{A}) \sqcup  \prg{U_{\iota(Q)}}(\alg{A})$.

 (3) If $\prg{E} = \prg{P}/\sigma$, $\sigma$ is a subsignature of visible function
 symbols, and $\prg{T_P} = \prg{U_{\iota(P)}}$, by {\it Theorem~\ref{homomorphism}}~({\it c}),
 $ \prg{T_{P\setminus \sigma}} = \prg{T_{P}} \sqcap {\Bbb T}_{exp(\prg{P}) \setminus \sigma}
    = \prg{U_{\iota(P)}} \sqcap {\Bbb T}_{exp(\prg{P}) \setminus \sigma}$, and by the
 definition of $\iota$, $\prg{U_{\iota(P\setminus\sigma)}(A)} =
 \prg{U_{\iota(P)\setminus\sigma}(A)}$, but again by {\it Theorem~\ref{homomorphism}}~({\it
 c}),
 $$
    \prg{T}_{(\iota(\prg{P})_V\cup\iota(\prg{P})_B)\setminus\sigma}
      =
     \prg{T}_{\iota(\prg{P})_V\cup\iota(\prg{P})_B} \sqcap
     {\Bbb T}_{exp(\iota(\prg{P})_V\cup\iota(\prg{P})_B)\setminus\sigma}
     =
     \prg{T}_{\iota(\prg{P})_V\cup\iota(\prg{P})_B} \sqcap
     {\Bbb T}_{exp(\prg{P})\setminus\sigma},
 $$
 because $exp(\prg{P}) = exp(\iota(\prg{P})_V\cup\iota(\prg{P})_B)$. So,
 $\prg{U_{\iota(P\setminus\sigma)}(A)} =
 \prg{U_{\iota(P)}} \sqcap {\Bbb T}_{exp(\prg{P}) \setminus \sigma}$.

 (4) If $\prg{E} = \rho(\prg{P})$, $\rho$ is a visible signature renaming, and
 $\prg{T_P} = \prg{U_{\iota(P)}}$, by {\it Theorem~\ref{homomorphism}}~({\it d}),
 $\prg{T_{\rho(P)}} = \prg{T}_{\rho^{-1}} {\footnotesize \circ} \prg{T_P}
 {\footnotesize \circ} \prg{T}_{\rho} = \prg{T}_{\rho^{-1}} {\footnotesize \circ}
 \prg{U_{\iota(P)}} {\footnotesize \circ}  \prg{T}_{\rho}$. Also, by definition of
 $\iota$,
 \[
  \prg{U_{\iota(\rho(P))}}(\alg{A}) = \prg{U_{\rho(\iota(P))}}(\alg{A}) =
  \prg{T}_{\rho(\iota(\prg{P})_V\cup\iota(\prg{P})_B)}
  (\prg{T}^\omega_{\iota(\prg{P})_H}(\bot_{\overline{\Sigma}}) \sqcup \overline{\alg{A}})
  |_\Sigma
 \]
 and, again by {\it Theorem~\ref{homomorphism}}~({\it d}), the above expression is equal
 to
 \[
  \prg{T}_{\rho^{-1}} (\prg{T}_{\iota(\prg{P})_V\cup\iota(\prg{P})_B}(\prg{T}_{\rho}(
  \prg{T}^\omega_{\iota(\prg{P})_H}(\bot_{\overline{\Sigma}}) \sqcup \overline{\alg{A}})))
  |_\Sigma,
 \]
 but $\prg{T}_{\rho}$ and $\prg{T}_{\rho^{-1}}$ only modify the interpretation of visible
 function symbols and $\prg{T}_{\rho}(\overline{\alg{A}}) =
 \overline{\prg{T}_{\rho}(\alg{A})}$ and $\prg{T}_{\rho^{-1}}(\alg{B})|_{\Sigma} =
 \prg{T}_{\rho^{-1}}(\alg{B}|_{\Sigma})$ for $\alg{B}\in
 \mathbf{CTAlg}_{\overline{\Sigma}}$.
 So, the above expression is equal to
 $
   \prg{T}_{\rho^{-1}}(\prg{T}_{\iota(\prg{P})_V\cup\iota(\prg{P})_B}(
   \prg{T}^\omega_{\iota(\prg{P})_H}(\bot_{\overline{\Sigma}})
   \sqcup \overline{\prg{T}_{\rho}(\alg{A})})) |_\Sigma =
   \prg{T}_{\rho^{-1}}(\prg{U_{\iota(P)}}(\prg{T}_{\rho}(\alg{A}))) \mbox{.}
 $

 (5) If $\prg{E} = \overline{\prg{P}}$ and $\prg{T_P} = \prg{U_{\iota(P)}}$, by
 {\it Theorem~\ref{homomorphism}}~({\it b}) $\prg{T_{\overline{P}}}$ applies every term
 algebra into $\prg{T^\omega_P}(\bot_\Sigma) = \prg{U^\omega_{\iota(P)}}(\bot_\Sigma)$,
 and by the definition of $\iota$, $\prg{U_{\iota(\overline{P})}} =
 \prg{U_{\iota(P)^\ast}}$; so, we have to prove
 $\prg{U_{\iota(P)^\ast}}(\alg{A}) = \prg{U^\omega_{\iota(P)}}(\bot_\Sigma)$
 for all $\alg{A}\in \mathbf{CTAlg}_\Sigma$.
 Let
 $\iota(\prg{P}) = <\sigma_p,\sigma_e,\prg{R}_V \cup \prg{R}_B \cup \prg{R}_H>$ and
 $\iota(\prg{P})^\ast = <\emptyset,\sigma_e,\prg{R}^\ast_B \cup\prg{R}^\ast_H>$, where
 $\prg{R}^\ast_B = \{f(\overline{X}) \to \prg{P}\mbox{.}f(\overline{X})\,|\, f/n\in \sigma_e\}$
 and $\prg{R}^\ast_H = \tau(\prg{P})(\prg{R}_V \cup \prg{R}_B)\cup \prg{R}_H$. For all
 $\alg{A}\in \mathbf{CTAlg}_\Sigma$,
 \[
   \prg{U_{\iota(P)^\ast}} =
   \prg{T}_{\prg{R}^\ast_B}(\prg{T}^\omega_{\prg{R}^\ast_H}
     (\bot_{\overline{\Sigma}}) \sqcup \overline{\alg{A}}) |_\Sigma =
   \prg{T}_{\prg{R}^\ast_B}(\prg{T}^\omega_{\prg{R}^\ast_H}
     (\bot_{\overline{\Sigma}})) |_\Sigma,
 \]
 because $\prg{T}_{\prg{R}^\ast_B}$ only uses information about labeled function symbols
 and $\overline{\alg{A}}$ has no information about such symbols, and
 \[
    \prg{U^\omega_{\iota(P)}}(\bot_\Sigma) =
    (\prg{T}_{\prg{R}_V \cup \prg{R}_B}\circ (Id \sqcup \prg{T}_{\overline{\prg{R}_H}}))^\omega
    (\bot_\Sigma) |_\Sigma,
 \]
 and, as $\prg{T}_{\prg{R}^\ast_B}$ and $\prg{T}_{\prg{R}_V \cup \prg{R}_B}$ only produce
 information about symbol functions of $\sigma_e$, we only have to prove
 $\prg{T}_{\prg{R}^\ast_B}(\prg{T}^\omega_{\prg{R}^\ast_H}(\bot_{\overline{\Sigma}})) =
  (\prg{T}_{\prg{R}_V \cup \prg{R}_B}\circ (Id \sqcup \prg{T}_{\overline{\prg{R}_H}}))^\omega
    (\bot_\Sigma)$.
 But it can be proved that
 $\prg{T}_{\prg{R}^\ast_B}(\prg{T}^i_{\prg{R}^\ast_H}(\bot_{\overline{\Sigma}})) \sqsubseteq
 (\prg{T}_{\prg{R}_V \cup \prg{R}_B}\circ (Id \sqcup \prg{T}_{\overline{\prg{R}_H}}))^i
 (\bot_\Sigma) \sqsubseteq
 \prg{T}_{\prg{R}^\ast_B}(\prg{T}^\omega_{\prg{R}^\ast_H}(\bot_{\overline{\Sigma}}))$
 for all $i \geq 0$.
 \end{proof}

 From this theorem we obtain that for two equivalent module expressions $\prg{P}$ and
$\prg{Q}$ (i.e. $\prg{P}$ and $\prg{Q}$ have the same components
but, possibly, different expressions with the operations),
$\prg{U}_{\iota(\prg{P})} = \prg{U}_{\iota(\prg{Q})}$ although it
is possible that $\iota(\prg{P})$ differs from $\iota(\prg{Q})$
due to the occurrence of closure operations. Also, the models of a
program module $\prg{P}$ will be the pre-fixpoints of
$\prg{U}_{\iota(\prg{P})}$ and we can define the visible semantics
of structured modules based on this operator. In particular we
obtain the deletion semantics by considering, for each structured
module $\prg{P} =
<\sigma_p,\sigma_e,\prg{R}_V\cup\prg{R}_B\cup\prg{R}_H>$, the
indexed family of sets of pre-fixpoints of
$\prg{U}_{\prg{P}\setminus(\sigma_e\setminus f)}$ for each $f/n\in
\sigma_e$.

% -----------------------------------------------------------------------------------
\subsection{Local Constructor Symbols}

To simplify the theoretical study of programs composition in
CRWL-programming, and to capture the idea of module as open
program, we have assumed that constructor symbols are common to
all programs. However, as it was discussed in
Section~\ref{modules}, this assumption prevents to hide
constructor symbols, what is not acceptable from a practical point
of view.

We can hide constructor symbols by labeling them as we have done
with function symbols to protect them against user manipulations.
Labeled constructor symbols can only be manipulated in the
internal representation of the closure of the module corresponding
to their label. Outside this module, function symbols defined on
labeled constructor symbols can only be applied to variable
symbols or to other function applications that can be reduced to
this labeled constructor symbols. To realize this idea we only
need to modify our closure implementation extending it to manage
constructor symbols also. So, we define \textit{closure hiding a
subsignature $C$ of constructor symbols} for a module $\prg{P}$ as
a (non plain) module $\overline{\prg{P}}_C$ such that
$\iota(\overline{\prg{P}}_C) = \prg{P}^\ast_C$ where
$\prg{P}^\ast_C$ is obtained as $\prg{P}^\ast$ but now the
renaming $\tau(\prg{P})$ also transforms each visible constructor
symbol $c$ of $C$ into ${\cal P}\mbox{.}c$.

\begin{example}
Let us suppose a module \verb#LNat# for lists of natural numbers
which exports the function symbols \verb#isnat/1#, \verb#_<_/2#
and \verb#_++_/2#, and consider the following module for binary
search trees of natural numbers where tree constructors
\verb#nil/0# and \verb#mktree/3# are used.

{\footnotesize
\begin{verbatim}
  BST =
  <{isnat/1, _<_/2, _++_/2}, {empty/0, insert/2, inorder/1},
   {empty -> nil .
    insert(N,nil) -> mktree(N,nil,nil) <= isnat(N) >< true .
    insert(N,mktree(M,T1,T2)) -> mktree(M,T1,T2) <= N >< M, isnat(N) >< true .
    insert(N,mktree(M,T1,T2)) -> mktree(M,insert(N,T1),T2) <= N<M >< true .
    insert(N,mktree(M,T1,T2)) -> mktree(M,T1,insert(N,T2)) <= M<N >< true .
    inorder(nil) -> [] .
    inorder(mktree(M,T1,T2)) -> inorder(T1)++[M|inorder(T2)] <= isnat(M) >< true .}>
\end{verbatim}
}

\noindent We may hide the tree constructors by considering
$\overline{(\verb#LNat#\cup \verb#BST#)}_{\{nil,mktree\}}$. This
module will have the following representation:

{\footnotesize
\begin{verbatim}
  <{}, {isnat/1, _<_/2, _++_/2, empty/0, insert/2, inorder/1},
   { ...                                        % bridge rules of LNat
    empty -> BST.empty .                        % bridge rules of BST
    insert(N,T1) -> BST.insert(N,T1) .
    inorder(T1) -> BST.inorder(T1) .

    ...                                         % hidden part of LNat
    BST.empty -> BST.nil .                      % hidden part of BST
    BST.insert(N,BST.nil) -> BST.mktree(N,BST.nil,BST.nil)
                                        <= BST.isnat(N) >< true .
    BST.insert(N,BST.mktree(M,T1,T2)) -> BST.mktree(M,T1,T2)
                                        <= N >< M, BST.isnat(N) >< true .
    BST.insert(N,BST.mktree(M,T1,T2)) -> BST.mktree(M,BST.insert(N,T1),T2)
                                        <= N NBST.< M >< true .
    BST.insert(N,BST.mktree(M,T1,T2)) -> BST.mktree(M,T1,BST.insert(N,T2))
                                        <= M MBST.< N >< true .
    BST.inorder(BST.nil) -> [] .
    BST.inorder(BST.mktree(M,T1,T2)) -> BST.inorder(T1) BST.++ [M|BST.inorder(T2)]
                                        <= BST.isnat(M) >< true .}>
\end{verbatim}
}

 \noindent
And we can use this module, without access to hidden constructor
symbols, by only using the exported signature and visible
constructor symbols, as in the following module for sorting lists:

{\footnotesize
\begin{verbatim}
 LSort =
 <{empty/0, insert/2, inorder/1},
  {listTotree/1, lsort/1},
  {listTotree([]) -> empty .
   listTotree([N|L]) -> insert(N,listTotree(L)) .
   lsort(L) -> inorder(listTotree(L)) .} >
\end{verbatim}
}

\noindent
 that has to be joined to $\overline{(\verb#LNat#\cup\verb#BST#)}_{\{nil, mktree\}}$ to obtain
 $\overline{(\verb#LNat#\cup\verb#BST#)}_{\{nil, mktree\}} \cup \verb#LSort#$.
\end{example}

The behaviour of a structured module $\prg{P} =
<\sigma_p,\sigma_e,\prg{R}_V\cup\prg{R}_B\cup\prg{R}_H>$ with
hidden constructor symbols wrt the visible signature can be
expressed with the aid of the algebra transformer
\[
   \prg{U_P}\colon \mathbf{CTAlg}_{\Sigma} \to \mathbf{CTAlg}_{\Sigma}
\]
defined for each $\alg{A}$ as
 $
  \prg{U_P}(\alg{A}) =
  \prg{T}_{\prg{R}_V\cup\prg{R}_B}(\prg{T}^\omega_{\prg{R}_H}(\bot_{\overline{\Sigma}})
    \sqcup \overline{\alg{A}}) |_\Sigma
 $,
where now $\overline{\alg{A}}$ is the extension of $\alg{A}$ to an
algebra of $\mathbf{CTAlg}_{\Omega\times\Sigma}$ obtained by
adding functions $\prg{P}\mbox{.}f^{\overline{\alg{A}}}$, defined
as $\prg{P}\mbox{.}f^{\overline{\alg{A}}}(\overline{t}) =
\ideal{\bot}$, for each $f/n \in FS_\Sigma$ and $\prg{P}\in
\Omega$, and defining $f^{\overline{\alg{A}}}(\overline{t}) =
f^{\alg{A}}(\overline{t}^\ast)$ where tuple $\overline{t}^\ast$ is
obtained from $\overline{t}$ by changing each term beginning with
a labeled constructor term for $\bot$, for each $f/n \in
FS_\Sigma$, and $\alg{B}|_\Sigma$ means the reduct of the algebra
$\alg{B}\in \mathbf{CTAlg}_{\Omega\times\Sigma}$ obtained by
restricting the carrier to $\mathbf{CTerm}_\bot$ and forgetting
all functions denoting labeled function symbols.

Obviously, the representation of the closure wrt the functional
signature is a particular case of closure hiding a set of
constructor symbols when this set is empty.

% ======================================================================
\section{Discussion}
\label{discussion}
Research in component-based software development is currently
becoming a very active area for the logic programming community. In
fact, we can find several proposals in the field of computational
logic for dealing with the design and development of large software
systems. Other related fields, like functional-logic programming
are now proving that the integration of logic variables and
functions may increase the expressive power of a programming
language. A number of attempts are being made in this direction
\cite{funLog,curry} to achieve a consensus on the characteristics a
functional-logic language has to present.

The current work tries to contribute to all these efforts by presenting a notion of
module in the context of functional-logic programming, and by providing a number of
operations (satisfying some expected algebraic properties) expressive enough to model
typical modularization issues like export/import relationships, hiding information,
inheritance, and a sort of abstraction. We have chosen the Constructor-based Conditional
Rewriting Logic \cite{mario:jlp} to develop our proposal and, in this context, we have
explored a rather wide range of semantics for program modules and we have studied some of
their relevant properties, in particular, those concerning compositionality and full
abstraction wrt the observation function $Ob(\prg{P})=\alg{M_{P}}$ and the set
$\{\cup,\overline{(\cdot)}^{\sigma},(\cdot)\!\setminus\!\sigma,\rho(\cdot)\}$ of module
operations.
%
%\begin{table}
% \caption{Semantics and Equivalence Relationships}
% \label{semantics:equivalence}
% \begin{tabular}{lllc}
% %{| p{5cm} || p{1.8cm} | p{1.8cm} | p{1.4cm} |}
% \hline\hline
%                       & Semantics             & Equivalence       & Definition \\
% \hline
% Least Model Semantics & $\ \ \modelsem\cdot$
%                       &\hspace{4mm} $\equiv_{LM}$ &\ref{model:semantics} \\
% \hline
% $\alg{T}$-Semantics   & $\ \ \semantics\cdot$
%                       &\hspace{4mm} $\equiv_{T}$  &\ref{Tsemantics} \\
% \hline
% Term Model Semantics  & $\ \ \modelsetsem\cdot$
%                       &\hspace{4mm} $\equiv_{M}$  &\ref{modelsequivalence} \\
% \hline
% Consistent Term Model Semantics   & $\ \ \cmodelsetsem\cdot$
%                       &\hspace{4mm} $\equiv_{CM}$ & \ref{consistentmodelsequivalence}\\
% \hline
% ($\alg{T}\sqcup Id$)-Semantics & $\ \ \tidsemantics\cdot$
%                       &\hspace{4mm}$\equiv_{T\sqcup Id}$ & --- \\
% \hline
% Deletion Semantics    & $\ \ \dcmodelsetsem\cdot$
%                       &\hspace{4mm} $\equiv_{D}$   & \ref{deletionsemantics}\\
% \hline \hline
% \end{tabular}
%\end{table}
%
Although these features are interesting enough from a theoretical point of
view, they present a special significance when module reusing, module refining
or module transforming are involved. The least model semantics,
$\modelsem\cdot$, is a fully abstract semantics, which is only compositional
wrt $\{\overline{(\cdot)}^{\sigma},\rho(\cdot)\}$, but only for injective
function renamings $\rho$. On the contrary, the $\alg{T}$-semantics,
$\semantics\cdot$, is compositional (wrt all operations), but is not fully
abstract. The third proposal, the loose model-theoretic semantics,
$\modelsetsem\cdot$, is also compositional (except for the deletion operation),
although the full abstraction property is not satisfied. A fully abstract
semantics,\ $\cmodelsetsem\cdot$, may be obtained by considering a consistency
property on term algebras, which is also compositional wrt the union, closure
and renaming operations. To recover the compositionality wrt deletion we need a
finer semantics able to capture the ``independent'' meaning of each function in
a module; this is the case of the deletion semantics,\ $\dcmodelsetsem\cdot$,
which still is fully abstract and compositional wrt all operations. We have
also studied the ($\alg{T}\sqcup Id$)-semantics,\ $\tidsemantics\cdot$, but we
have not included this study in this paper because it exhibits the same
properties as the $\alg{T}$-semantics. Table~\ref{cfa} summarizes the
properties satisfied by each one of the analyzed semantics.
%
%$$
%  \prg{P} \equiv_{T} \prg{Q} \ \Rightarrow\ \prg{P} \equiv_{T\sqcup Id} \prg{Q}
%                             \ \Rightarrow\ \prg{P} \equiv_{D}\ \prg{Q}
%                             \ \Rightarrow\ \prg{P} \equiv_{CM} \prg{Q}
%                             \ \Rightarrow\ \prg{P} \equiv_{LM} \prg{Q}.
%$$
%\begin{figure}
%\setlength{\unitlength}{0.0080in}%
%\centering
%\begin{picture}(300,160)(60,580)
%\thinlines
%\put(70,580){\makebox(280,150)[tr]{$\equiv_{LM}$}}
%\put(70,580){\makebox(220,120)[tr]{$\equiv_{CM}$}}
%\put(70,580){\makebox(160,90)[tr]{$\equiv_D$}}
%\put(70,580){\makebox(100,60)[tr]{$\equiv_{T\sqcup Id}$}}
%\put(70,580){\makebox(50,30)[tr]{$\equiv_T$}}
%
%\put(212,652){\oval(300,160){}}
%\put(184,639){\oval(240,130){}}
%\put(156,626){\oval(180,100){}}
%\put(128,613){\oval(120,70){}}
%\put(105,605){\oval(60,40){}}
%\end{picture}
%$$
%  \equiv_{T} \ \longrightarrow\  \equiv_{T\sqcup Id}
%             \ \longrightarrow\  \equiv_{D}\
%             \ \longrightarrow\  \equiv_{CM}
%             \ \longrightarrow\  \equiv_{LM}
%$$
%$$
%  \begin{array}{rcl}
%          &  \equiv_T            &    \\
%\swarrow  &                      & \searrow \\
%\equiv_{T\sqcup Id}  &                      & \equiv_{D} \\
%\searrow  &                      & \swarrow \\
%          & \equiv_{CM}          & \\
%          &  \downarrow            & \\
%          & \equiv_{LM}          &
%  \end{array}
%$$
%\caption{The Semantics Hierarchy}
%\label{semanticsHierarchy}
%\end{figure}
 It is possible to establish a semantics hierarchy ranging from the
model-theoretic semantics to the $\alg{T}$-semantics on the basis of the following order
for the equivalence relationships induced by these semantics
$$
   \equiv_{T}\, \sqsubseteq\, \equiv_{T\sqcup Id}\, \sqsubseteq\, \equiv_{D}\,
   \sqsubseteq\,
    \equiv_{CM}\, \sqsubseteq\, \equiv_{C}
$$
where they are ordered upon their strength. The $\alg{T}$-equivalence relation,
$\equiv_{T}$, is the strongest one, and it is contained obviously into the
($\alg{T}\sqcup Id$)-equivalence relation, $\equiv_{T\sqcup Id}$. Taking into account
that this equivalence relation is compositional but not fully abstract, it will be
contained in $\equiv_{D}$, which is also contained in the consistent term-model
equivalence, $\equiv_{CM}$. Obviously, the least term-model equivalence, $\equiv_{LM}$,
is the weakest one.

In order to establish some conclusions about the compositionality and the full
abstraction of all these semantics, we are going to discuss the information
exhibited in Table~\ref{cfa}. In this table, we can observe a sort of
dependency between fulfilling compositionality/full abstraction and the
strength of the equivalence relationship defined by the semantics, in such a
way that the strongest ones are compositional whereas the weakest ones are
fully abstract. The best semantics must be an intermediate semantics satisfying
both properties; in our case, the semantics \ $\dcmodelsetsem\cdot$. A similar
study was already made by Brogi in \cite{brogi:TCS} in the field of logic
programming, but he did not deal with variables, and avoided the complexity
inherent to the non-ground term algebras. Another difference (apart from the
context) with respect to the current work is the set of operations we are
considering, which does not coincide with the set of inter-module operations
defined by Brogi. One of the most significative operations described by him is
the intersection of programs. This operation makes the $(\alg{T}\sqcup
Id$)-semantics compositional and fully abstract in a logic programming context.
However, the difficult justification of this operation in our framework (the
functional-logic programming paradigm) has inclined us to think in an
alternative: the deletion operation. We believe that this operation is more
natural (as a composing mechanism) than program intersection. This has an
inconvenience: the ($\alg{T}\sqcup Id$)-semantics is not fully abstract
(although it is compositional) wrt our operations. In fact, the intersection of
programs is a very powerful tool to distinguish programs (more than the
deletion operation), and it can be used to delete a single rule, whereas our
deletion operation only can be used to delete a whole set of rules defining a
function. Nevertheless, we have found a fully abstract and compositional
semantics, also for the deletion operation, which completes the results
provided by this work. \\

\begin{table}
\caption{Compositionality (C) and Full Abstraction (FA)} \label{cfa}
  \begin{tabular}{l cc cc cc cc}
  \hline\hline
   & \multicolumn{2}{c}{\makebox[2cm]{$\cup,\overline{(\cdot)}^{\sigma},(\cdot)\!\setminus\!\sigma,\rho(\cdot)$}}
   & \multicolumn{2}{c}{\makebox[2cm]{$\cup,\overline{(\cdot)}^{\sigma},\rho(\cdot)$}}
   & \multicolumn{2}{c}{\makebox[2cm]{$\overline{(\cdot)}^{\sigma},\rho(\cdot)$}}
   & \multicolumn{2}{c}{\makebox[2cm]{$\overline{(\cdot)}^{\sigma}$}} \\

%\hline
%   & \makebox[1.5cm]{\tiny FA} & \makebox[1.5cm]{\tiny C}
%   & \makebox[1.5cm]{\tiny FA} & \makebox[1.5cm]{\tiny C}
%   & \makebox[1.5cm]{\tiny FA} & \makebox[1.5cm]{\tiny C}\\
%\hline\hline
%$\semantics\cdot$     &         & $\bullet$  &         & $\bullet$
%&         & $\bullet$ \\ \hline
%$\tidsemantics\cdot$  &         & $\bullet$  &         & $\bullet$
%&         & $\bullet$ \\ \hline
%$\dcmodelsetsem\cdot$   & $\circ$    & $\bullet$           &         &
%$\bullet$ & & $\bullet$ \\ \hline $\cmodelsetsem\cdot$  & $\circ$ &
%& $\circ$ & $\bullet$  &         & $\bullet$ \\ \hline
%$\modelsem\cdot$ & $\circ$ &            & $\circ$ & & $\circ$ &
%$\bullet$ \\ \hline
  \hline
  $\semantics\cdot$    & \makebox[1cm]{} & \makebox[1cm]{C}
                       & \makebox[1cm]{} & \makebox[1cm]{C}
                       & \makebox[1cm]{} & \makebox[1cm]{C}
                       & \makebox[1cm]{} & \makebox[1cm]{C} \\
  \hline
  $\tidsemantics\cdot$   &    & C &    & C &    & C &    & C \\ \hline
  $\dcmodelsetsem\cdot$  &\hspace{4mm} FA & C &    & C &    & C &    & C \\ \hline
  $\cmodelsetsem\cdot$   &\hspace{4mm} FA &   &\hspace{4mm} FA & C &    & C &    & C \\ \hline
  $\modelsem\cdot$       &\hspace{4mm} FA &   &\hspace{4mm} FA &   &\hspace{4mm} FA &   &\hspace{4mm} FA & C \\ \hline\hline
  \end{tabular}
\end{table}

% $$$$$$$$$$$$$$$$$$$$$$$$$$$$$$$ 7 $$$$$$$$$$$$$$$$$$$$$$$$$$$$$$$$$$$

\section*{Acknowledgments}
We would like to thank Mario Rodr\'{\i}guez Artalejo and Ana Gil
Luezas for their helpful comments and suggestions on the initial
versions of this work. We would also like to thank Narciso
Mart\'{\i}-Oliet for their insightful comments and suggestions,
that greatly helped us improving the quality and presentation of
the paper. This work has been partially supported by the Spanish
project TIC98-0445-C03-03.

% Bibliography $$$$$$$$$$$$$$$$$$$$$$$$$$$$$$$$$$$$$$$$$$$$$$$$$$$$$$$$

\end{document}